\documentclass[11pt,preprint,aps,showkeys,onecolumn]{revtex4} 
\usepackage{amsmath}
\usepackage{amsfonts}
\usepackage{amssymb}
\usepackage{natbib}
\usepackage{graphicx}
\usepackage{epstopdf}
\usepackage{subcaption}
\usepackage{mathrsfs}
\usepackage{dcolumn}
\usepackage{bm}
\usepackage[section]{placeins}
\usepackage{hyperref}
\usepackage{xcolor}
\usepackage{tensor}
\begin{document}
\title{Poisson spot and wave scattering of scalar fields by conformal anomaly black holes:
probing the near-horizon geometry}
\author{Zeming Zhuang$^{1}$}
\email{zhuangzm$@$stu.ujn.edu.cn}
\author{Hongsheng Zhang$^{1}$}
\email{corresponding author: sps\_zhanghs$@$ujn.edu.cn}
\affiliation{1.School of Physics and Technology, University of Jinan, 336 West Road of Nan Xinzhuang, Jinan, Shandong 250022, China}
\author{Zong-Kuan Guo$^{2}$}
\email{guozk@itp.ac.cn }
\affiliation{2.Institute of Theoretical Physics, Chinese Academy of Sciences, Beijing 100190, China}
 \author{Rong-Gen Cai$^{3,4}$}
\email{cairg@itp.ac.cn }
\affiliation{3.Zhejiang Key Laboratory of Extreme Universe, Ningbo University, Ningbo 315211, China}
\affiliation{4.Institute of Fundamental Physics and Quantum Technology, Ningbo University, Ningbo 315211, China}

\begin{abstract}

We investigate the scattering and diffraction of scalar plane waves by static conformal anomaly black
holes. We precisely examine the truncation requirements for the partial-wave series (PWS) method at finite
distances and compute the full waveform diagrams of scalar waves. The waveforms display a clear diffraction
pattern: the field on the negative $z$-axis reduces to the incident plane wave, whereas on the positive $z$-axis a
distorted plane wave coexists with a pronounced scattered spherical wave, and a bright Poisson spot appears
on the axis at $\theta = 0$, surrounded by concentric alternating bright and dark diffraction rings in the paraxial
region. We further obtain the intensity of the on-axis Poisson spot as a function of the radial coordinate.
The intensity is most sensitive to the spacetime geometry in the immediate vicinity of the black hole: very
close to the horizon, the on-axis intensity of the conformal anomaly black hole differs markedly from that
of the RN black hole, whereas in the far-field region the two intensity curves converge. We trace this near-
horizon difference to the metric functions in that region: different values of $f'(r_+)$ make the same radial
interval correspond to different phase increments, compressing and shifting the on-axis interference fringes to
different extents, while the conformal anomaly correction falls off as $O(\tilde{\alpha}M^2/r^4)$ and the two intensity curves
gradually coincide. By analyzing the role of the potential barrier, we study the absorption cross section, and
employ the series reduction method to accelerate the convergence of the PWS for computing the differential
scattering cross section. We find that the low-frequency and high-frequency absorption cross sections are
strongly correlated with the $\ell = 0$ part and the $\ell$-dependent part of the potential barrier, respectively. The
differential scattering cross section and glory scattering are primarily attributed to scattering within a finite
region that excludes a small neighborhood of the outer event horizon. Charge and the conformal anomaly
parameter $\tilde{\alpha}$ have opposite effects on the width of the glory peak, but the same effect on its height.
\end{abstract}
\maketitle

\section{Introduction} \label{sec1}
The direct detection of gravitational waves (GWs) has opened a new observational window on strong-field gravity, compact objects, and cosmology\cite{LIGOScientific:2016aoc,LIGOScientific:2016lio}. With the continued operation of ground-based interferometers such as LIGO\cite{LIGOScientific:2014pky}, Virgo\cite{VIRGO:2014yos}, and KAGRA\cite{KAGRA:2020tym}, and with the planned space-based missions Taiji\cite{Hu:2017nsr,Ruan:2020ijmpa,Taiji:web}, TianQin\cite{Luo:2016cqg,Mei:2020ptep,TianQin:web}, and LISA\cite{LISA:2017pwj,LISA:web}, together with third-generation ground-based detectors such as the Einstein Telescope\cite{Punturo:2010zz,Hild:2010id} and Cosmic Explorer\cite{Reitze:2019iox,Evans:2021gyd,LIGOScientific:2016wof}, both the sensitivity and the frequency coverage of GW observations will be greatly improved\cite{LIGOScientific:2018mvr,LIGOScientific:2020ibl,LIGOScientific:2021usb,KAGRA:2021vkt,LIGOScientific:2025slb,LIGOScientific:2026wfs}. Two aspects of this development are directly relevant to the present work. First, in the ringdown phase of a binary black hole merger the quasinormal-mode frequencies and damping times of the remnant are governed by the potential-barrier scattering properties of the black hole, which makes this phase a key window for testing general relativity (GR) and for searching for new physics\cite{Leaver:1985ax,Nollert:1999ji,Berti:2009kk,Huang:2015cha,Lin:2019qyx,CAO:2018ddz,Chen:2026ack,Li:2024zwt,Li:2025eig}. Second, GWs can be lensed and scattered by foreground compact objects during propagation, so that scattering effects may leave additional imprints on the observed waveforms and amplitudes\cite{Fan:2016swi,Fan:2020sou,Hou:2019wdg,Liao:2019aqq,Yin:2023kzr,Li:2024oke,Li:2025gou,Yang:2025kfj,Zhang:2025xqh}. A quantitative understanding of wave scattering in curved spacetime is therefore not only a fundamental problem in theoretical physics, but also a prerequisite for the accurate extraction of physical parameters from GW data.

Most descriptions of black hole imaging are formulated in geometric optics: radiation propagates along null geodesics, the horizon acts as a perfect absorber, and a distant observer therefore sees a shadow. This picture discards the phase of the wave. When the wavelength becomes comparable to the size of the scatterer, the particle-like picture of scattering must be replaced by a wave picture whose essential physics is coherent superposition. In optics, the direct evidence for coherent superposition came from the double-slit and single-slit experiments. For GWs this experimental route is blocked by physics itself: the gravitational interaction is so weak that ordinary matter is essentially transparent to GWs, so that one can neither fabricate a ``single slit'' for GWs nor find a natural mirror that reflects them. A black hole is therefore almost the only object in the real universe that can actually block a GW: the event horizon imposes a purely ingoing (perfectly absorbing) boundary condition, and the part of the wave that is not absorbed but sweeps past the black hole interferes with the edge wave emerging from the geometric shadow. The interference is constructive on the symmetry axis, producing a bright Poisson spot (the spot of Arago) surrounded by concentric diffraction fringes. Shadow and Poisson spot are thus two faces of the same physics, and the latter is the sharpest theoretical evidence that a wave can bend around a black hole. It should be stressed that this wave-optics aspect is not accessible to present detection strategies, which rely on matched filtering and hence extract only the energy and phase correlation of the signal rather than resolving interference fringes in space; coherent diffraction by a black hole therefore remains a theoretical prediction awaiting observational confirmation.

From a theoretical perspective, the scattering of classical fields by black holes has a long and well-established history\cite{Sanchez:1977vz,Matzner:1985rjn,Dolan:2006vj,Dolan:2008kf,Dolan:2009zza,Chen:2011jgd,Crispino:2009xt,Crispino:2009ki,Crispino:2015gua,Huang:2020bdf,Richarte:2021fbi,Xavier:2023ljy,Chen:2023ese,Li:2025jpa,Li:2025yoz,Li:2025lvl,Li:2026ayn,Tang:2026wmb,Tang:2026mem,Bian:2026syn,Sporea:2017zxe,deOliveira:2019tlk}. The partial-wave series (PWS) method is one of the most fundamental and rigorous tools for such problems: the incident and scattered waves are expanded in angular-momentum modes, the radial (Regge-Wheeler) equation is solved mode by mode, and the scattering amplitude is obtained as a series over $\ell$. For the differential scattering cross section (DCS) the convergence of this series is slow, and a large number of partial waves is needed for adequate accuracy. Several techniques have been developed to accelerate the convergence, notably the Sommerfeld-Watson transformation\cite{Kubyshin:1982xz,Pumplin:1968ca,Lopez:1970cv} and the series reduction method (SRM)\cite{Yennie:1954zz,Crispino:2009ki,Li:2025lvl}. These methods not only improve the computational efficiency but also expose important physical structures in the DCS, such as glory scattering\cite{Matzner:1985rjn} and rainbow scattering\cite{Stratton:2019deq,Leite:2019uql}. More recently, the PWS method has been pushed from asymptotic scattering amplitudes to the wave field itself: it was shown in Ref.\cite{Li:2025lvl} that, in order to reconstruct the wave function at a finite distance $r$, the PWS has to be summed at least up to $\ell \sim kr$, and that the resulting scalar waveform of a Schwarzschild black hole exhibits a clear Poisson spot in the paraxial region, in close analogy with optical diffraction.

This progress raises a question that, to our knowledge, has not been addressed: how do the finite-distance diffraction waveforms, and the cross sections associated with them, respond to a modification of the {\it near-horizon} geometry? Answering it requires a background that can be continuously deformed away from GR while remaining analytically tractable, and an observable that is genuinely sensitive to the near field rather than to the asymptotic charges $(M,Q)$ alone. The conformal anomaly black hole provides precisely such a background. In quantum field theory in curved spacetime the conformal symmetry of massless fields is broken at the quantum level, so that the expectation value of the energy-momentum tensor acquires a nonzero trace\cite{Capper:1974ic,Duff:1977ay,Duff:1993wm,Birrell:1982ix}. Feeding this trace anomaly back into the semiclassical Einstein equations yields, in the static spherically symmetric case, an exact solution\cite{Cai:2009ua} whose asymptotically flat (``$-$'') branch is characterized by the mass $M$, a $U(1)$ charge $Q$, and a parameter $\tilde{\alpha}$ measuring the strength of the anomaly; it reduces to the Reissner-Nordstr\"om (RN) metric at large $r$. The solution may thus be regarded as a natural one-parameter generalization of the RN black hole. Two features make it particularly well suited to the question posed above. First, the anomaly correction to the metric function falls off as $O(\tilde{\alpha}M^{2}/r^{4})$, much faster than the mass and charge terms, so the deformation is concentrated in the strong-field region and the solution remains indistinguishable from the RN one asymptotically. Second, the admissible charge range is relaxed from the RN bound $Q^{2}\leq M^{2}$ to $Q^{2}<2M^{2}$, which opens a window on scattering by black holes carrying charges that no RN solution can support.

In this paper we present a systematic wave-optics study of scalar-wave scattering by static conformal anomaly black holes, and compare the results in detail with those for the Schwarzschild and RN black holes. For the wave field, we first establish a practical truncation criterion for the PWS at finite distance. The analysis of the effective potential shows that the barrier height grows as $\ell^{2}$ while its width is set almost entirely by $\ell$ and is insensitive to the metric and to the parameters; high-$\ell$ modes are therefore reflected before reaching the strong-field region and decay rapidly inside the barrier. This yields the criterion $\ell_{\max}\gtrsim kr$, which we verify numerically. On this basis we compute the full scalar waveforms at finite distance and find a diffraction pattern common to all the backgrounds considered: on the negative $z$-axis the field reduces to the incident plane wave, whereas on the positive $z$-axis a distorted plane wave coexists with a pronounced scattered spherical wave, and a bright Poisson spot appears on the axis at $\theta=0$, surrounded by concentric alternating bright and dark rings in the paraxial region. The novelty emerges when the on-axis intensity is followed as a function of the radial coordinate: it is most sensitive to the spacetime geometry in the immediate vicinity of the horizon and essentially blind to it far away. Very close to $r_{+}$ the on-axis intensity of a conformal anomaly black hole differs markedly from that of the RN black hole with the same $M$ and $Q$, while in the far field the two intensity curves converge. We trace this difference to the metric itself: near the horizon the tortoise coordinate behaves as $r_{\ast}\simeq \ln (r-r_{+})/f'(r_{+})$, so that different values of $f'(r_{+})$ make the same radial interval correspond to different phase increments, compressing and shifting the on-axis interference fringes by different amounts; and since the anomaly correction decays as $O(\tilde{\alpha}M^{2}/r^{4})$, the two curves must eventually coincide. The Poisson spot thus acts as a genuine near-horizon probe, in contrast to asymptotic observables that only resolve $M$ and $Q$.

For the cross sections, we find that the parameter dependence becomes transparent once the effective potential is decomposed into its $\ell=0$ part and its $\ell$-dependent part, because these two parts control two different frequency regimes. The absorption cross section (ACS) is governed near the low-frequency limit by the $\ell=0$ barrier — consistently with the universal result $\sigma_{\text{abs}}\to A_{H}$ — and at high frequency by the $\ell$-dependent part, consistently with the geodesic capture cross section $\sigma_{\text{abs}}\to\pi r_{c}^{2}/f(r_{c})$. We also identify a characteristic downward ``dip'' that occurs precisely when the $\ell=0$ barrier develops a peak higher than the $\ell$-dependent part. Moreover, because the horizon radius is determined by $Q_{M}^{2}=Q^{2}-2\tilde{\alpha}$ rather than by $\tilde{\alpha}$ alone, we are able to compare black holes at fixed horizon area and thereby disentangle the effect of the anomaly from that of the charge: the two act in opposite directions on the ACS, with the charge dominating when they are increased jointly. For the DCS we employ the SRM to accelerate the convergence of the angular factor series, which allows us to resolve the glory peak: increasing $Q$ broadens and lowers it, whereas increasing $\tilde{\alpha}$ narrows and lowers it. Two observations then localize the origin of glory scattering: in the extremal limit $\tilde{\alpha}\to\tilde{\alpha}_{\max}$ the barrier becomes infinitely high at the horizon, a configuration akin to the Newtonian and Coulomb potentials, which nevertheless exhibit no glory; and for small $\tilde{\alpha}$ the DCS reduces continuously to that of the RN black hole. We conclude that glory scattering is generated by the scattering of the wave in a finite region that excludes a small neighborhood of the outer event horizon, rather than by the near-horizon region itself.

Although the explicit calculation is carried out for a massless scalar field, the setup is a direct theoretical preparation for GW scattering. In a static spherically symmetric background the radial perturbation equations for fields of any spin reduce to a one-dimensional Schr\"odinger-type (Regge-Wheeler) equation, the differences residing only in the coefficient of the centrifugal term and in the details of the effective potential, whereas the boundary condition at the horizon — a purely ingoing wave — is universal. The coherent-superposition phenomena identified here, in particular the diffraction fringes and the Poisson spot, are therefore expected to carry over qualitatively to gravitational perturbations.

The remainder of this paper is organized as follows. Section\ref{sec:2} reviews the PWS method for scalar plane-wave scattering by a Schwarzschild black hole and introduces the SRM. Section\ref{sec:3} presents the conformal anomaly black hole solution and the admissible range of its parameters. Section\ref{sec:4} establishes the truncation criterion for the PWS and investigates the scalar waveforms at finite distance, including the intensity of the on-axis Poisson spot. Section\ref{sec:5} computes and analyzes the ACS and the DCS, and compares the conformal anomaly black hole with the RN black hole. Section\ref{sec:6} summarizes our conclusions and outlines directions for future work. Throughout, the PWS method is our basic numerical tool, and the SRM is used only where the slow convergence of the angular factor series demands it.

\section{Scalar plane wave scattering of Schwarzschild black hole}
\label{sec:2}
In this section, we review the rigorous calculation of the scattering of scalar plane waves by a Schwarzschild black hole\cite{Li:2025lvl}.
The PWS method is employed in the study of black holes. To calculate the scattering amplitude, the plane wave incident from negative infinity along the $z$ axis must also be expressed in the form of a PWS expansion as follows
\begin{equation}
\label{eq:1}
\tilde{\psi}_{\text{inc}}(k, \boldsymbol{r}) = e^{ikr\cos\theta} = \sum_{\ell=0}^{\infty} \tilde{R}_{\ell}(k, r) \mathrm{P}_{\ell}(\cos\theta)\,,
\end{equation}
where $\mathrm{P}_{\ell}(\cos\theta)$ denotes the $\ell$-th order Legendre polynomial of the first kind. Since the scattering process does not break the rotational symmetry about the  $z$-axis, the wave function is independent of the azimuthal angle $\phi$ . Using the orthonormality of the Legendre functions, the expression for the radial function can be derived as:
\begin{equation}
\label{eq:2}
\tilde{R}_\ell(k, r) = i^\ell (2\ell + 1) j_\ell(kr)\,,
\end{equation}
where $j_\ell(kr)$ denotes the $\ell$-th order spherical Bessel function of the first kind.

Using the asymptotic expansion of the spherical Bessel function $j_\ell(z)$ for  $z \gg \ell$, the asymptotic expansion of the radial function at infinity (under the condition  $kr \gg \ell$ ) can be written as:
\begin{equation}
\label{eq:3}
\tilde{R}_\ell(k, r \to \infty) \to (-1)^{\ell+1} \frac{2\ell + 1}{2ikr} \left[e^{-ikr} - (-1)^\ell e^{ikr} \right]\,.
\end{equation}
Therefore, the asymptotic expansion of the incident plane wave as r$ \to \infty$ is:
\begin{equation}
\label{eq:4}
\tilde{\psi}_{\text{inc}}(k, \boldsymbol{r}) = \sum_{\ell=0}^{\infty} (-1)^{\ell+1} \frac{2\ell+1}{2ikr} \left[ e^{-ikr} - (-1)^{\ell} e^{ikr} \right] \mathrm{P}_{\ell}(\cos\theta)\,,
\end{equation}
where the condition $\ell \ll kr$ must be satisfied.

The line element of a static spherically symmetric metric is written as
\begin{equation}
\label{eq:5}
\mathrm{d}s^2 = -f(r)\mathrm{d}t^2 + \frac{1}{f(r)}\mathrm{d}r^2 + r^2(\mathrm{d}\theta^2 + \sin^2\theta\mathrm{d}\varphi^2) \,.
\end{equation}
For the Schwarzschild metric, the metric function satisfies $f(r) = 1 - \frac {2M}{r}$, where $M$ is the mass of the Schwarzschild black hole. The massless Klein-Gordon(KG) equation is
\begin{equation}
\label{eq:6}
\Box^2 \psi(t, \boldsymbol{r}) = \frac{1}{\sqrt{-g}} \partial_\mu \left[ \sqrt{-g} g^{\mu\nu} \partial_\nu
\right] \psi(t, \boldsymbol{r}) = 0 \,.
\end{equation}
In parallel with Eq.\eqref{eq:1} the wave function of the frequency domain $\tilde{\psi}(k, \boldsymbol{r})$ can be expanded in the form of the PWS as
\begin{equation}
\label{eq:7}
\tilde{\psi}(k, \mathbf{r}) = \sum_{\ell=0}^{\infty} \tilde{R}_{\ell}(k, r) \mathrm{P}_{\ell}(\cos\theta) \,.
\end{equation}
Under the static spherically symmetric metric, Eq.\eqref{eq:6} can be written as the form of the spin-0 Regge-Wheeler(RW) equation by using Eq.\eqref{eq:7}
\begin{equation}
\label{eq:8}
\left[ \frac{\mathrm{d}^2}{\mathrm{d}r_\ast^2} + k^2 - V_\ell(r) \right] u_\ell(k, r) = 0 \,,
\end{equation}
where the radial function is defined as $u_\ell(k, r) = r \tilde{R}_\ell(k, r)$, the tortoise coordinate $r_\ast$ is defined by
\begin{equation}
\label{eq:9}
\mathrm{d}r_\ast = \frac{1}{f(r)} \mathrm{d}r \,,
\end{equation}
and the potential function $V_\ell(r)$ is given by
\begin{equation}
\label{eq:10}
V_\ell(r) = f(r) \left[ \frac{f'(r)}{r} + \frac{\ell (\ell + 1)}{r^2} \right]
\end{equation}
where $f'(r) = \frac{\mathrm{d}f(r)}{\mathrm{d}r}$.

For the Schwarzschild metric, when $r \to \infty$ and the condition $r \gg \ell$ is satisfied, the asymptotic infinity boundary condition for the radial Eq.\eqref{eq:8} can be approximated as
\begin{equation}
\label{eq:11}
u_\ell(k, r) \left|_{\substack{r \to \infty \\ r \gg \ell}} \right. \to A_{\ell} e^{-ikr_\ast} + B_{\ell} e^{ikr_\ast} \,.
\end{equation}
When $r \to r_+$, the boundary condition for the radial Eq.\eqref{eq:8} can be approximated as
\begin{equation}
\label{eq:12}
u_\ell(k, r) \left|_{\substack{r \to r_+}} \right. \to C_\ell e^{-ikr_\ast} \,.
\end{equation}

Consider the physical process in which a plane wave incident from $z \to -\infty$ and propagating along the positive $z$-direction is scattered by a black hole. To determine the incident coefficient $A_\ell$, reflection coefficient $B_\ell$, and transmission coefficient $C_\ell$, it is necessary to numerically solve Eq.\eqref{eq:8}. In the numerical solution, the coefficient at the inner boundary $r \to r_+$ is set as $C_{\ell {\text{Num}}} = 1$. The radial function $u_{\ell {\text{Num}}}$ is then evolved to the outer boundary where $r$ is sufficiently large for the $r$ of $V_\ell (r) = k^2$, and matched with the asymptotic boundary condition\eqref{eq:11} to obtain $A_{\ell {\text{Num}}}$ and $B_{\ell {\text{Num}}}$. By matching the incident coefficient $A_\ell$ with the incident coefficient of the plane wave in Eq.\eqref{eq:3}, we obtain
\begin{equation}
\label{eq:13}
A_\ell = (-1)^{l+1} \frac{2 l + 1}{2 i k} \,,\quad B_\ell = \frac{B_{\ell {\text{Num}}}}{A_{\ell {\text{Num}}}} A_\ell \,,\quad C_\ell = \frac{1}{A_{\ell {\text{Num}}}} A_\ell \,,\quad u_\ell = \frac{1}{A_{\ell {\text{Num}}}} A_\ell u_{\ell {\text{Num}}} \,.
\end{equation}
In the limit $r \to \infty$, the numerical radial function can be written as
\begin{equation}
\label{eq:14}
u_\ell (k,r) = \frac{1}{A_{\ell {\text{Num}}}} A_\ell u_{\ell {\text{Num}}} \to A_\ell \left( e^{-ikr_\ast} - (-1)^l e^{ikr_\ast} + (-1)^{l} e^{ikr_\ast} + \frac{B_{\ell {\text{Num}}}}{A_{\ell {\text{Num}}}} e^{ikr_\ast} \right) \,.
\end{equation}
The wave function can be expressed using the PWS as
\begin{equation}
\label{eq:15}
\tilde{\psi}(k, \boldsymbol{r}) = \frac{1}{2ikr} \sum_{\ell=0}^{\infty} (-1)^{l+1} (2l+1) \frac{u_{\ell {\text{Num}}}}{A_{\ell {\text{Num}}}} \mathrm{P}_{\ell}(\cos\theta) \,.
\end{equation}
In the limit $r \to \infty$, the wave function can be written as
\begin{equation}
\label{eq:16}
\tilde{\psi}(k, \boldsymbol{r}) \to e^{ikr_\ast \cos\theta} + f(\theta) \frac{e^{ikr_\ast}}{r} \,.
\end{equation}
Here, the first term represents the incident plane wave, and the second term represents the scattered spherical wave. The angular factor is defined as
\begin{equation}
\label{eq:17}
f(\theta) = \frac{1}{2ik} \sum_{\ell=0}^{\infty} (2\ell + 1) \{e^{2i\delta_\ell} - 1\} \mathrm{P}_\ell(\cos\theta) \,,
\end{equation}
where the phase shift $e^{2i\delta_\ell}$ is defined as
\begin{equation}
\label{eq:18}
e^{2i\delta_\ell} = (-1)^{l+1} \frac{B_{\ell {\text{Num}}}}{A_{\ell {\text{Num}}}} \,.
\end{equation}
The DCS can be written as
\begin{equation}
\label{eq:19}
\frac{\mathrm{d}\sigma}{\mathrm{d}\Omega} = |f(\theta)|^2 \,.
\end{equation}

The DCS converges very slowly with respect to the angular momentum quantum number $\ell$. In particle physics, the SRM is commonly used to accelerate the convergence of the DCS. The SRM will be introduced below. The Eq.\eqref{eq:17} can be expressed in the following form
\begin{equation}
\label{eq:20}
f(\theta) = \sum_{\ell=0}^{\infty} a_{\ell} \mathrm{P}_{\ell}(\cos\theta) \,.
\end{equation}
The $n$-th reduced series is defined as
\begin{equation}
\label{eq:21}
f^{(n)}(\theta) = \sum_{\ell=0}^{\infty} a_\ell^{(n)} \mathrm{P}_\ell(\cos\theta) = (1 - \cos\theta)^n \sum_{\ell=0}^{\infty} a_\ell \mathrm{P}_\ell(\cos\theta) \,.
\end{equation}
Using the recurrence relation of Legendre functions $\mathrm{P}_\ell(\cos\theta)$, the recurrence relation for the $n$-th coefficient $a_\ell^{(n)}$ can be derived as
\begin{equation}
\label{eq:22}
a_{\ell}^{(n+1)} = a_{\ell}^{(n)} - \frac{\ell}{2\ell - 1} a_{\ell-1}^{(n)} - \frac{\ell + 1}{2\ell + 3} a_{\ell+1}^{(n)} \,,
\end{equation}
where $a_{-1}^{(n)} = 0$. The series $f(\theta)$ is rewritten as
\begin{equation}
\label{eq:23}
\hat{f}(\theta) = (1 - \cos\theta)^{-n} f^{(n)}(\theta) \,.
\end{equation}

In Ref.\cite{Li:2025lvl}, it was demonstrated that to compute the wave function at a finite distance $r$, the PWS should be summed at least up to $\ell \sim kr$. That work also shows that in the paraxial region ($\theta \sim 0$) at a finite distance, the scalar wave obtained via the SRM significantly differs from the scalar wave calculated exactly using the PWS method.

\section{Black Holes in Gravity with Conformal Anomaly}
\label{sec:3}
In quantum field theory, the conformal symmetry of massless fields is broken at the quantum level, leading to a nonzero trace anomaly in the energy–momentum tensor, known as the conformal anomaly. In four-dimensional curved spacetime, this effect backreacts on the background geometry and must be incorporated into the semi‑classical Einstein equations
\begin{equation}
\label{eq:24}
R_{ab} - \frac{1}{2} R g_{ab} = 8\pi \langle T_{ab} \rangle \,,
\end{equation}
where $\langle T_{ab} \rangle$ is the effective energy–momentum tensor of the quantum field. Reference\cite{Cai:2009ua}, under the assumption of a static and spherically symmetric metric, incorporates the trace expression of the conformal anomaly
\begin{equation}
\label{eq:25}
g^{ab}\langle T_{ab} \rangle = \lambda I_{(4)} - \alpha E_{(4)} \,,
\end{equation}
where $\lambda$ and $\alpha$ are two positive constants related to the degrees of freedom of the quantum field, $I_{(4)} = C_{abcd} C^{abcd}$ is Weyl tensor term, and    $E_{(4)} = R^2 - 4R_{ab}R^{ab} + R_{abcd}R^{abcd}$ is Gauss-Bonnet term. For simplicity, this trace anomaly considers only the Gauss–Bonnet term, and does not take into account the squared term of the Weyl tensor. Further, the assumption $\langle T_t^t(r) \rangle = \langle T_r^r(r) \rangle$ is introduced and generalized to the entire space. Finally, an exact analytical solution is obtained, with the metric function given by
\begin{equation}
\label{eq:26}
f(r) = 1 - \frac{r^2}{4\tilde{\alpha}} \left( 1 \pm \sqrt{1 - \frac{16\tilde{\alpha}M}{r^3} + \frac{8\tilde{\alpha}Q^2}{r^4}} \right) \,,
\end{equation}
where $\tilde{\alpha} = 8\pi \alpha$, $M$ is the mass of the black hole, and $Q$ is an integration constant that can be interpreted as the square of a certain $U(1)$ conserved charge. When the ``$-$'' sign is adopted, the solution is asymptotically flat as $r \to \infty$ and reduces to the RN metric
\begin{equation}
\label{eq:27}
f(r) \approx 1 - \frac{2M}{r} + \frac{Q^2}{r^2} \,.
\end{equation}
When the ``$+$'' sign is chosen, the solution asymptotes to de Sitter space, which is deemed potentially unstable. In the subsequent scalar‑wave scattering calculations in this paper, we will adopt the metric of the ``$-$'' branch as the background geometry.

To facilitate subsequent calculations, we hereby determine the event horizon of the black hole for the metric of the ``$-$'' branch and further constrain the permissible range of the parameter $\tilde{\alpha}$. Using Eq.\eqref{eq:26} and taking the ``$-$'' sign, the corresponding event horizon of the black hole can be obtained as
\begin{equation}
\label{eq:28}
r_{\pm} = M \pm \sqrt{M^2 - Q^2 + 2 \tilde{\alpha}} \,.
\end{equation}

\begin{figure}[htbp]
\centering
\begin{subfigure}[b]{0.49\textwidth}
\centering
\includegraphics[width=\textwidth]{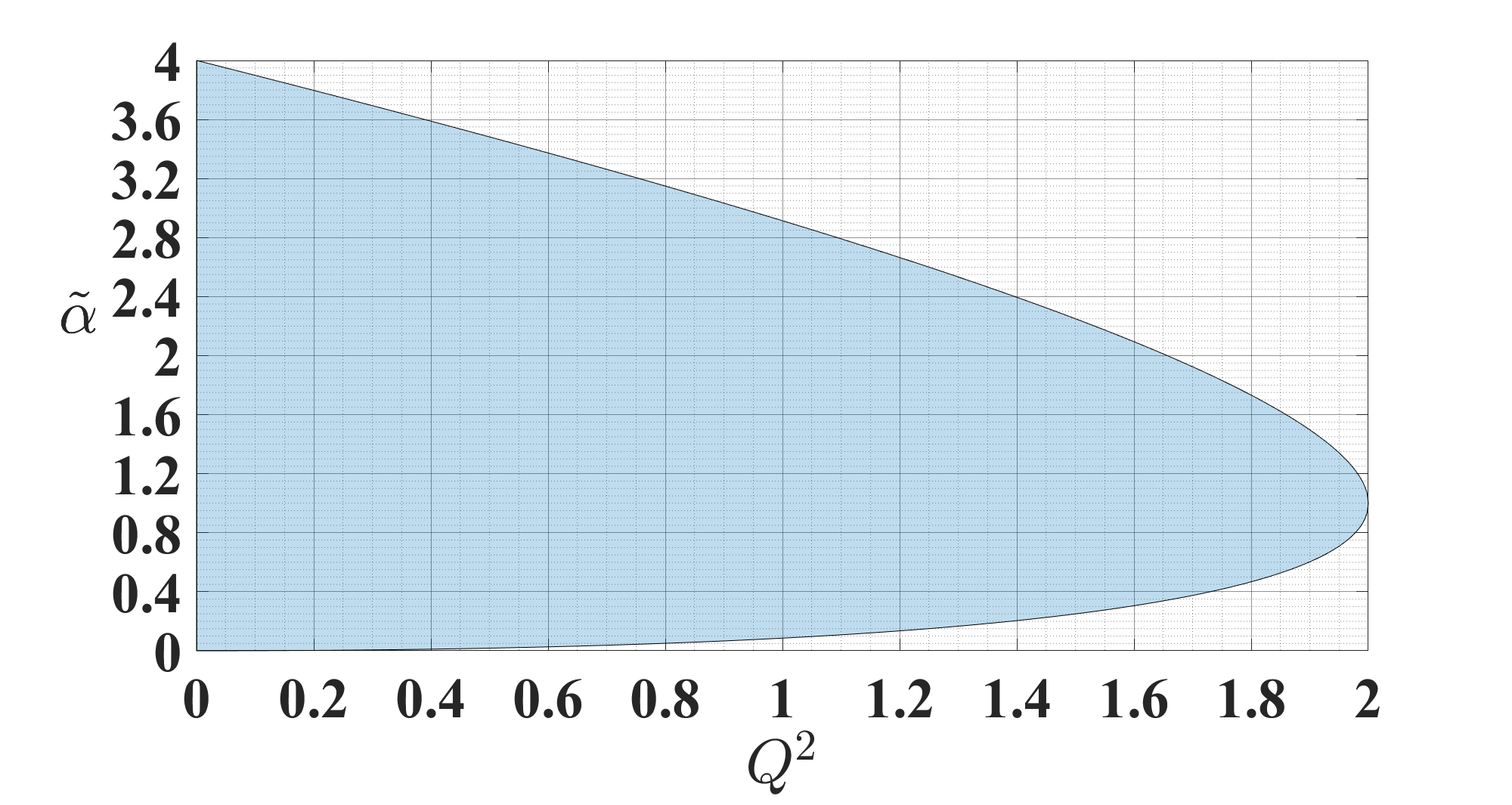}
\caption{}
\label{subfig:alpha-QM}
\end{subfigure}
\hfill
\begin{subfigure}[b]{0.49\textwidth}
\centering
\includegraphics[width=\textwidth]{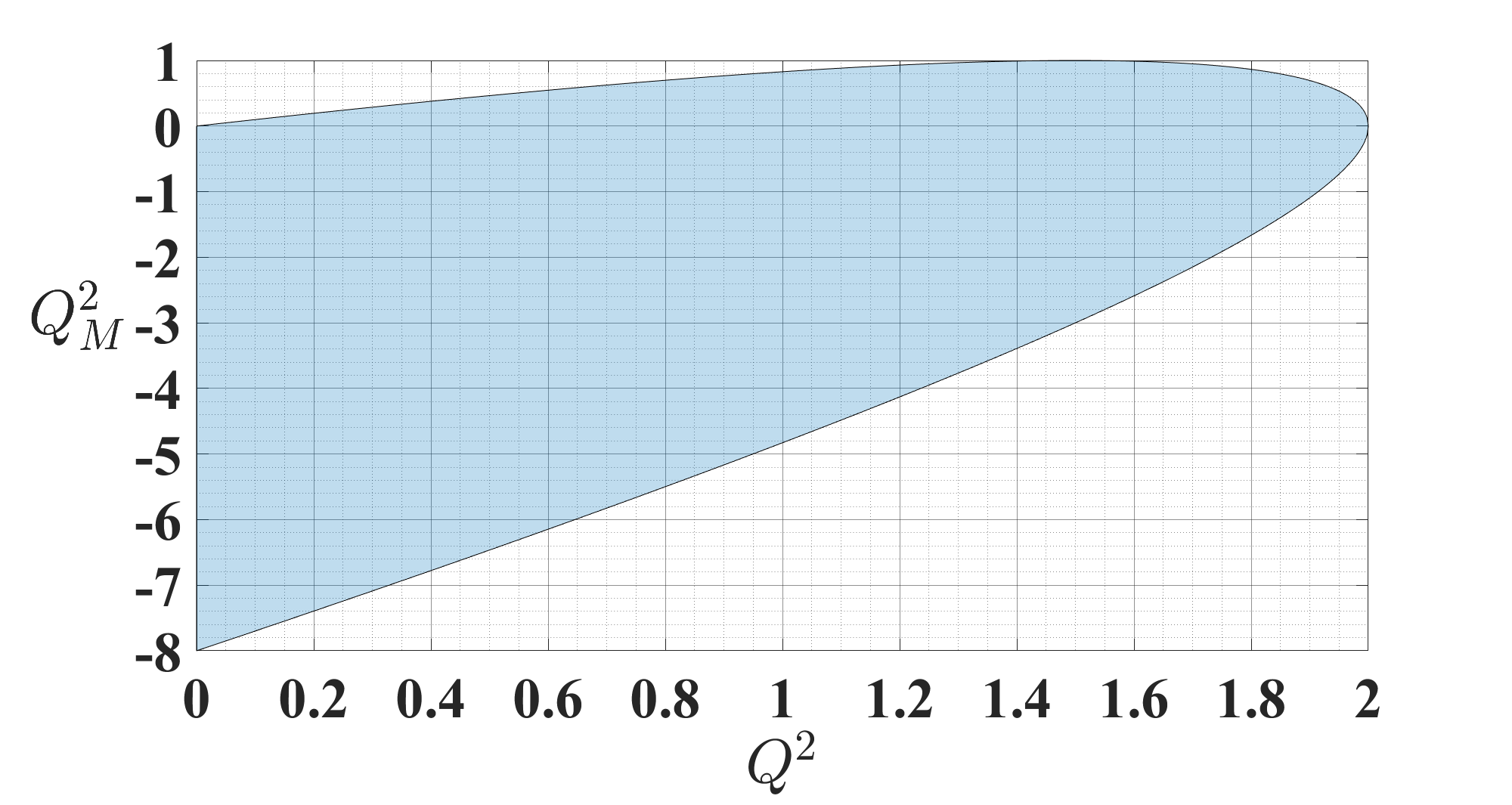}
\caption{}
\label{subfig:Q-QM}
\end{subfigure}
\caption{The allowable ranges of $\tilde{\alpha}$ and $Q_M$ as functions of $Q$. (a) and (b) show the allowable ranges of $\tilde{\alpha}$ and $Q_M^2$ as functions of $Q^2$, respectively. Mass of the black hole $M$ is set to $1$, and this value is used in all subsequent figures. The $x$-axis represents the dimensionless ratio $(Q/M)^2$, while the $y$-axis corresponds to the values of $\tilde{\alpha}$ and $Q_M^2$, respectively. The blue region represents the allowed values, and the upper and lower boundaries of the blue region are not included. It should be noted that, since the formulas relevant to scattering involve only the effects of $Q^2$ and $Q_M^2$, rather than $Q$ and $Q_M$ separately, we represent the parameters as $Q^2$ and $Q_M^2$ in this figure and the figures that follow.}
\label{FIG:alphaQ-QM}
\end{figure}

Setting $Q^2 = Q_M^2 + 2 \tilde{\alpha}$, the above expression then reduces to the event horizon of a RN black hole, where $Q_M$ corresponds to the charge of the RN black hole.  When the independent parameters are changed from $Q$ and $\tilde{\alpha}$ to $Q_M$ and $\tilde{\alpha}$, the black hole event horizon $r_{\pm}$ becomes formally independent of $\tilde{\alpha}$. We consider that $\tilde{\alpha}$ should satisfy three conditions: first, $\tilde{\alpha}$ is a positive number ($\tilde{\alpha} > 0$); second, the naked singularity condition requires ($2 \tilde{\alpha} > Q^2 - M^2$); and third, the ``$-$'' sign is chosen in the metric's ``$\pm$'' ($\tilde{\alpha} < r_+^2/4$). Taking these three conditions into account, we obtain
\begin{equation}
\label{eq:29}
\begin{split}
& Q^2 < 2M^2 \,,\\
& 2 M^2 - \frac{Q^2}{2} - M \sqrt{4 M^2 - 2 Q^2} < \tilde{\alpha} < 2 M^2 - \frac{Q^2}{2} + M \sqrt{4 M^2 - 2 Q^2} \,,\\
& 2 Q^2 - 4 M^2 - 2 M \sqrt{4 M^2 - 2 Q^2} < Q_M^2 < 2 Q^2 - 4 M^2 + 2 M \sqrt{4 M^2 - 2 Q^2} \,.
\end{split}
\end{equation}
It should be noted that when $Q = 0$, the RN black hole reduces to the Schwarzschild black hole. Therefore, the Schwarzschild black hole can be regarded as a special case of the RN black hole. 
When the inequality $\tilde{\alpha} < r_+^2/4$ becomes an equality, we have $f(r_+) = 0$ and $f'(r_+) \to \infty$ at the outer event horizon, implying that matter can fall onto the outer horizon. This differs significantly from typical black hole metrics. Therefore, in our consideration of scattering, we adopt only the strict inequality ($<$).
To visually illustrate the permissible ranges of the parameter $\tilde{\alpha}$ and the newly introduced parameter $Q_M^2$, we plot in FIG.\ref{FIG:alphaQ-QM} the allowable regions of $\tilde{\alpha}$ and $Q_M^2$ as functions of $Q^2$. From the figure, it can be seen that as $Q$ increases, the permissible ranges for both $\tilde{\alpha}$ and $Q_M$ progressively narrow.

\section{Exact Calculation of the Waveform}
\label{sec:4}
In this section, we focus on studying the waveforms of RN black holes and conformal anomaly black holes. Both belong to the category of static spherically symmetric black holes, and their asymptotic behavior is identical to that of the Schwarzschild black hole. We will precisely compute the scalar‑wave waveforms within a bounded region and generate the corresponding waveform diagrams. Before applying the PWS method to compute waveforms, it is necessary to determine the truncation condition for the PWS. This section will first examine the truncation of the PWS, and then investigate the scalar‑wave waveforms.

\subsection{Truncation of the PWS}
\begin{figure}[htbp]
\centering
\begin{subfigure}[b]{0.32\textwidth}
\centering
\includegraphics[width=\textwidth]{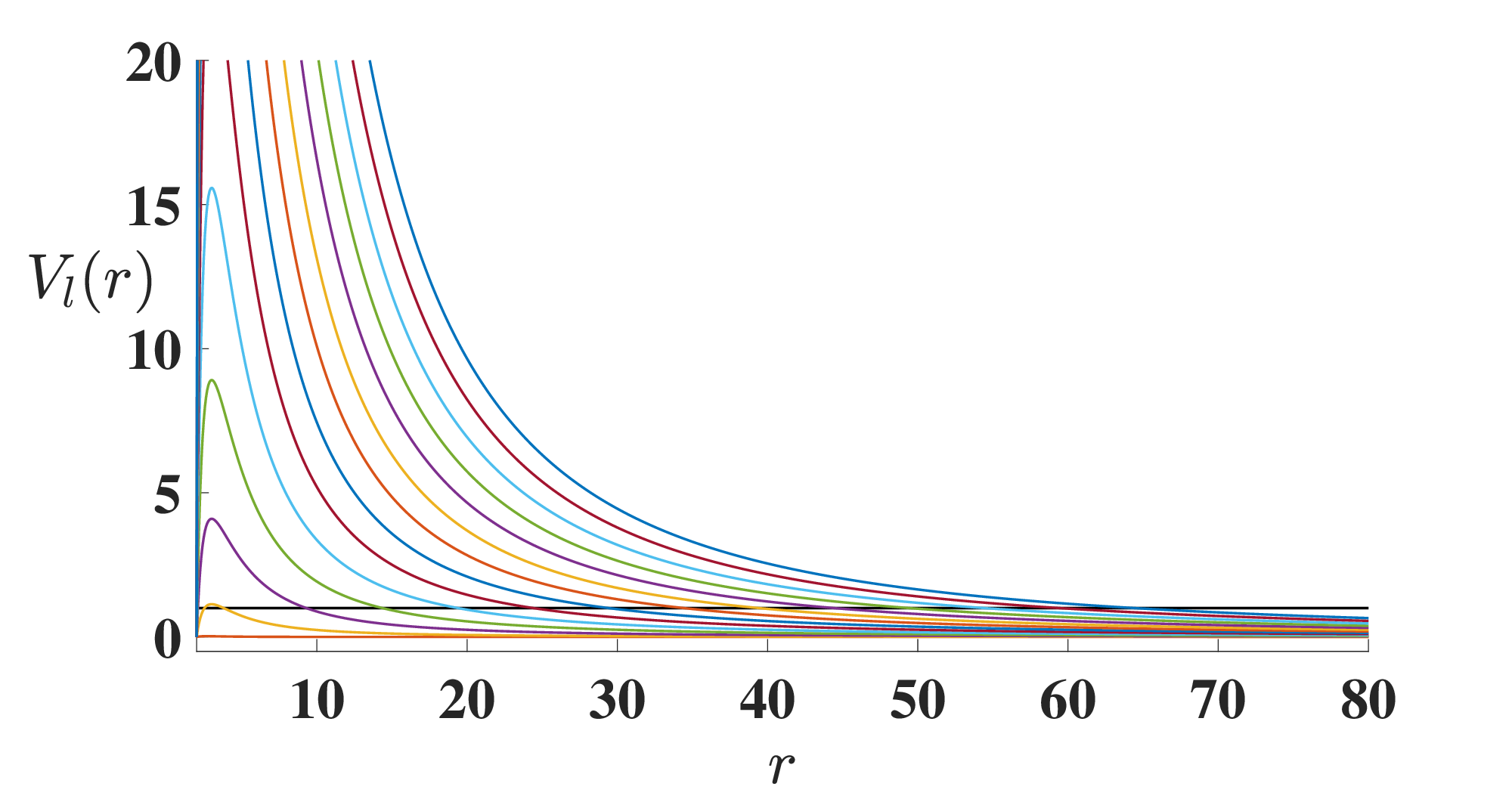}
\caption{}
\label{subfig:Vl-r a}
\end{subfigure}
\hspace{0\textwidth}  
\begin{subfigure}[b]{0.32\textwidth}
\centering
\includegraphics[width=\textwidth]{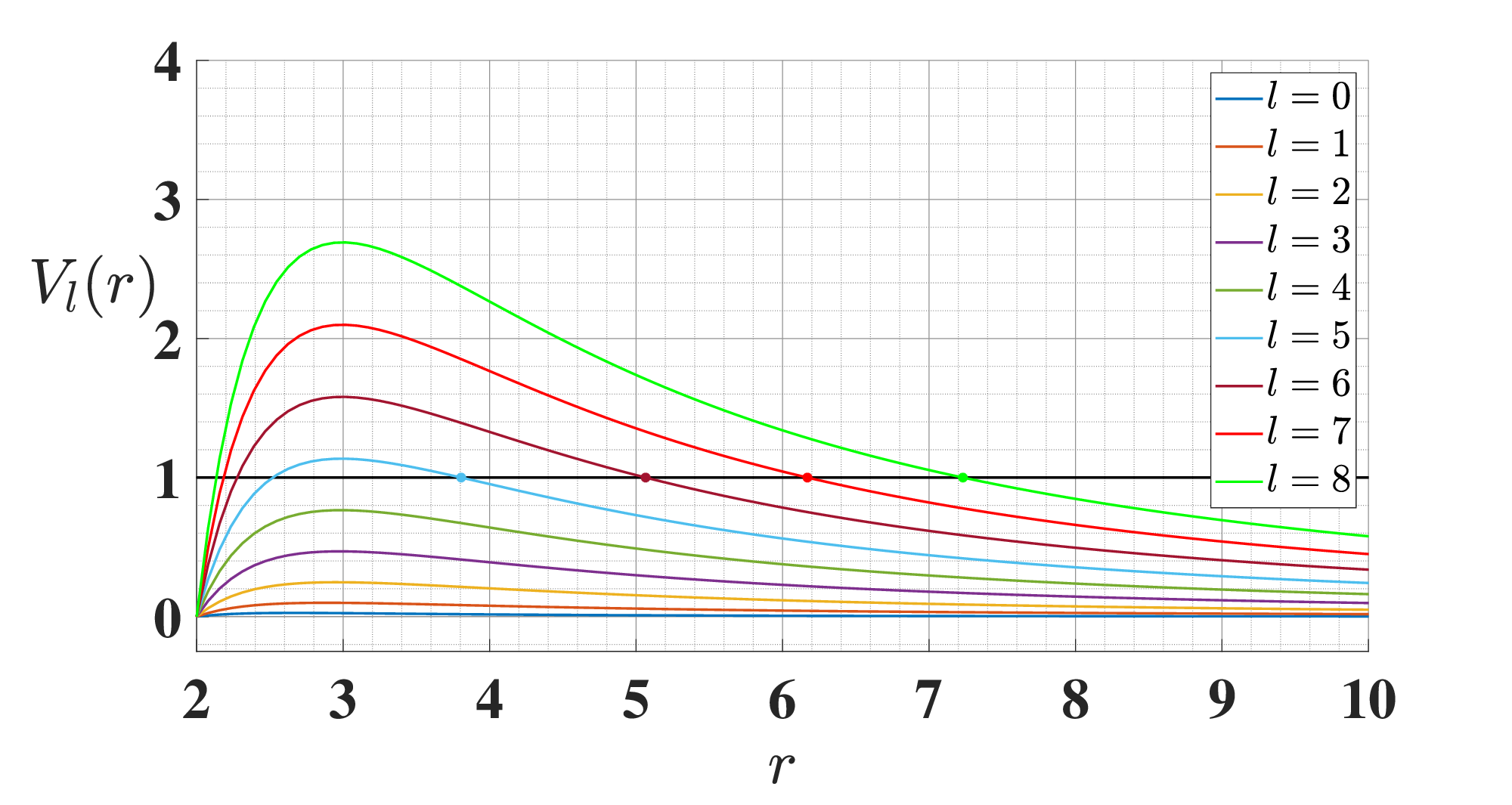}
\caption{}
\label{subfig:Vl-r b}
\end{subfigure}
\hspace{0\textwidth}  
\begin{subfigure}[b]{0.32\textwidth}
\centering
\includegraphics[width=\textwidth]{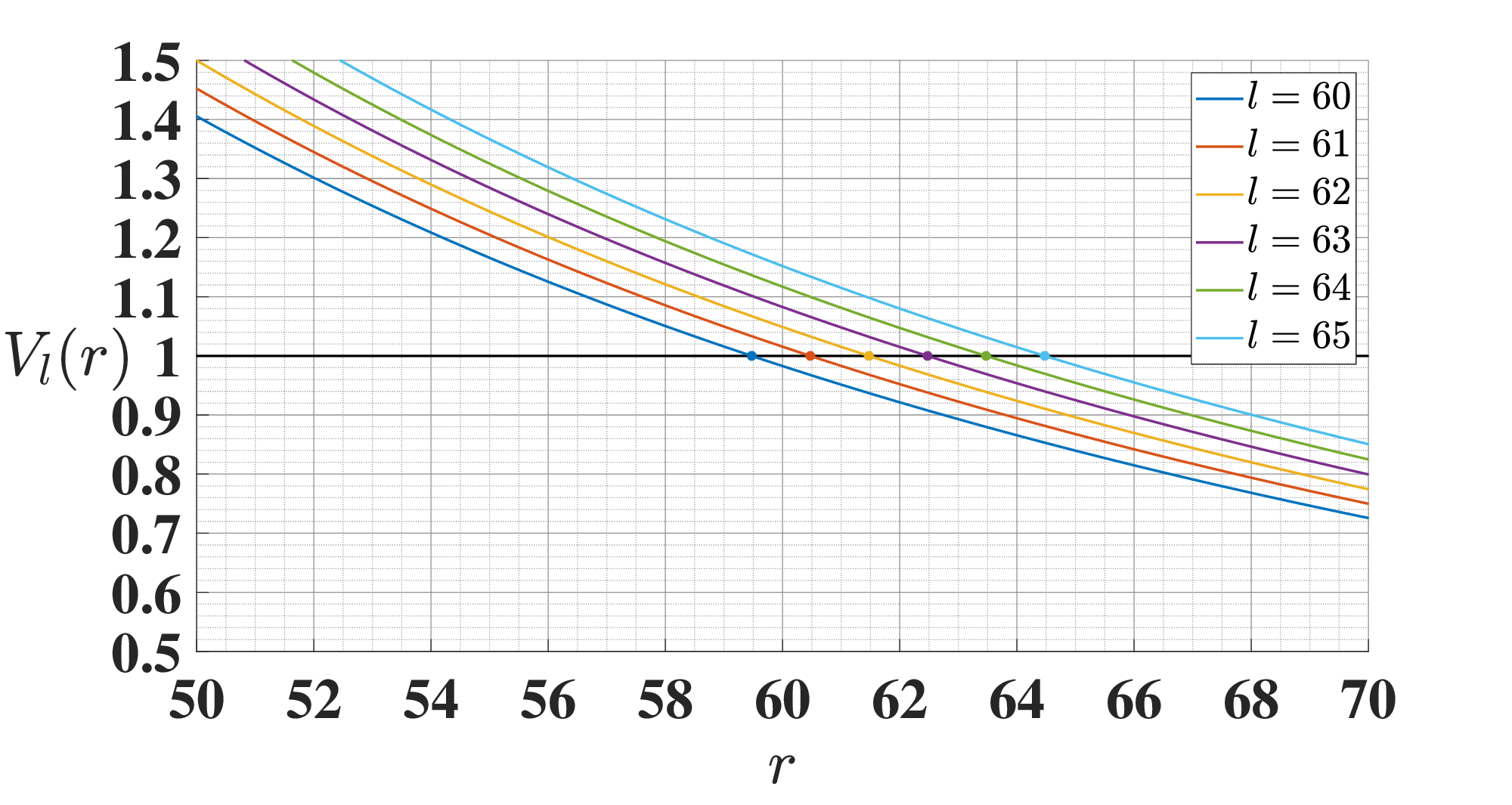}
\caption{}
\label{subfig:Vl-r c}
\end{subfigure}

\vspace{0\textwidth}  

\begin{subfigure}[b]{0.32\textwidth}
\centering
\includegraphics[width=\textwidth]{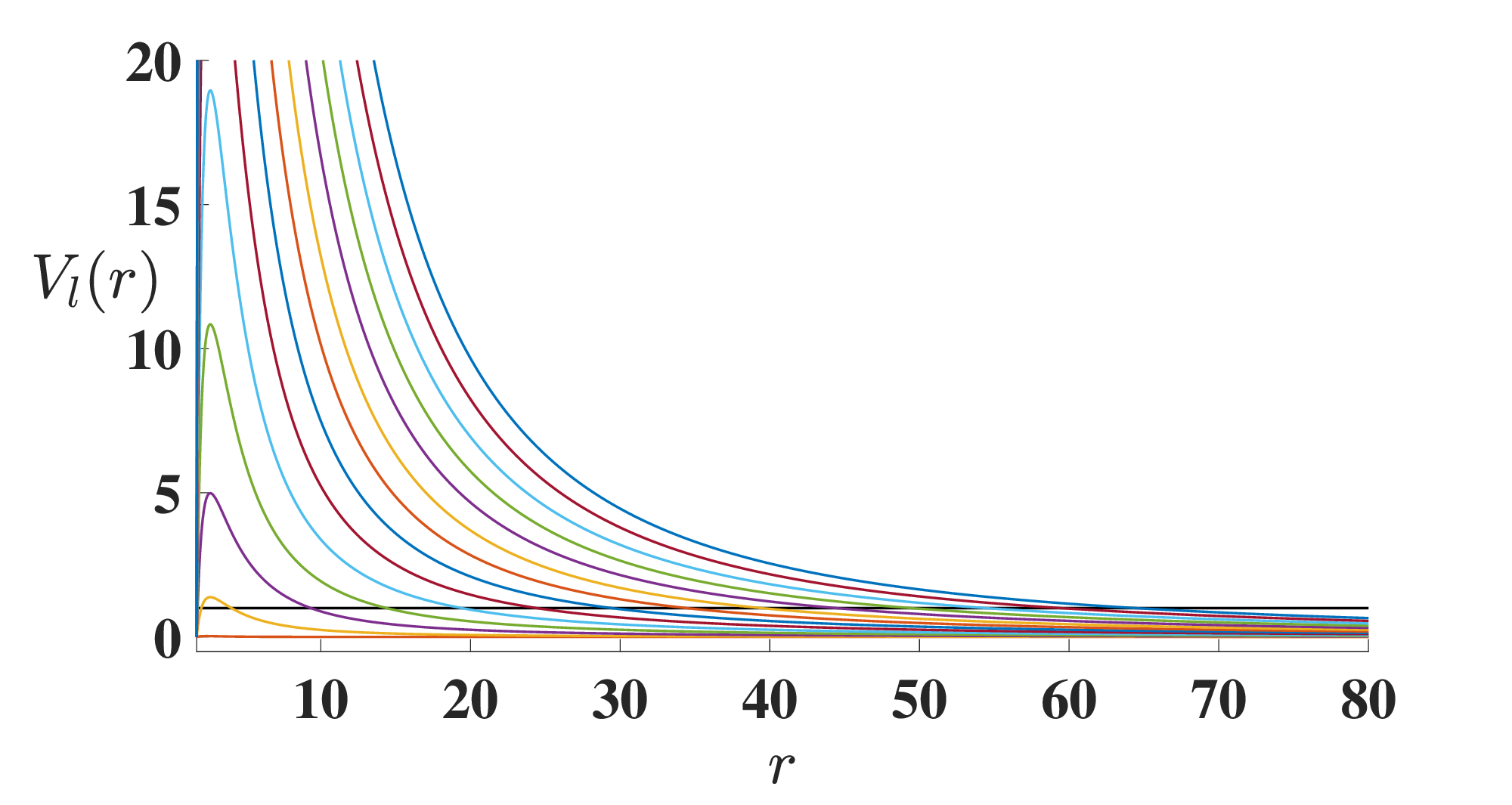}
\caption{}
\label{subfig:Vl-r d}
\end{subfigure}
\hspace{0\textwidth}  
\begin{subfigure}[b]{0.32\textwidth}
\centering
\includegraphics[width=\textwidth]{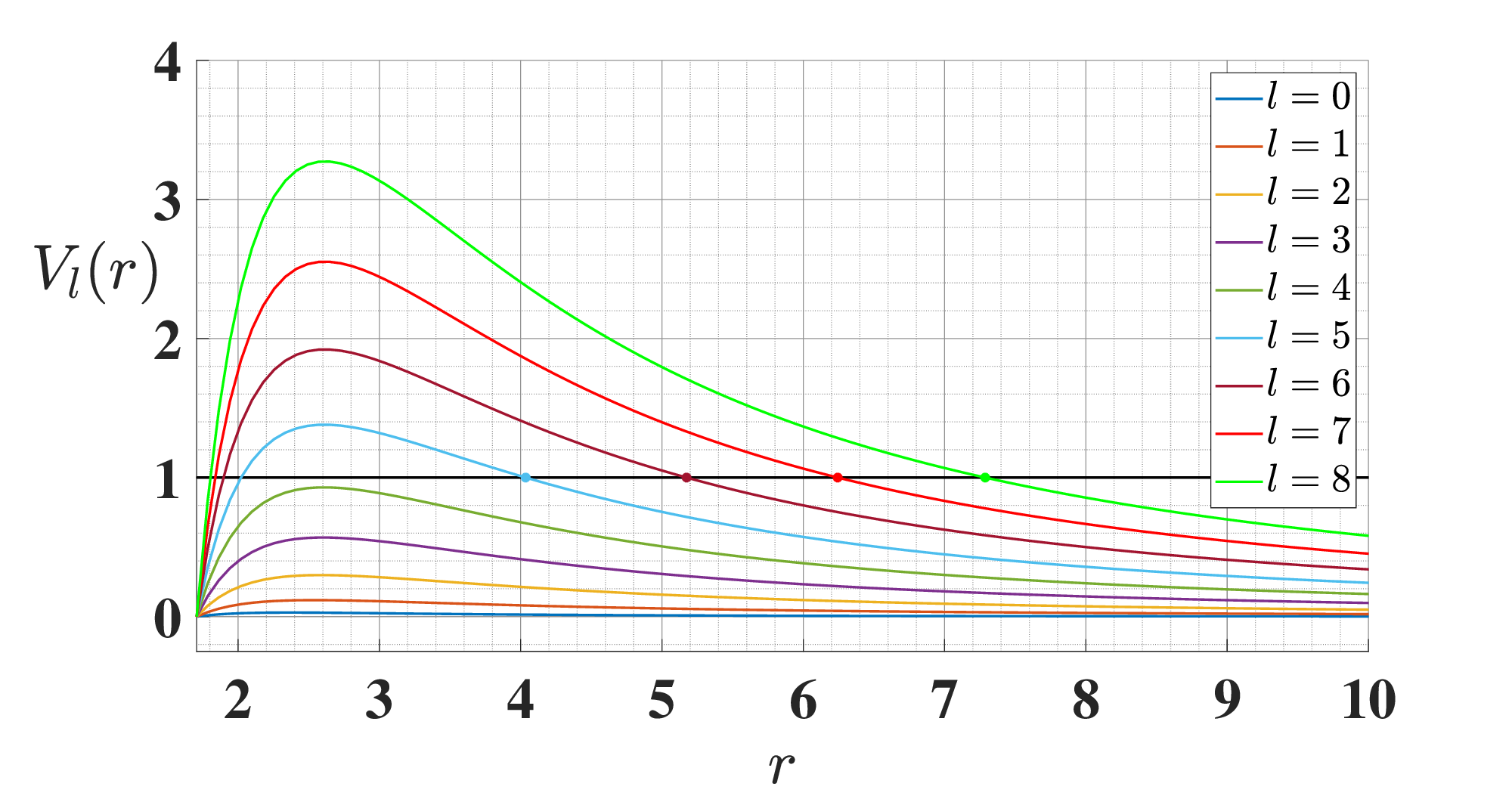}
\caption{}
\label{subfig:Vl-r e}
\end{subfigure}
\hspace{0\textwidth}  
\begin{subfigure}[b]{0.32\textwidth}
\centering
\includegraphics[width=\textwidth]{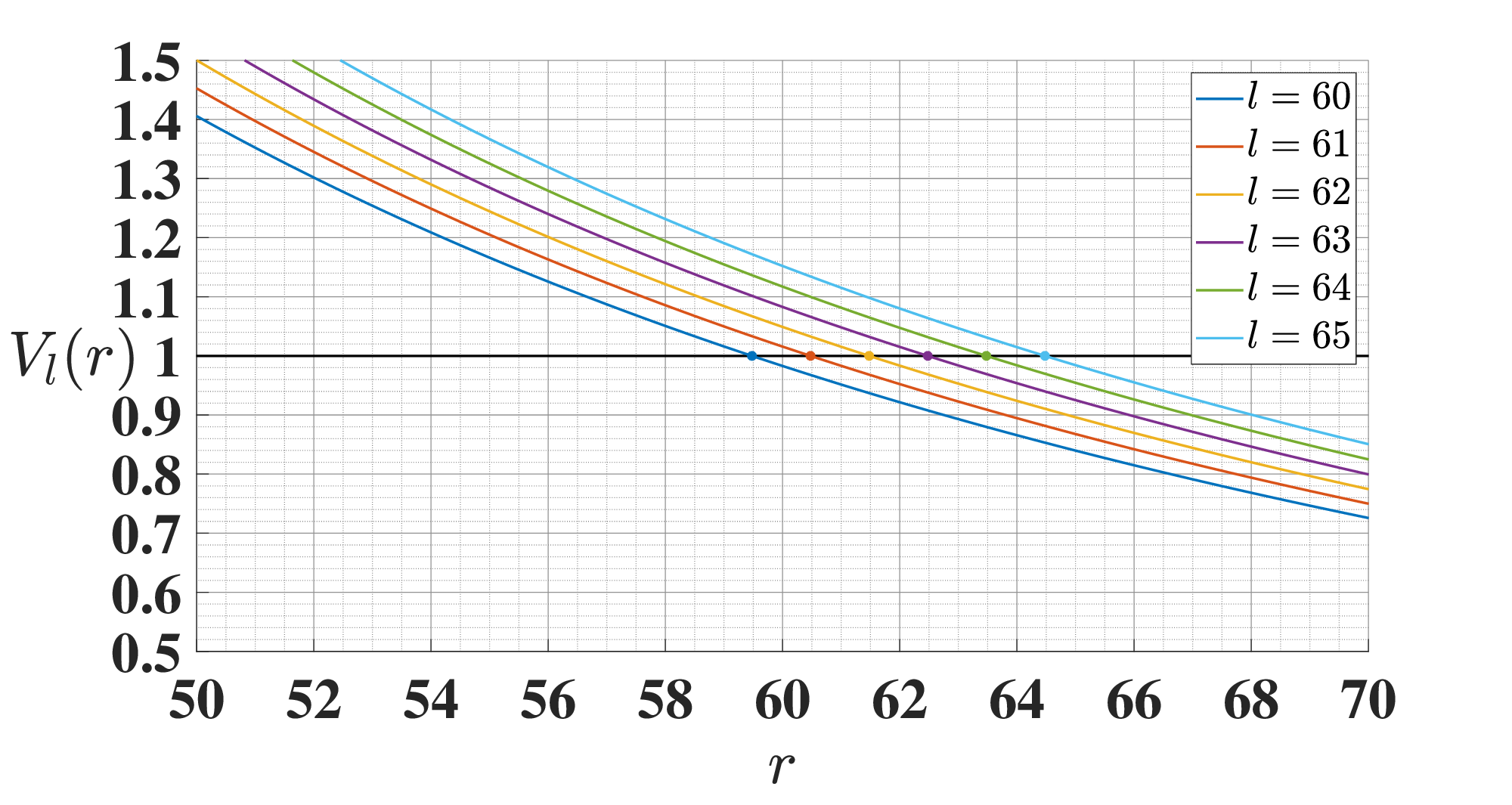}
\caption{}
\label{subfig:Vl-r f}
\end{subfigure}

\vspace{0\textwidth}  

\begin{subfigure}[b]{0.32\textwidth}
\centering
\includegraphics[width=\textwidth]{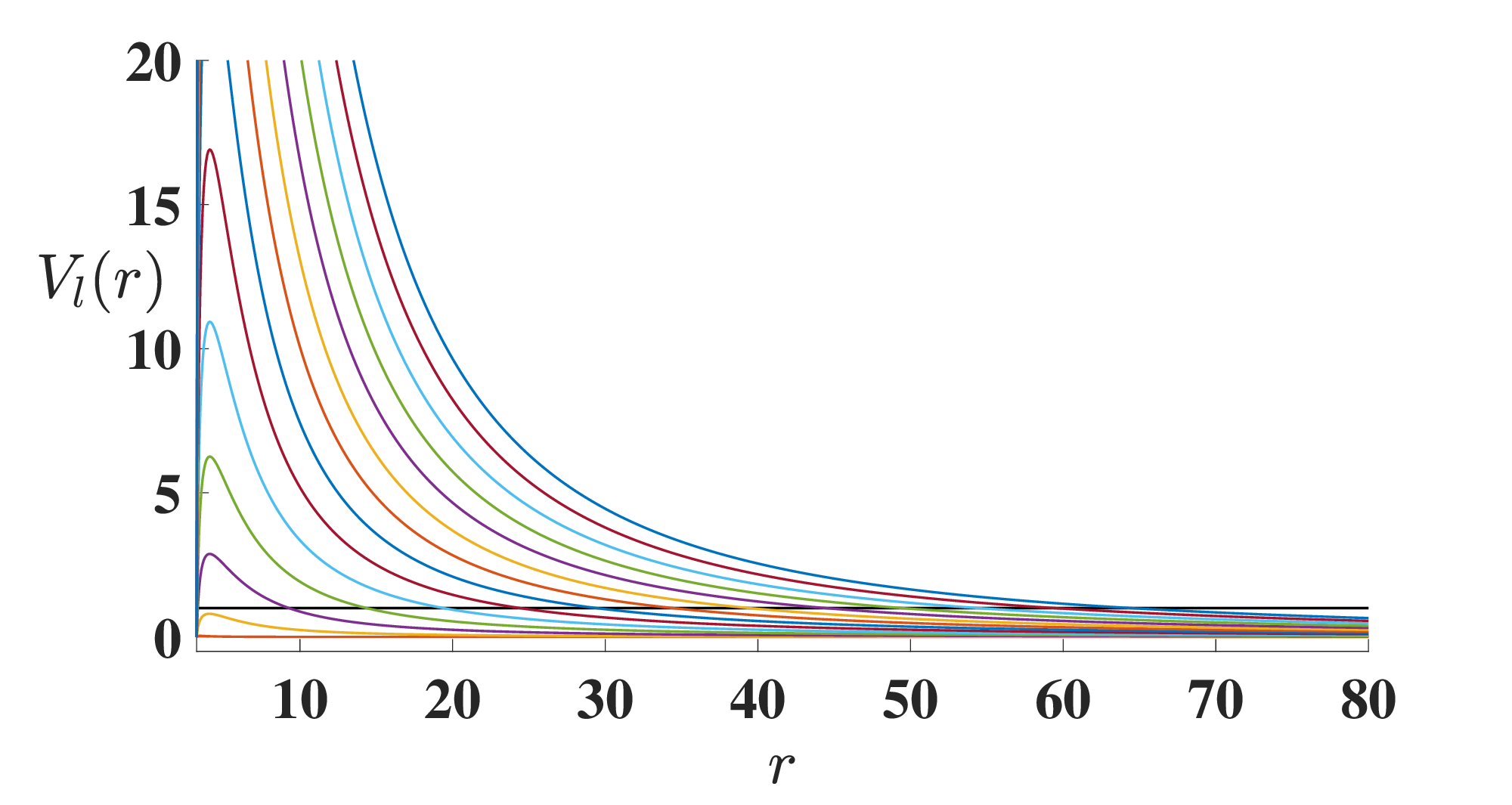}
\caption{}
\label{subfig:Vl-r g}
\end{subfigure}
\hspace{0\textwidth}  
\begin{subfigure}[b]{0.32\textwidth}
\centering
\includegraphics[width=\textwidth]{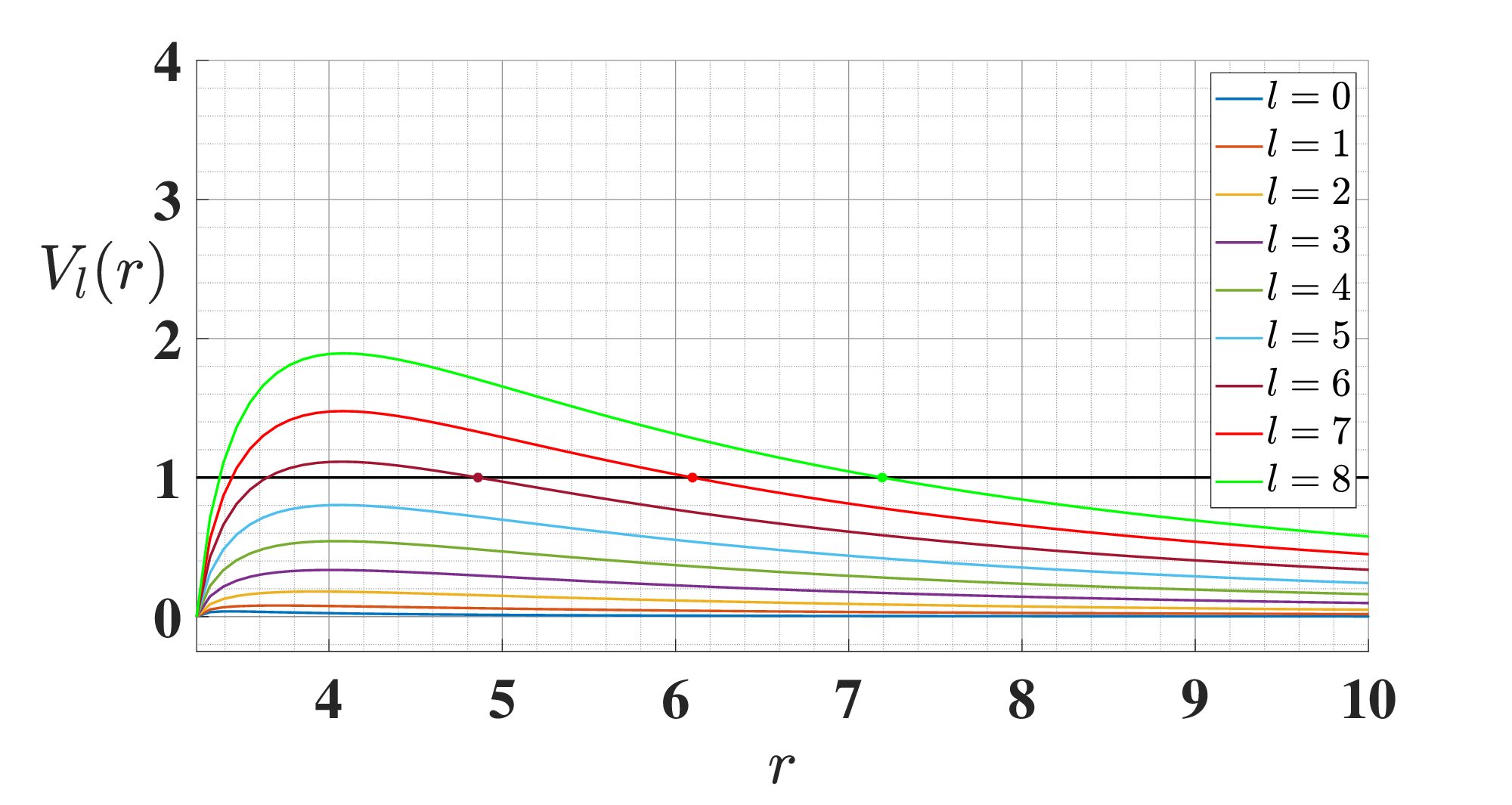}
\caption{}
\label{subfig:Vl-r h}
\end{subfigure}
\hspace{0\textwidth}  
\begin{subfigure}[b]{0.32\textwidth}
\centering
\includegraphics[width=\textwidth]{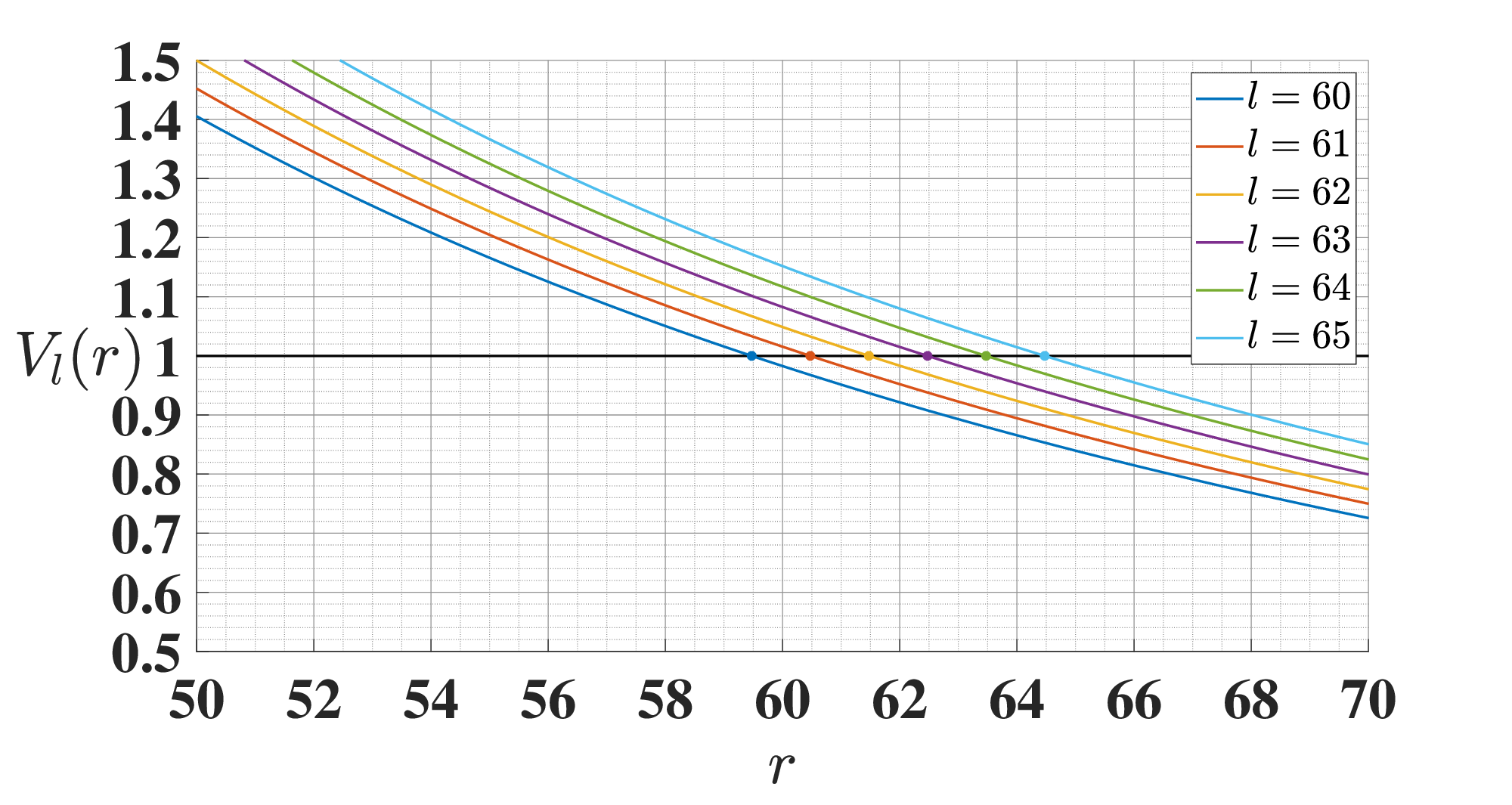}
\caption{}
\label{subfig:Vl-r i}
\end{subfigure}
\vspace{0\textwidth}  

\begin{subfigure}[b]{0.32\textwidth}
\centering
\includegraphics[width=\textwidth]{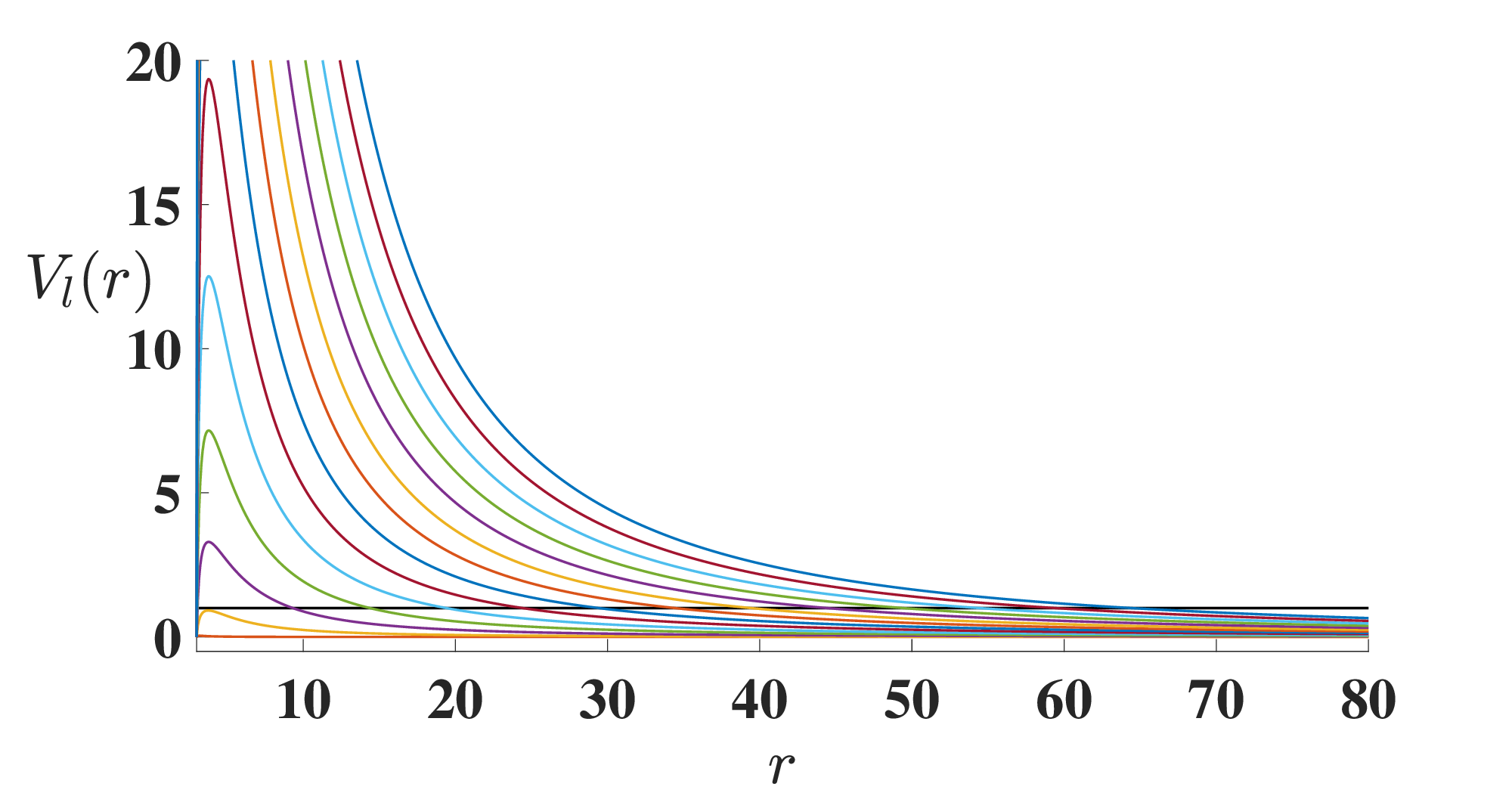}
\caption{}
\label{subfig:Vl-r j}
\end{subfigure}
\hspace{0\textwidth}  
\begin{subfigure}[b]{0.32\textwidth}
\centering
\includegraphics[width=\textwidth]{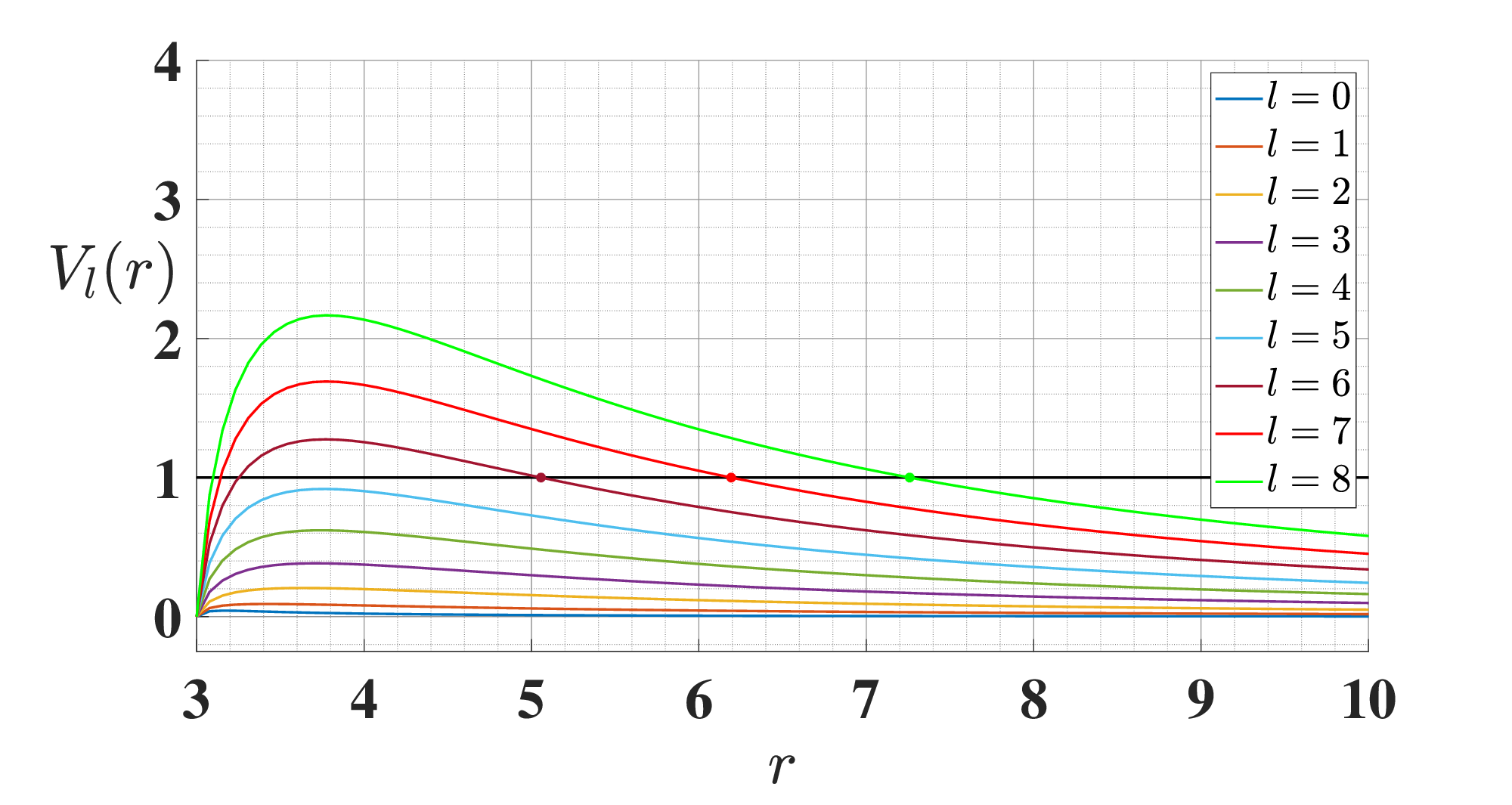}
\caption{}
\label{subfig:Vl-r k}
\end{subfigure}
\hspace{0\textwidth}  
\begin{subfigure}[b]{0.32\textwidth}
\centering
\includegraphics[width=\textwidth]{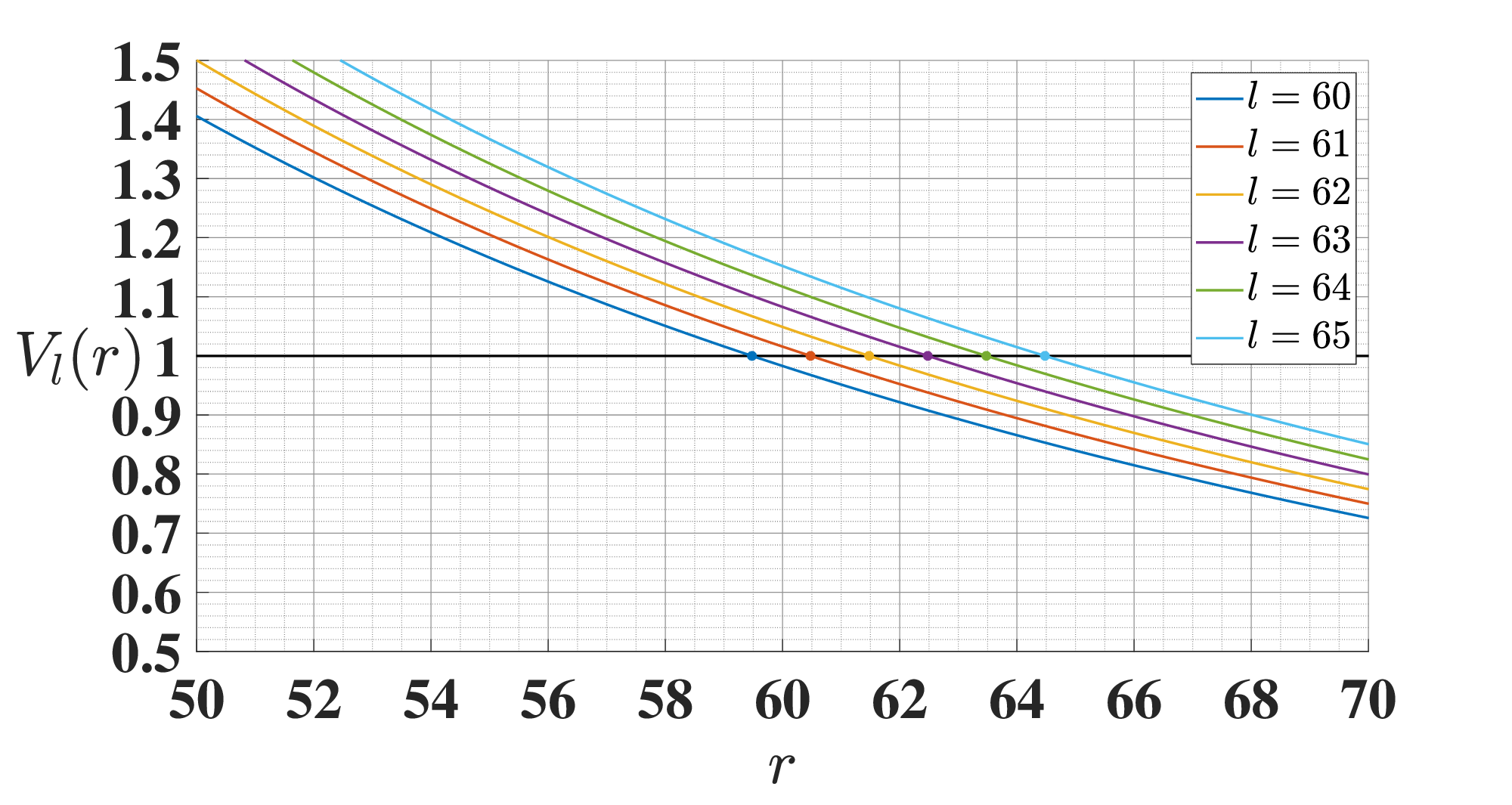}
\caption{}
\label{subfig:Vl-r l}
\end{subfigure}
\caption{The $V_\ell(r)-r$ curves. The first, second, third, and fourth rows represent the $V_\ell(r)-r$ curves for the Schwarzschild black hole, the RN black hole with $Q^2 = 0.5$, the conformal anomaly black hole with $Q^2=0$ and $\tilde{\alpha}=(\tilde{\alpha}_{\mathrm{max}} + \tilde{\alpha}_{\mathrm{min}})/2$, and the conformal anomaly black hole with $Q^2=0.5$ and $\tilde{\alpha}=(\tilde{\alpha}_{\mathrm{max}} + \tilde{\alpha}_{\mathrm{min}})/2$, respectively. The first column shows the global $V_\ell(r)-r$ curves from the event horizon to a finite distance. In each subfigure of the first column, the curves are arranged from bottom to top corresponding to $\ell = 0, 5, 10, \dots, 60, 65$. The second column displays the $V_\ell(r)-r$ curves in the vicinity of $V_\ell(r) = k^2$ for the low-$\ell$ modes. The second column displays the $V_\ell(r)-r$ curves in the vicinity of the larger root of $V_\ell(r) = k^2$ for the high-$\ell$ modes.}
\label{FIG:Vl-r}
\end{figure}

Before determining the truncation condition for the PWS, we first perform a qualitative analysis of the radial function based on the potential function given in Eq.\eqref{eq:10}. For the RN metric\eqref{eq:27} and the conformal anomaly metric\eqref{eq:26}, the potential function can be written respectively as
\begin{equation}
\label{eq:30}
\begin{split}
& V_{\text{RN} \, \ell}(r) = \left( 1 - \frac{2M}{r} + \frac{Q^2}{r^2} \right) \left[ \frac{2 M}{r^3} -\frac{2 Q^2}{r^4} + \frac{\ell (\ell + 1)}{r^2} \right] \,,\\
& V_{\text{Anomaly} \, \ell}(r) = \left[ 1 - \frac{r^2}{4\tilde{\alpha}} \left( 1 - \sqrt{1 - \frac{16\tilde{\alpha}M}{r^3} + \frac{8\tilde{\alpha}Q^2}{r^4}} \right) \right] \\
& \qquad \left[ \frac{6Mr^2-4Q^2}{r^4 \sqrt{1 - \frac{16\tilde{\alpha}M}{r^3} + \frac{8\tilde{\alpha}Q^2}{r^4}}} - \frac{1}{2\tilde{\alpha}} \left( 1 - \sqrt{1 - \frac{16\tilde{\alpha}M}{r^3} + \frac{8\tilde{\alpha}Q^2}{r^4}} \right) + \frac{\ell (\ell + 1)}{r^2}  \right] \,.
\end{split}
\end{equation}
In FIG.\ref{FIG:Vl-r}, we display the variation of the potential function as a function of the radial coordinate $r$ for various $\ell$ modes. Combining FIG.\ref{FIG:Vl-r} and Eq.\eqref{eq:30}, we find that for large-$\ell$ modes, the height of the potential barrier increases as $\ell^2$. For large-$\ell$ modes (approximately corresponding to $\ell > 10 r_+$), the value of $r$ far from the outer event horizon that satisfies $V_\ell(r) = k^2$ approximately satisfies $kr \approx \ell$. For different metrics and parameters, when $k=1$, the lowest few $\ell$-modes are all below $k^2$, meaning that the height of the potential barrier is lower than the energy of the scalar wave. In addition, by combining Eq.\eqref{eq:30} and FIG.\ref{FIG:Vl-r}, it can also be observed that as the mode number $\ell$ increases, the influence of the studied metrics and parameters on the height of the potential barrier remains almost unchanged, but their influence on the width of the barrier gradually decreases(i.e., the width of the potential barrier depends almost exclusively on $\ell$ and is independent of the metrics and parameters). 
Through numerical calculations and formula analysis, we have also found that for the RN metric, the position of the peak of the potential barrier shifts to the left as $Q$ increases. For the conformal anomaly metric, the peak position shifts to the left as $Q$ increases and to the right as $\tilde{\alpha}$ increases. However, the influence of $Q$ is more pronounced. Specifically, when $Q_M$ is fixed, the peak position shifts to the left as both $Q$ and $\tilde{\alpha}$ increase ($Q^2 = Q_M^2 + 2\tilde{\alpha}$). For the RN metric, when $\ell$ is sufficiently large (at the peak, only the $\ell(\ell+1)$ term is considered), the position of the potential barrier peak and the height of the barrier approximately satisfie
\begin{equation}
\label{eq:31}
\begin{split}
& r_{\mathrm{peak}} \approx \frac{1}{2} \left(3 M + \sqrt{9 M^2-8 Q^2}\right) \,,\\
& V_{\mathrm{RN} \ell}\left( r_{\mathrm{peak}} \right) \approx \frac{8 \left(M \sqrt{9 M^2-8 Q^2}+3 M^2-2 Q^2\right)}{\left(\sqrt{9 M^2-8 Q^2}+3 M\right)^4} \ell(\ell+1) \,.
\end{split}
\end{equation}
For the conformal anomaly metric, an approximate expression like the one above cannot be found. However, the influence pattern of $Q$ is the same as that for the RN metric, while the influence of $\tilde{\alpha}$ is opposite to that of $Q$.

\begin{figure}[htbp]
\centering
\begin{subfigure}[b]{0.49\textwidth}
\centering
\includegraphics[width=\textwidth]{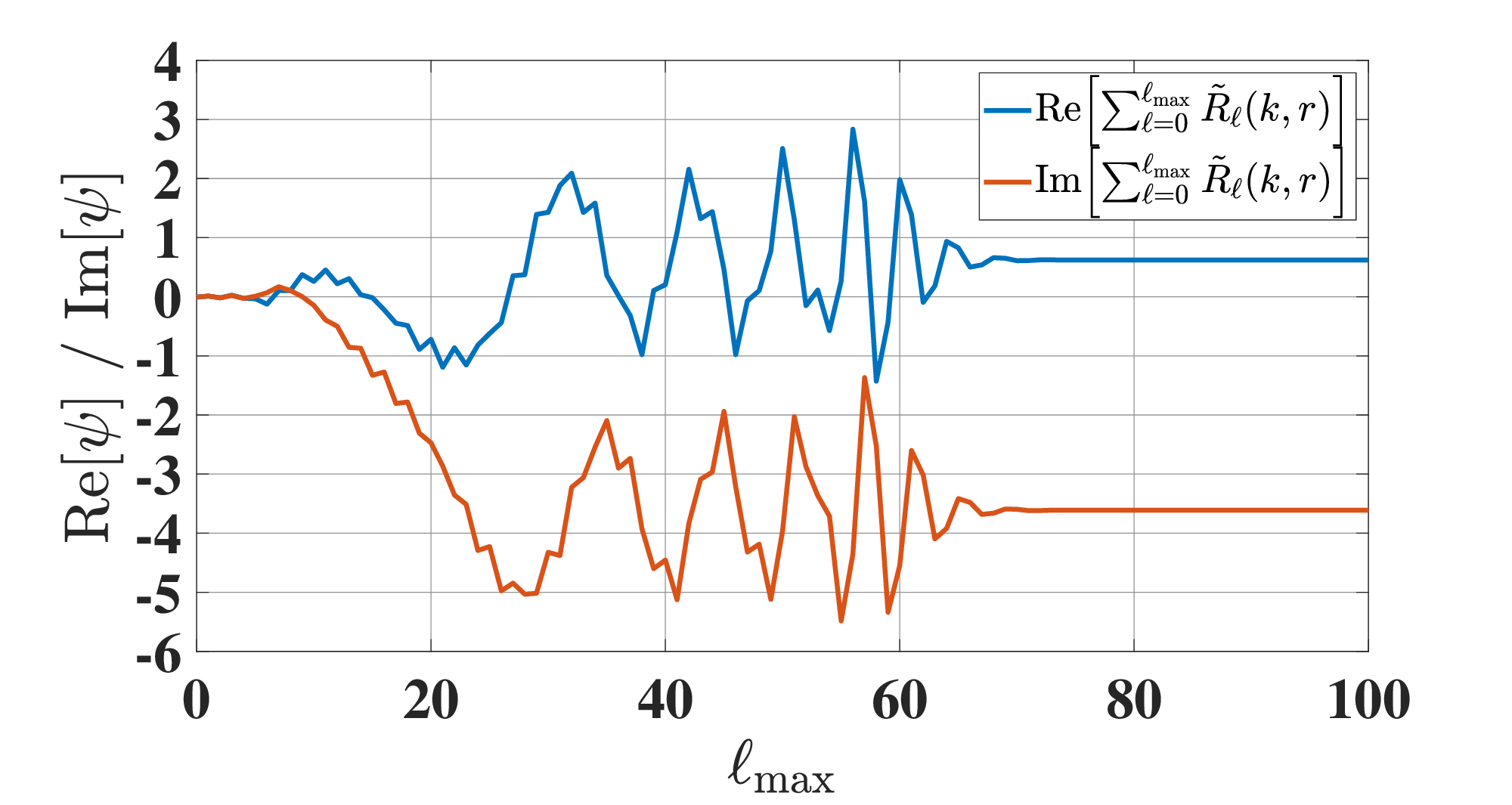}
\caption{}
\label{subfig:lMax_RN_Q_0}
\end{subfigure}
\hspace{0\textwidth}  
\begin{subfigure}[b]{0.49\textwidth}
\centering
\includegraphics[width=\textwidth]{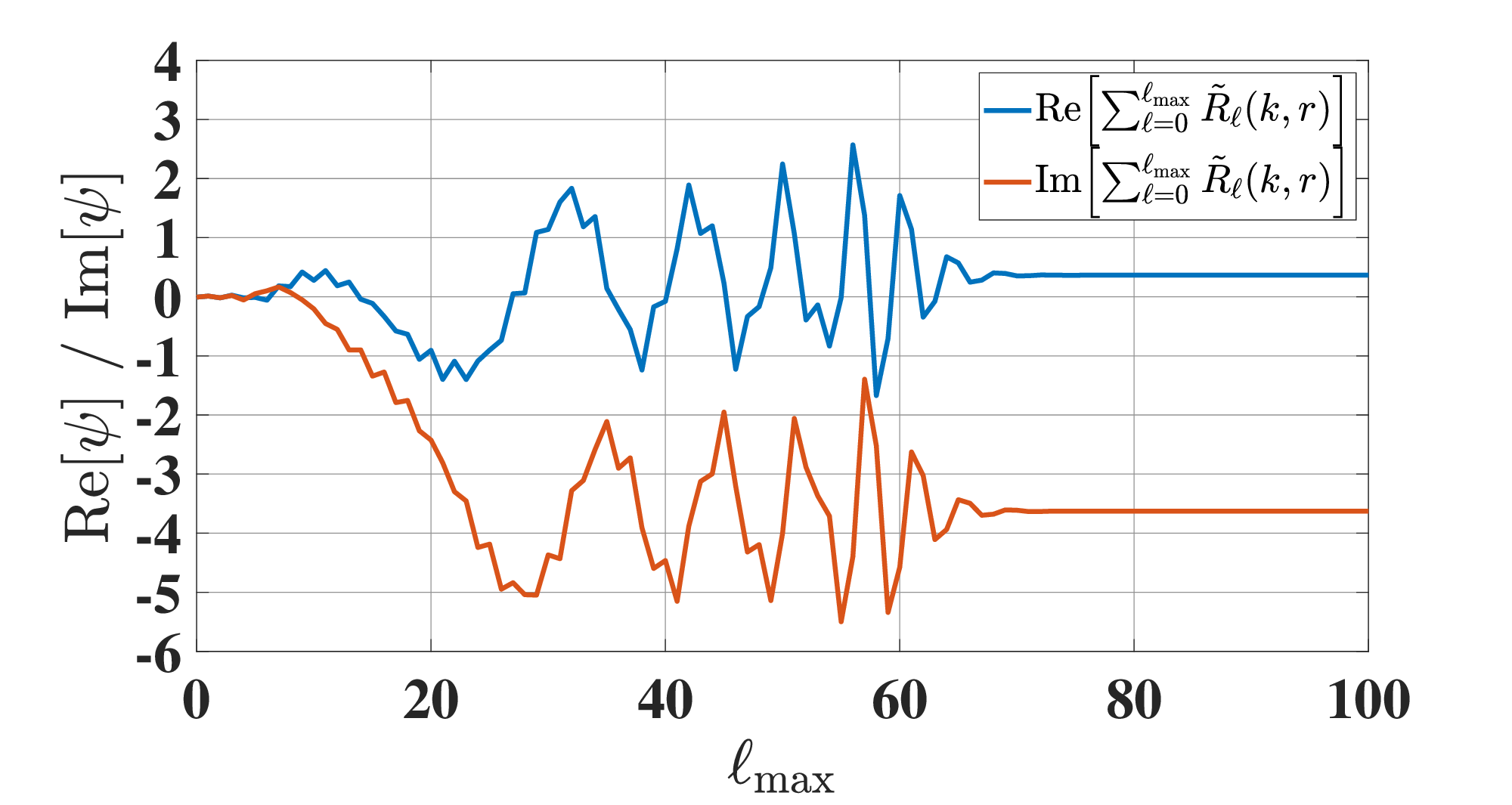}
\caption{}
\label{subfig:lMax_RN_Q_0.5}
\end{subfigure}

\vspace{0\textwidth}  

\begin{subfigure}[b]{0.49\textwidth}
\centering
\includegraphics[width=\textwidth]{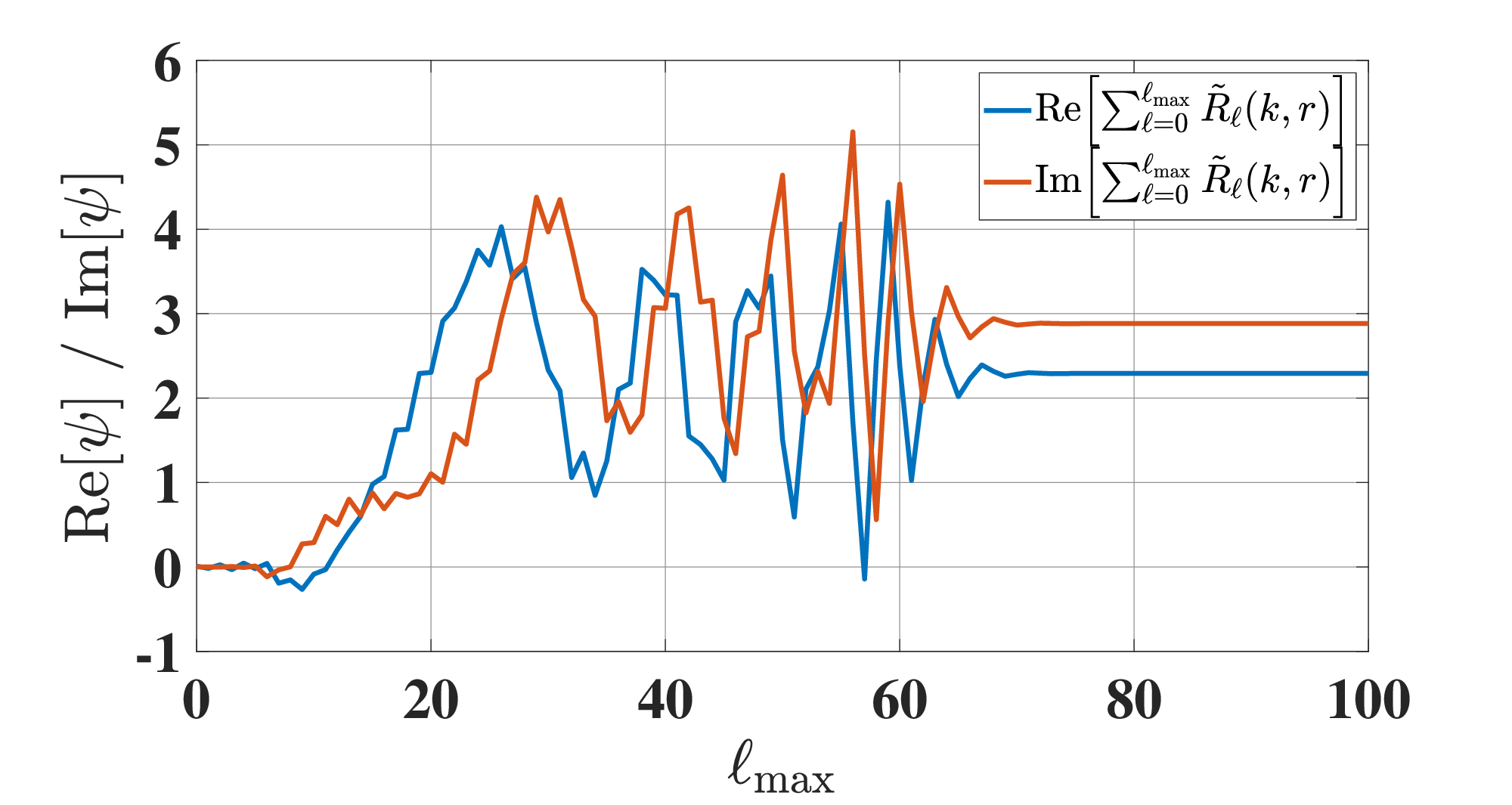}
\caption{}
\label{subfig:lMax_Anomaly_Q_0}
\end{subfigure}
\hspace{0\textwidth}  
\begin{subfigure}[b]{0.49\textwidth}
\centering
\includegraphics[width=\textwidth]{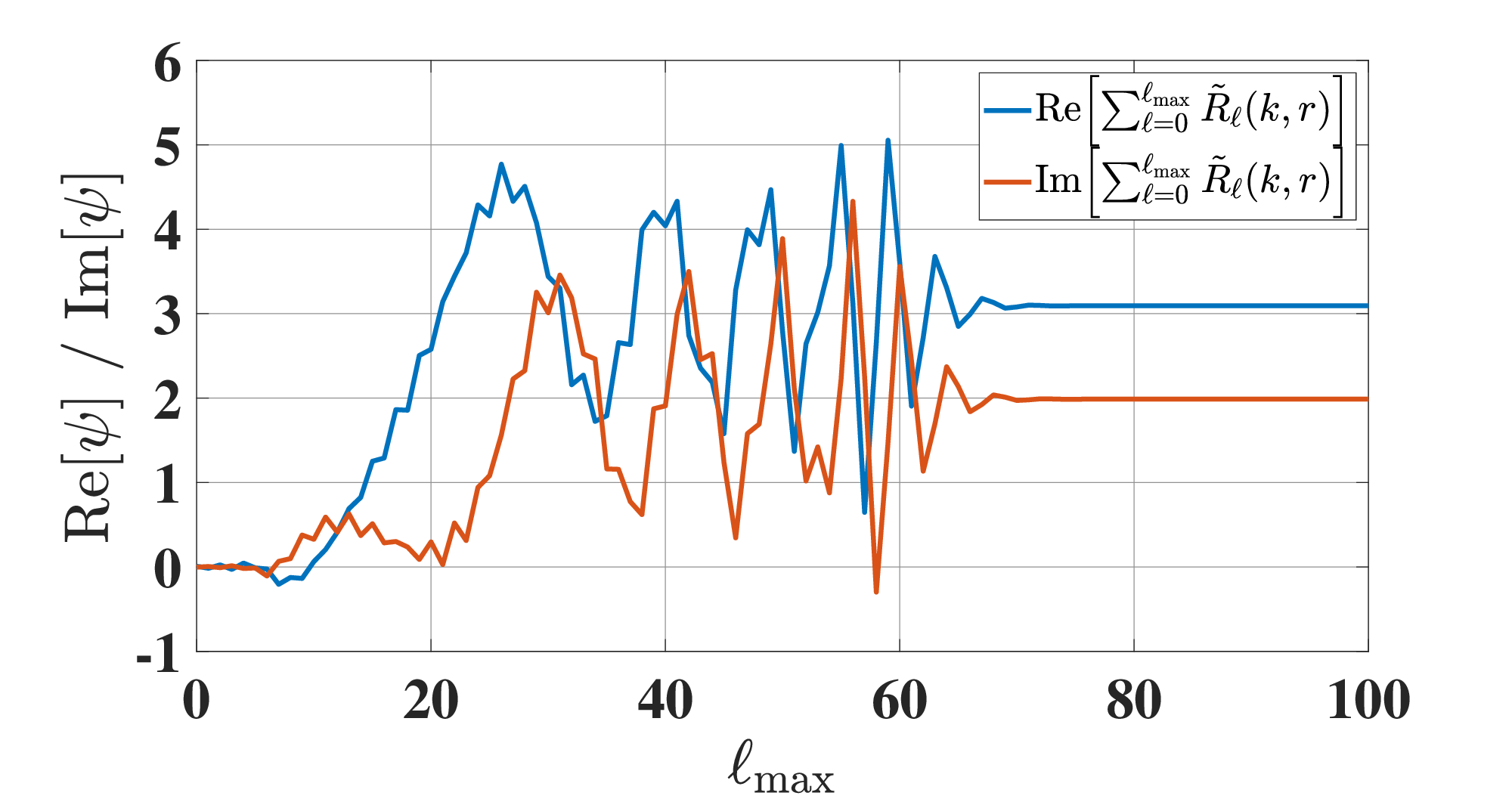}
\caption{}
\label{subfig:lMax_Anomaly_Q_0.5}
\end{subfigure}

\caption{Convergence of the PWS expansion for the total wave function. The observer is located at a finite distance of $r = 60M$ and the frequency of the wave function is taken as $k = 1$. Subfigures (a), (b), (c), and (d) correspond to the Schwarzschild black hole, the RN black hole with $Q^2 = 0.5$, the conformal anomaly black hole with $Q^2=0$ and $\tilde{\alpha}=(\tilde{\alpha}_{\mathrm{max}} + \tilde{\alpha}_{\mathrm{min}})/2$, and the conformal anomaly black hole with $Q^2=0.5$ and $\tilde{\alpha}=(\tilde{\alpha}_{\mathrm{max}} + \tilde{\alpha}_{\mathrm{min}})/2$, respectively. To more clearly demonstrate the contribution of each $\ell$-mode radial function, we set $\theta = 0$, at which point the Legendre function $\mathrm{P}_{\ell}(\cos\theta) = 1$.}
\label{FIG:lmax}
\end{figure}

For a plane wave incident from infinity, reflection and transmission will occur where the potential barrier exceeds $k^2$. The transmitted component will decay rapidly. Combining Eq.\eqref{eq:30}, Fig.\ref{FIG:Vl-r}, and the aforementioned analysis, it can be seen that the radial wave function of high-$\ell$ modes will decay rapidly and its amplitude is almost negligible inside the potential barrier(approximately $kr<\ell$). Considering the penetration depth of the incident wave in the potential barrier, the truncation in the PWS expansion for the value of the total wave function at radial coordinate $r$ should therefore be set with $\ell$ slightly greater than  $kr$. In FIG.\ref{FIG:lmax}, we demonstrate the convergence of the PWS expansion for the total wave function at a finite distance. This clearly shows that truncating the PWS at $\ell$ slightly greater than $kr$ is accurate, and contributions from higher $\ell$-modes to the total wave function are negligible. To generate FIG.\ref{FIG:lmax}, the total wave function must be computed numerically using the PWS method. In the following subsection, we will detail how to numerically calculate the total wave function with the PWS method and further investigate the waveforms.

\subsection{Investigation of Waveform}
Consider the physical process in which a plane wave incident from $z \to -\infty$ and propagating along the positive $z$-direction is scattered by a black hole. Due to the limitation of the PWS method, which cannot compute modes up to an infinite $\ell$, in this subsection we will investigate the waveform diagrams within a finite distance. Combining Eqs.\eqref{eq:8}–\eqref{eq:15} and considering the RN metric\eqref{eq:27} and the conformal anomaly metric\eqref{eq:26} under study, the numerical calculation of the total wave function at a finite distance can be performed. A special note on the calculation of the tortoise coordinate is needed. Eq.\eqref{eq:9} gives the differential expression for the tortoise coordinate, but the expression required in practical calculation is the functional form $r_\ast(r)$. For the RN black hole, the tortoise coordinate can be written as
\begin{equation}
\label{eq:32}
r_{\ast}\left( r \right) = r + \frac{r_+^2}{r_+ - r_-} \ln \frac{|r - r_{+}|}{2M} - \frac{r_-^2}{r_+ - r_-} \ln \frac{|r - r_{-}|}{2M} \,,
\end{equation}
where $r_+$ and $r_-$ satisfy the relation $r_{\pm} = M \pm \sqrt{M^2 - Q^2}$ and denote the outer and inner event horizons, respectively. For a general static spherically symmetric metric (including the conformal anomaly metric), an analytical expression for the tortoise coordinate cannot be obtained as it can for the RN metric. Numerical integration is hindered by the fact that the right-hand side of Eq.\eqref{eq:9} diverges at the outer event horizon. Therefore, for the tortoise coordinate integration of a general static spherically symmetric metric, a series expansion at the outer event horizon must be employed, retaining terms up to first order
\begin{equation}
\label{eq:33}
f \left( r \right) = f \left( r_+ \right) + f' \left( r_+ \right) \left( r - r_+ \right) + o \left( \left( r - r_+ \right)^2 \right) \,.
\end{equation}
When  $r_0 - r_+ \ll 1$, the tortoise coordinate $r_\ast(r_0)$ can be written as
\begin{equation}
\label{eq:34}
r_{\ast}\left( r_0 \right) = \frac{1}{f'\left( r_+ \right)} \left[ \ln\left( r-r_+ \right) - \ln 0_+ \right] + C = \frac{1}{f'\left( r_+ \right)} \ln\left( r-r_+ \right)\,.
\end{equation}
Because even with a first-order approximation, the value of the tortoise coordinate at $r_+$ should be $-\infty$, we may therefore consider setting the integration constant to $C = \frac{\ln 0_+}{f'\left( r_+ \right)}$. For  $r > r_0$, the formula for computing $r_\ast(r)$ using numerical integration is
\begin{equation}
\label{eq:35}
r_{\ast} = \frac{1}{f'\left( r_+ \right)} \ln\left( r-r_+ \right) + \int_{r_0}^r \frac{1}{f \left( r \right)} \mathrm{d} r \,.
\end{equation}
In numerical calculations, one may take $r_0 = r_+ + 10^{-5} r_+$.

\begin{figure}[htbp]
\centering
\begin{subfigure}[b]{0.24\textwidth}
\centering
\includegraphics[width=\textwidth]{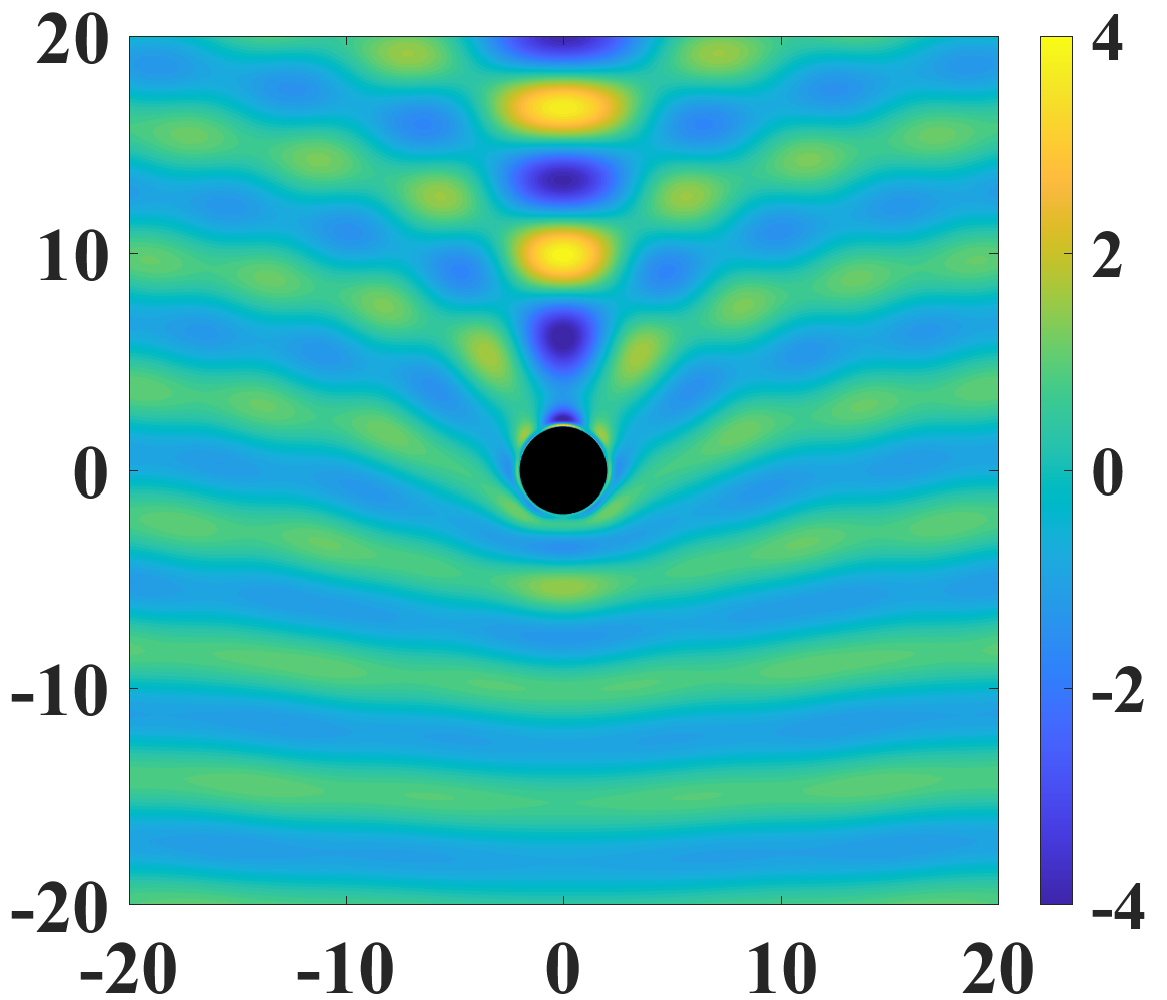}
\caption{$\mathrm{Re}\left( \psi|_{k=1} \right)$}
\label{subfig:Waveform a}
\end{subfigure}
\hspace{0\textwidth}  
\begin{subfigure}[b]{0.24\textwidth}
\centering
\includegraphics[width=\textwidth]{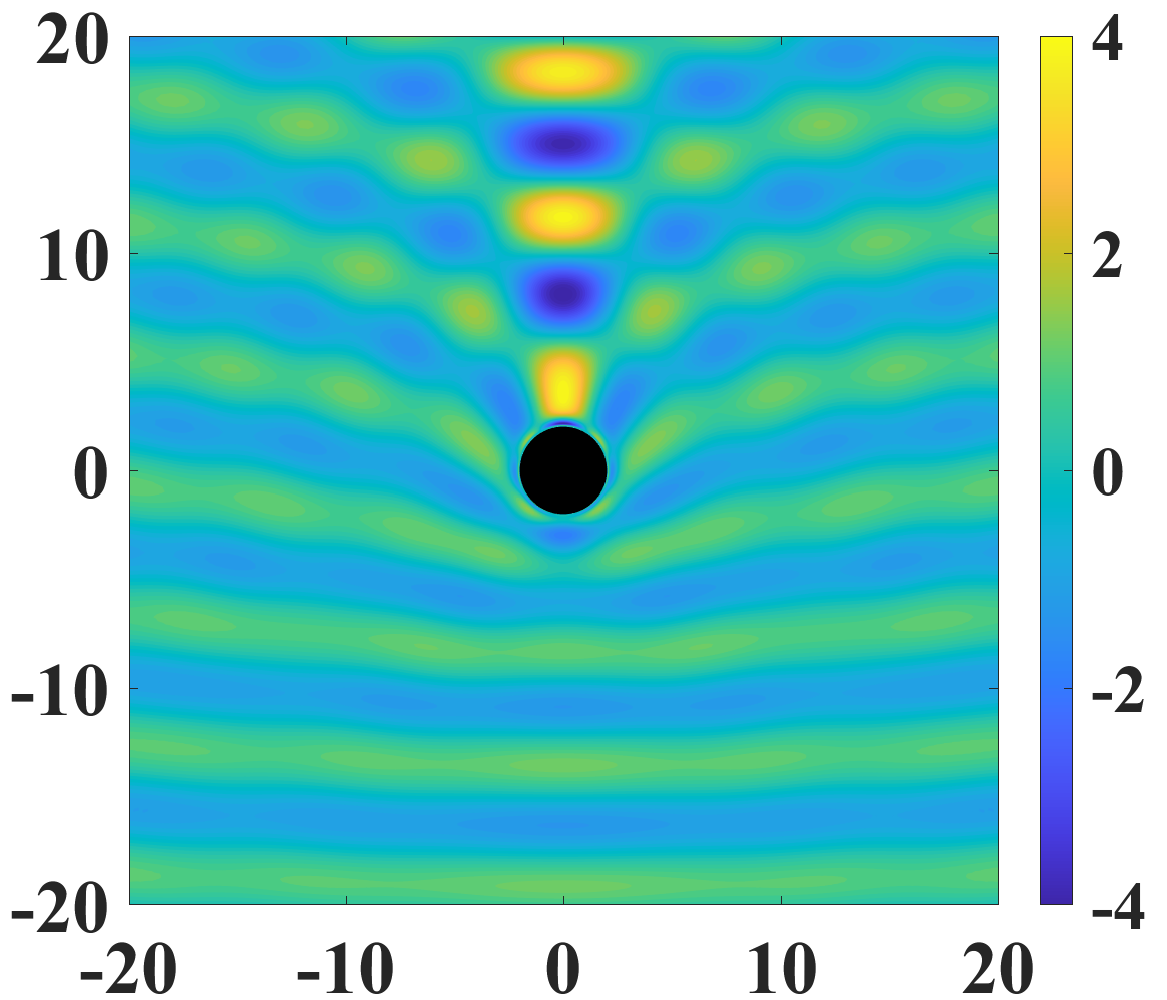}
\caption{$\mathrm{Im}\left( \psi|_{k=1} \right)$}
\label{subfig:Waveform a}
\end{subfigure}
\hspace{0\textwidth}  
\begin{subfigure}[b]{0.24\textwidth}
\centering
\includegraphics[width=\textwidth]{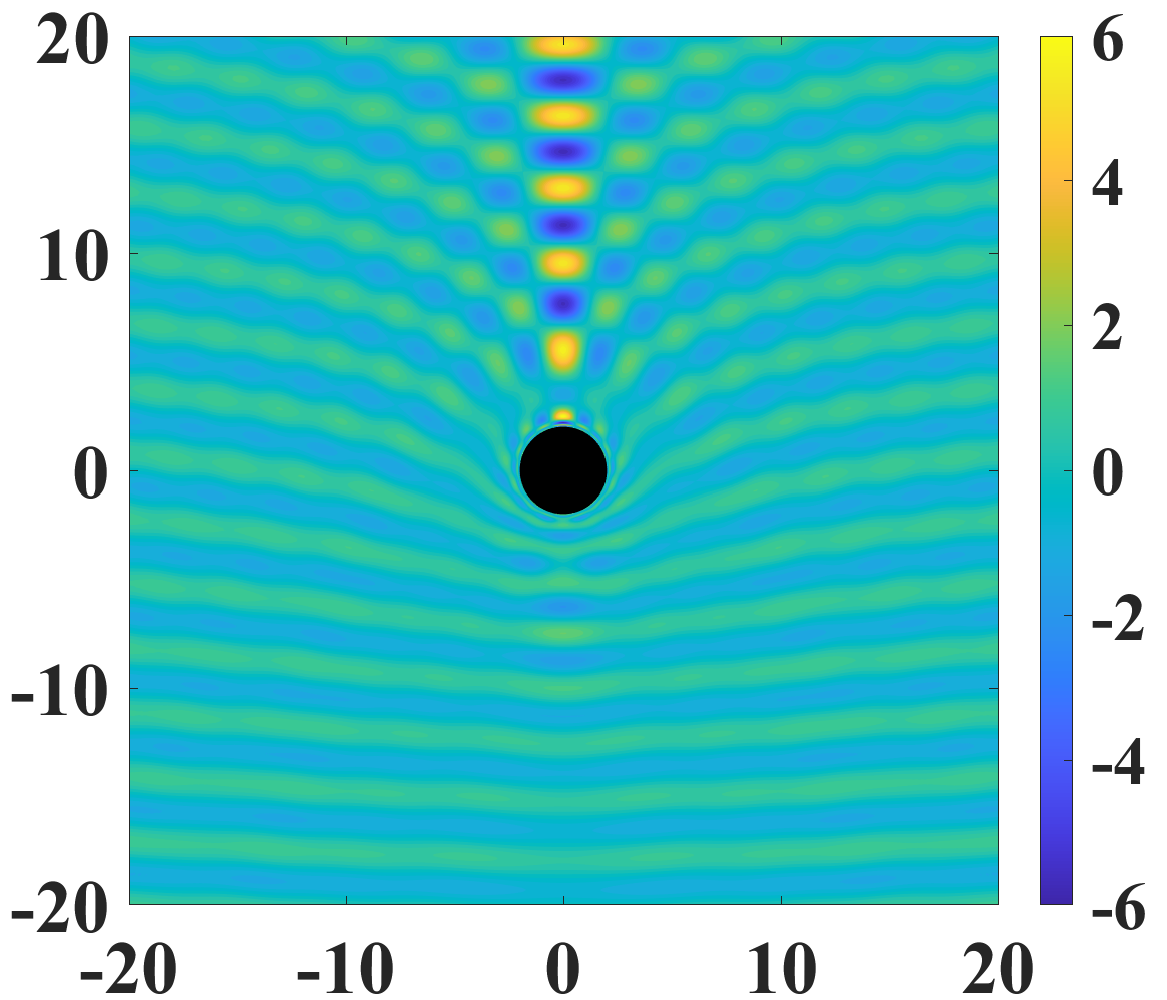}
\caption{$\mathrm{Re}\left( \psi|_{k=2} \right)$}
\label{subfig:Waveform c}
\end{subfigure}
\hspace{0\textwidth}  
\begin{subfigure}[b]{0.24\textwidth}
\centering
\includegraphics[width=\textwidth]{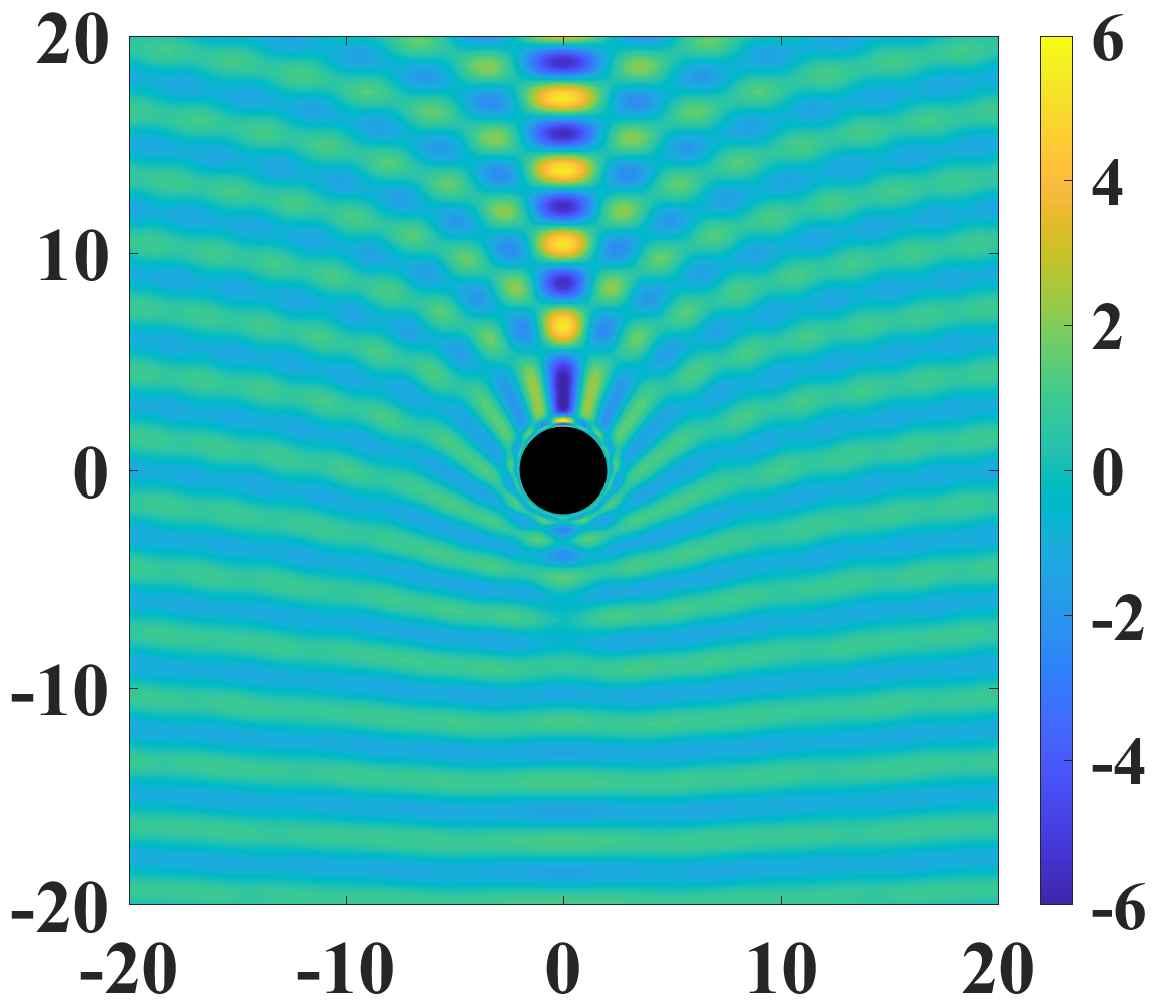}
\caption{$\mathrm{Im}\left( \psi|_{k=2} \right)$}
\label{subfig:Waveform d}
\end{subfigure}

\vspace{0\textwidth}  

\begin{subfigure}[b]{0.24\textwidth}
\centering
\includegraphics[width=\textwidth]{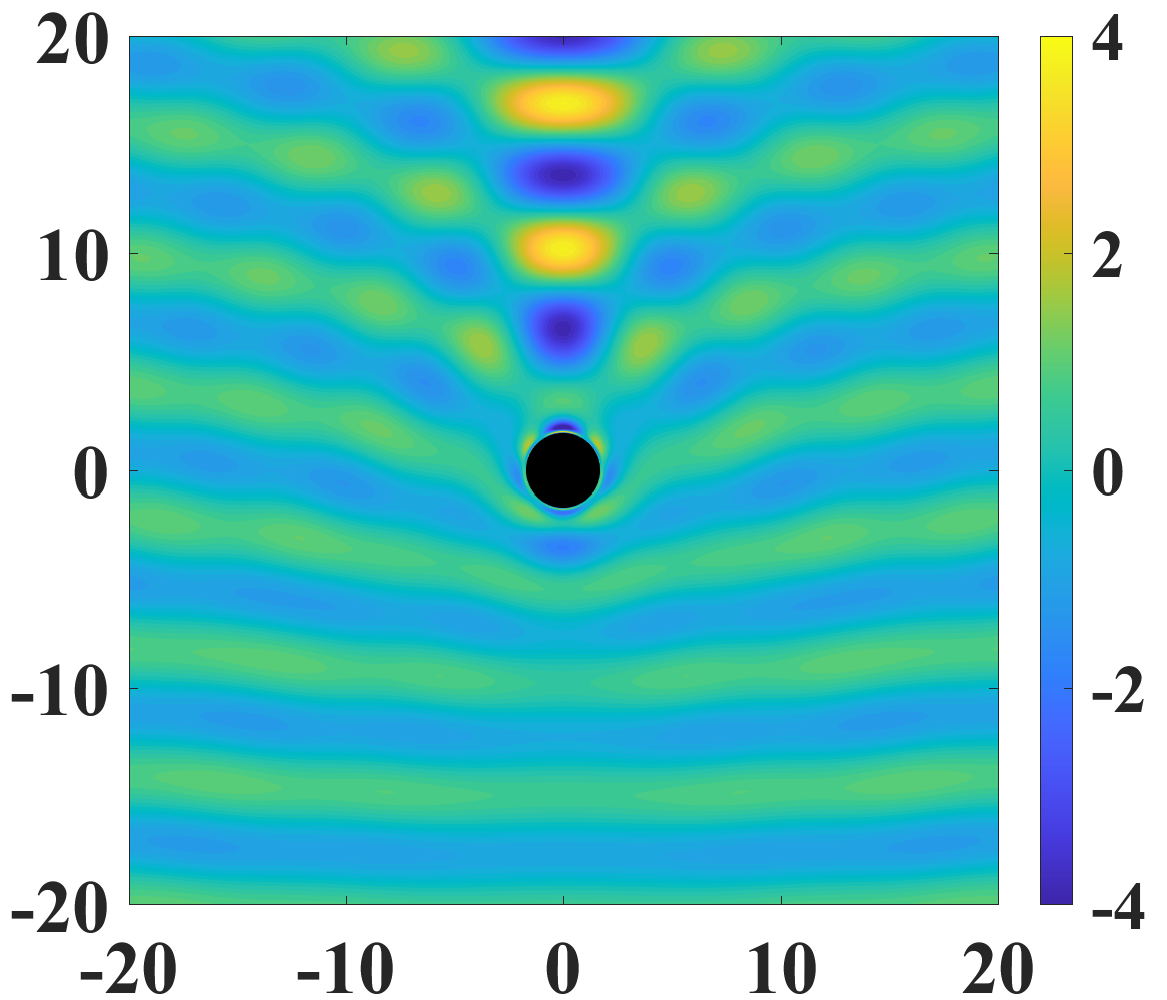}
\caption{$\mathrm{Re}\left( \psi|_{k=1} \right)$}
\label{subfig:Waveform e}
\end{subfigure}
\hspace{0\textwidth}  
\begin{subfigure}[b]{0.24\textwidth}
\centering
\includegraphics[width=\textwidth]{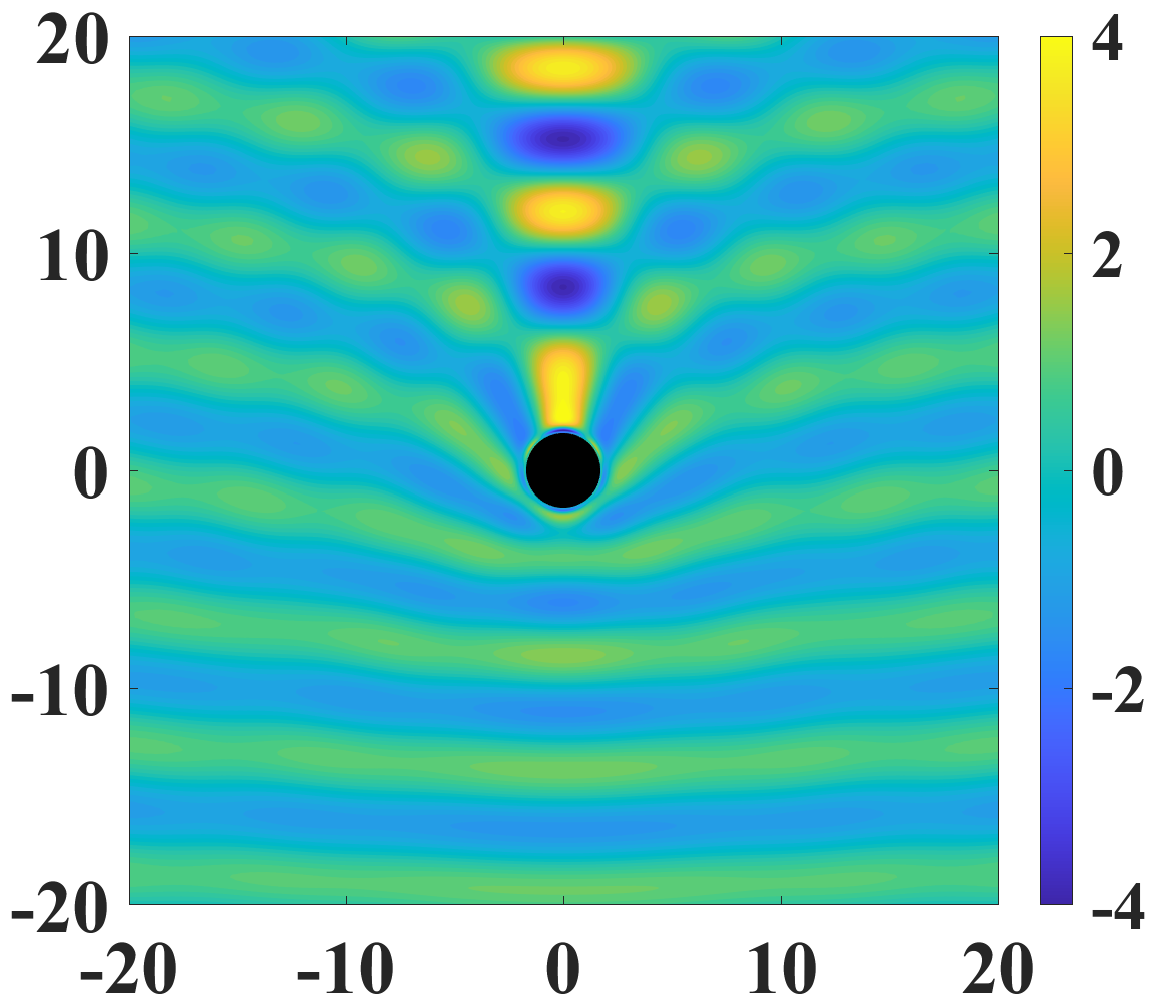}
\caption{$\mathrm{Im}\left( \psi|_{k=1} \right)$}
\label{subfig:Waveform f}
\end{subfigure}
\hspace{0\textwidth}  
\begin{subfigure}[b]{0.24\textwidth}
\centering
\includegraphics[width=\textwidth]{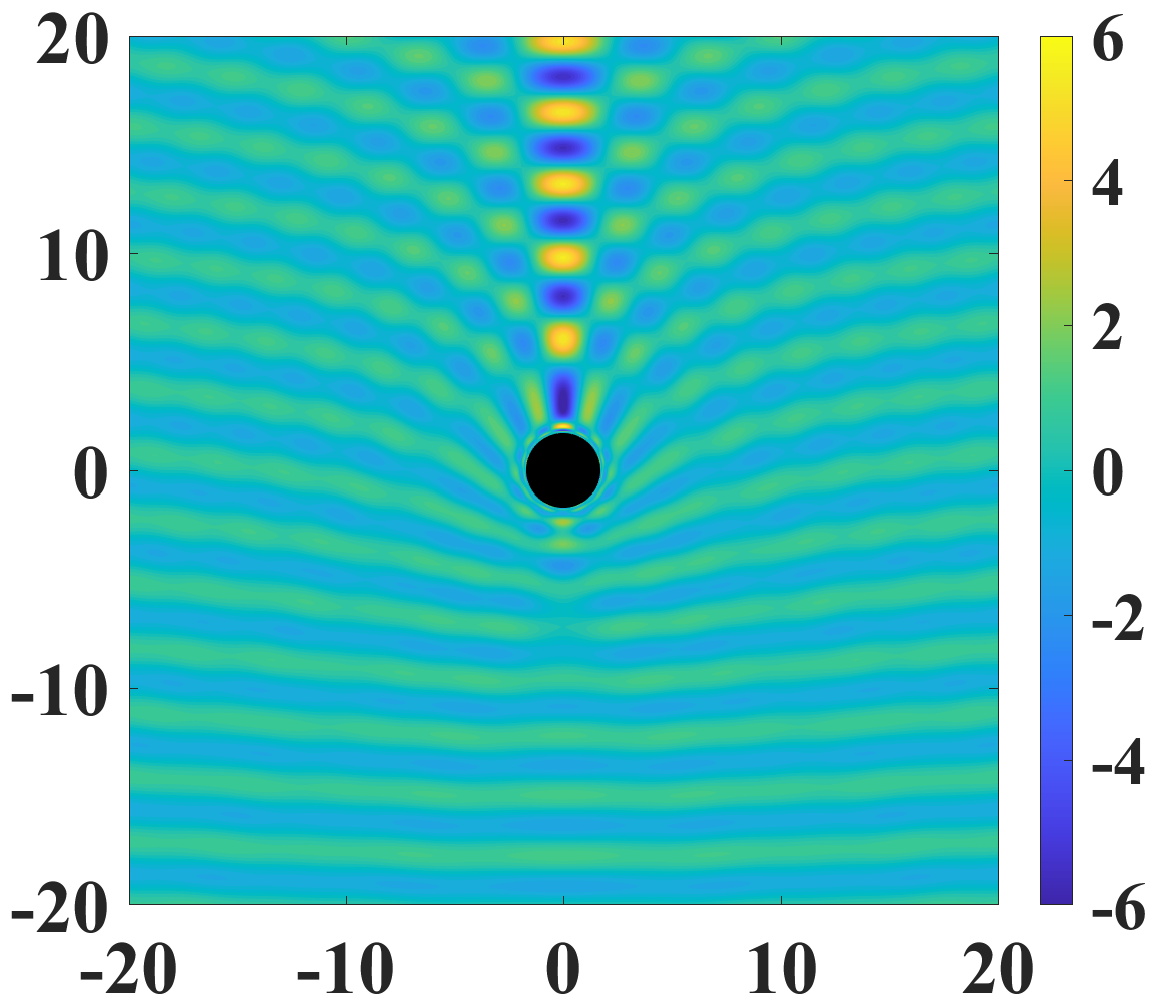}
\caption{$\mathrm{Re}\left( \psi|_{k=2} \right)$}
\label{subfig:Waveform g}
\end{subfigure}
\hspace{0\textwidth}  
\begin{subfigure}[b]{0.24\textwidth}
\centering
\includegraphics[width=\textwidth]{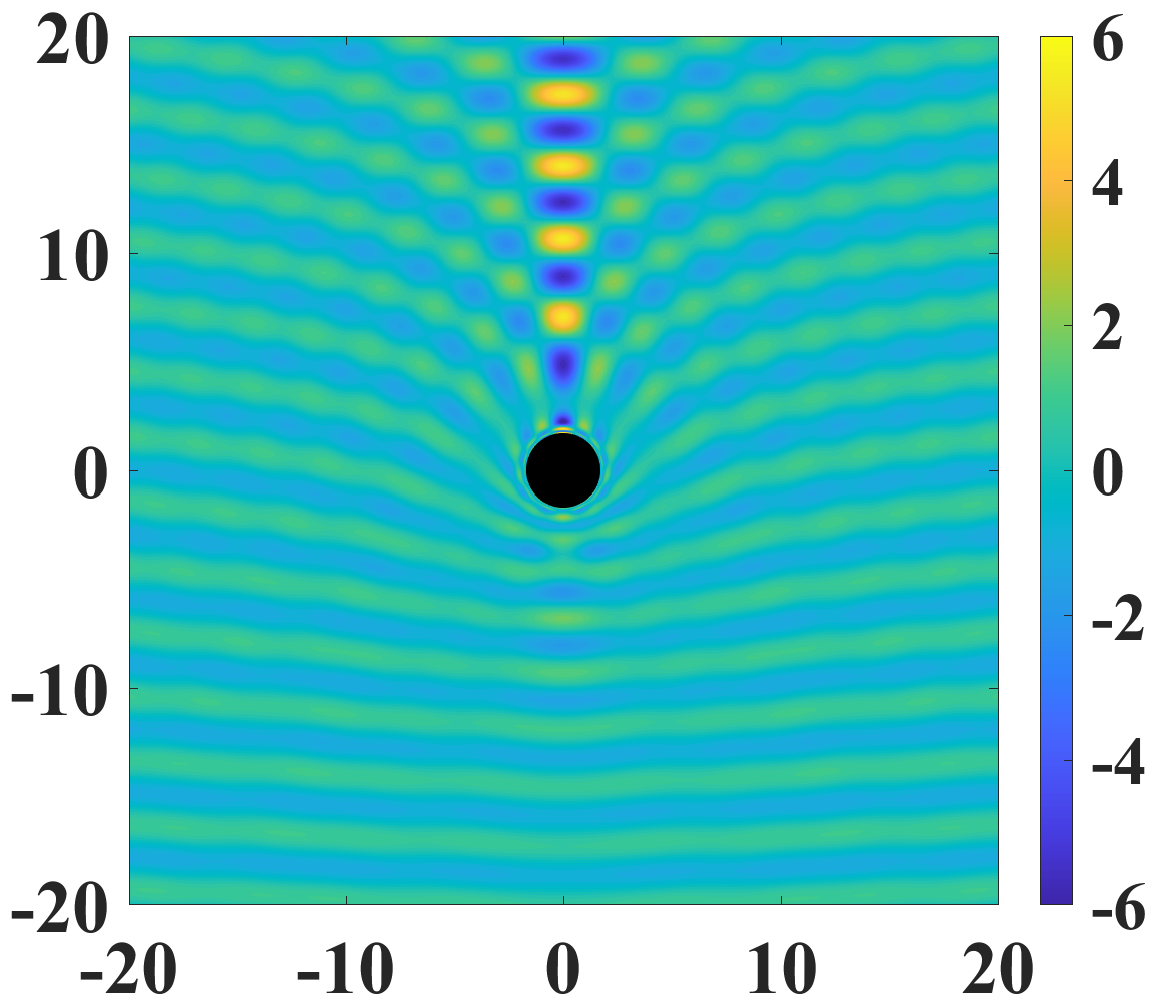}
\caption{$\mathrm{Im}\left( \psi|_{k=2} \right)$}
\label{subfig:Waveform h}
\end{subfigure}

\caption{
The effect of charge: Waveforms of the Schwarzschild and RN black holes. The first and second rows represent the waveform diagrams for the Schwarzschild black hole and the RN black hole with $Q^2 = 0.5$, respectively.}
\label{FIG:Waveform1}
\end{figure}

\begin{figure}[htbp]
\centering
\begin{subfigure}[b]{0.24\textwidth}
\centering
\includegraphics[width=\textwidth]{Fig/Re_RN_Q_0_k1_Waveform.eps}
\caption{$\mathrm{Re}\left( \psi|_{k=1} \right)$}
\label{subfig:Waveform a}
\end{subfigure}
\hspace{0\textwidth}  
\begin{subfigure}[b]{0.24\textwidth}
\centering
\includegraphics[width=\textwidth]{Fig/Im_RN_Q_0_k1_Waveform.eps}
\caption{$\mathrm{Im}\left( \psi|_{k=1} \right)$}
\label{subfig:Waveform a}
\end{subfigure}
\hspace{0\textwidth}  
\begin{subfigure}[b]{0.24\textwidth}
\centering
\includegraphics[width=\textwidth]{Fig/Re_RN_Q_0_k2_Waveform.eps}
\caption{$\mathrm{Re}\left( \psi|_{k=2} \right)$}
\label{subfig:Waveform c}
\end{subfigure}
\hspace{0\textwidth}  
\begin{subfigure}[b]{0.24\textwidth}
\centering
\includegraphics[width=\textwidth]{Fig/Im_RN_Q_0_k2_Waveform.eps}
\caption{$\mathrm{Im}\left( \psi|_{k=2} \right)$}
\label{subfig:Waveform d}
\end{subfigure}

\vspace{0\textwidth}  

\begin{subfigure}[b]{0.24\textwidth}
\centering
\includegraphics[width=\textwidth]{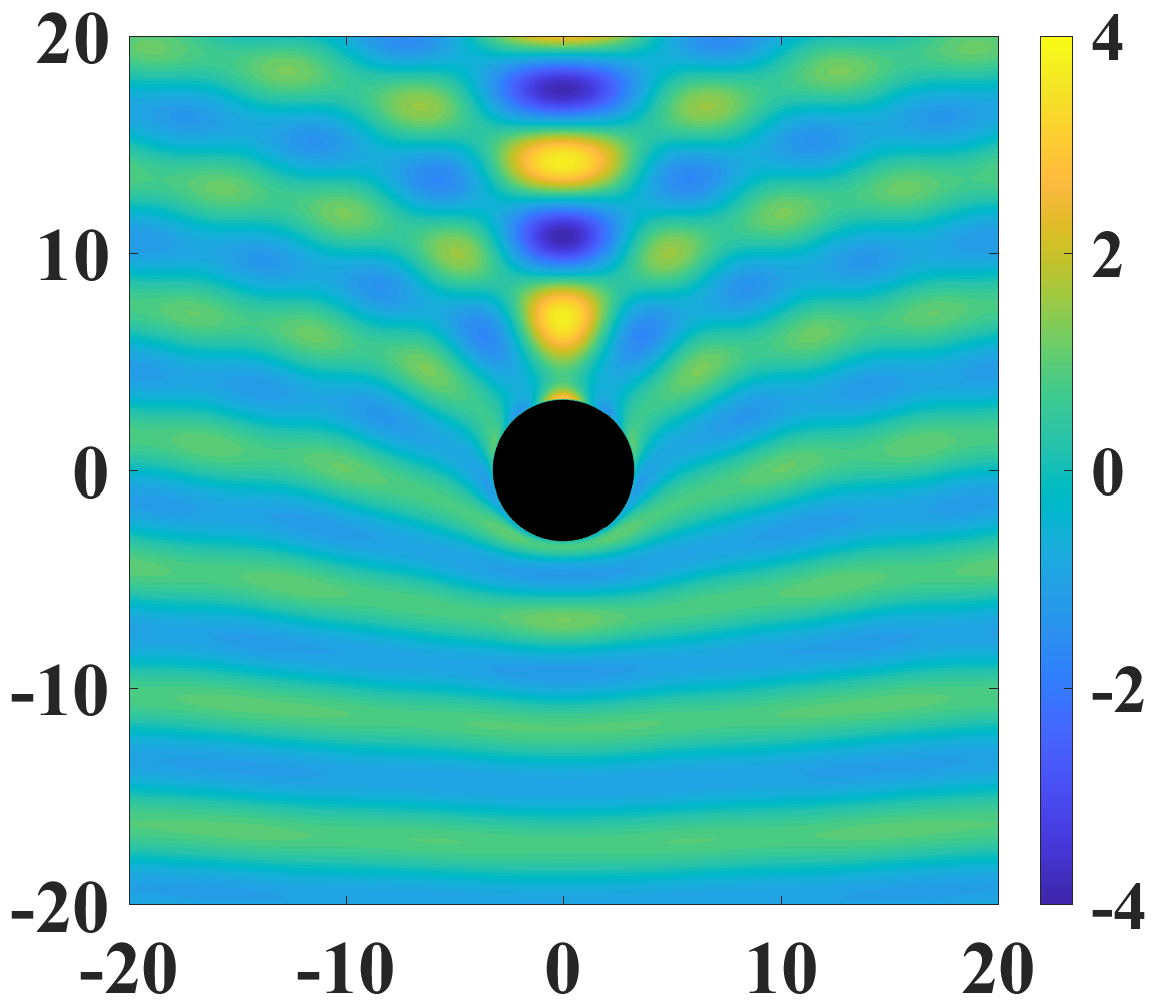}
\caption{$\mathrm{Re}\left( \psi|_{k=1} \right)$}
\label{subfig:Waveform e}
\end{subfigure}
\hspace{0\textwidth}  
\begin{subfigure}[b]{0.24\textwidth}
\centering
\includegraphics[width=\textwidth]{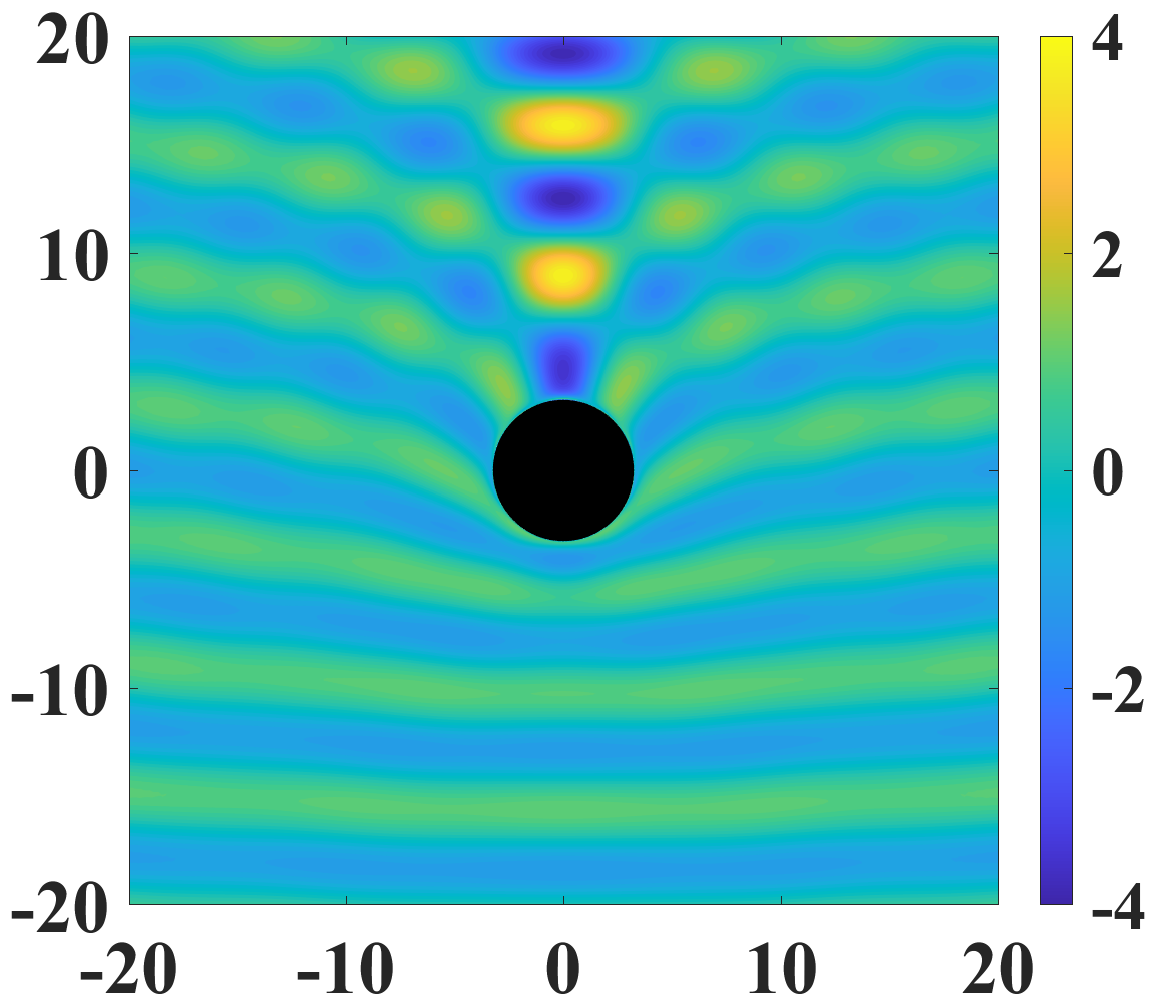}
\caption{$\mathrm{Im}\left( \psi|_{k=1} \right)$}
\label{subfig:Waveform f}
\end{subfigure}
\hspace{0\textwidth}  
\begin{subfigure}[b]{0.24\textwidth}
\centering
\includegraphics[width=\textwidth]{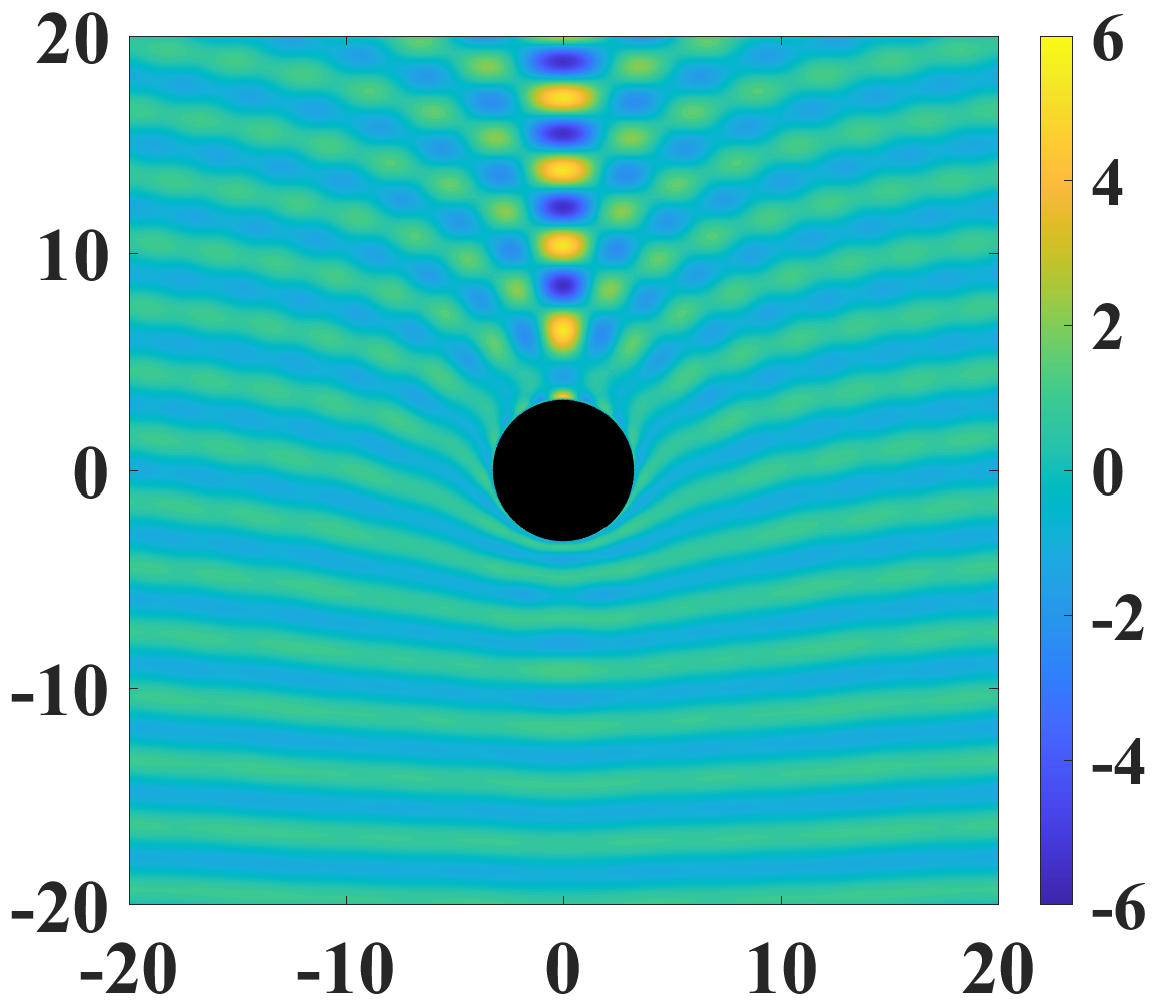}
\caption{$\mathrm{Re}\left( \psi|_{k=2} \right)$}
\label{subfig:Waveform g}
\end{subfigure}
\hspace{0\textwidth}  
\begin{subfigure}[b]{0.24\textwidth}
\centering
\includegraphics[width=\textwidth]{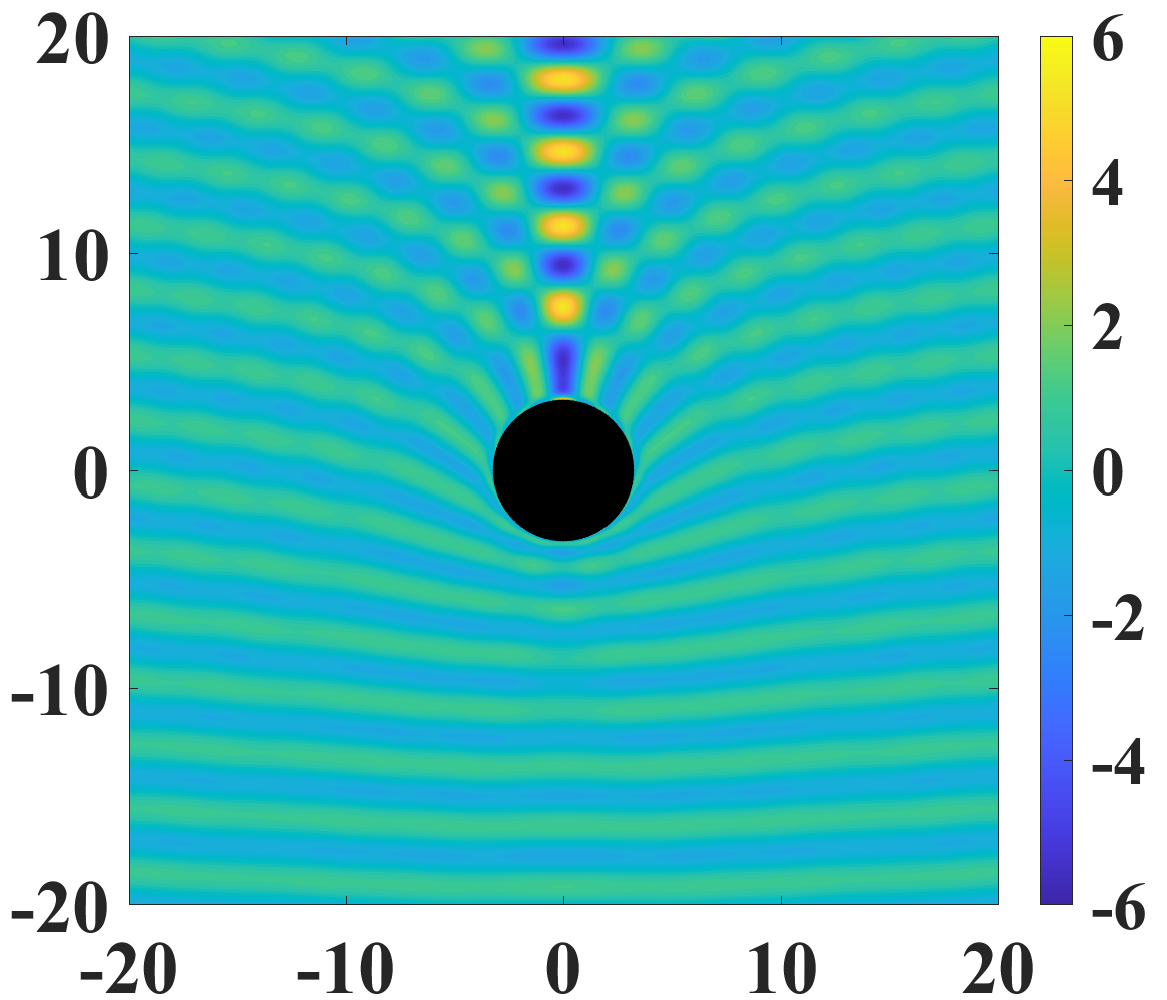}
\caption{$\mathrm{Im}\left( \psi|_{k=2} \right)$}
\label{subfig:Waveform h}
\end{subfigure}

\caption{The effect of conformal anomaly charge: Waveforms of the Schwarzschild and conformal anomaly black holes. The first and second rows represent the waveform diagrams for the Schwarzschild black hole and the conformal anomaly black hole with $Q^2=0$ and $\tilde{\alpha}=(\tilde{\alpha}_{\mathrm{max}} + \tilde{\alpha}_{\mathrm{min}})/2$, respectively.}
\label{FIG:Waveform2}
\end{figure}

\begin{figure}[htbp]
\centering
\begin{subfigure}[b]{0.24\textwidth}
\centering
\includegraphics[width=\textwidth]{Fig/Re_RN_Q_0.5_k1_Waveform.eps}
\caption{$\mathrm{Re}\left( \psi|_{k=1} \right)$}
\label{subfig:Waveform a}
\end{subfigure}
\hspace{0\textwidth}  
\begin{subfigure}[b]{0.24\textwidth}
\centering
\includegraphics[width=\textwidth]{Fig/Im_RN_Q_0.5_k1_Waveform.eps}
\caption{$\mathrm{Im}\left( \psi|_{k=1} \right)$}
\label{subfig:Waveform b}
\end{subfigure}
\hspace{0\textwidth}  
\begin{subfigure}[b]{0.24\textwidth}
\centering
\includegraphics[width=\textwidth]{Fig/Re_RN_Q_0.5_k2_Waveform.eps}
\caption{$\mathrm{Re}\left( \psi|_{k=2} \right)$}
\label{subfig:Waveform c}
\end{subfigure}
\hspace{0\textwidth}  
\begin{subfigure}[b]{0.24\textwidth}
\centering
\includegraphics[width=\textwidth]{Fig/Im_RN_Q_0.5_k2_Waveform.eps}
\caption{$\mathrm{Im}\left( \psi|_{k=2} \right)$}
\label{subfig:Waveform d}
\end{subfigure}

\vspace{0\textwidth}  

\begin{subfigure}[b]{0.24\textwidth}
\centering
\includegraphics[width=\textwidth]{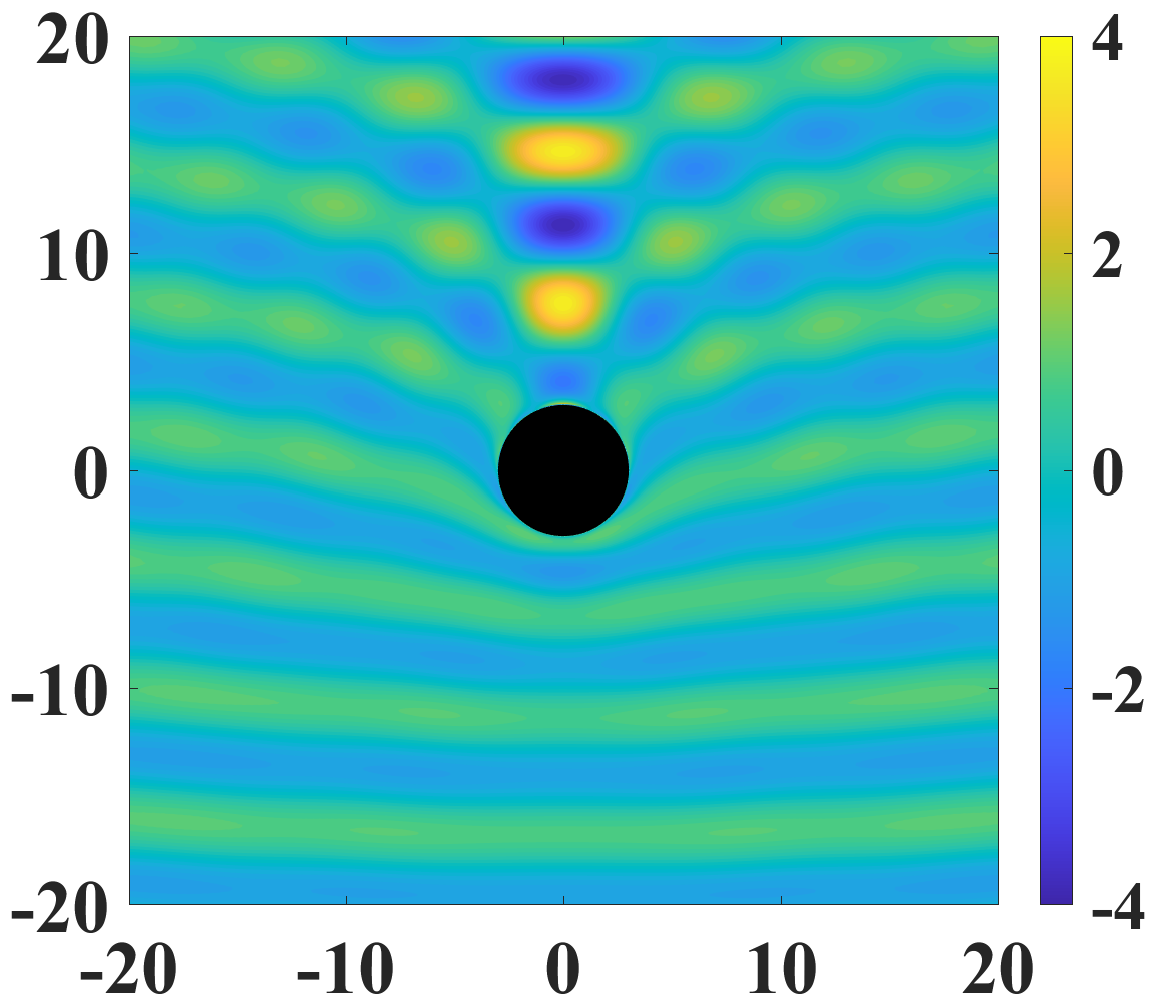}
\caption{$\mathrm{Re}\left( \psi|_{k=1} \right)$}
\label{subfig:Waveform e}
\end{subfigure}
\hspace{0\textwidth}  
\begin{subfigure}[b]{0.24\textwidth}
\centering
\includegraphics[width=\textwidth]{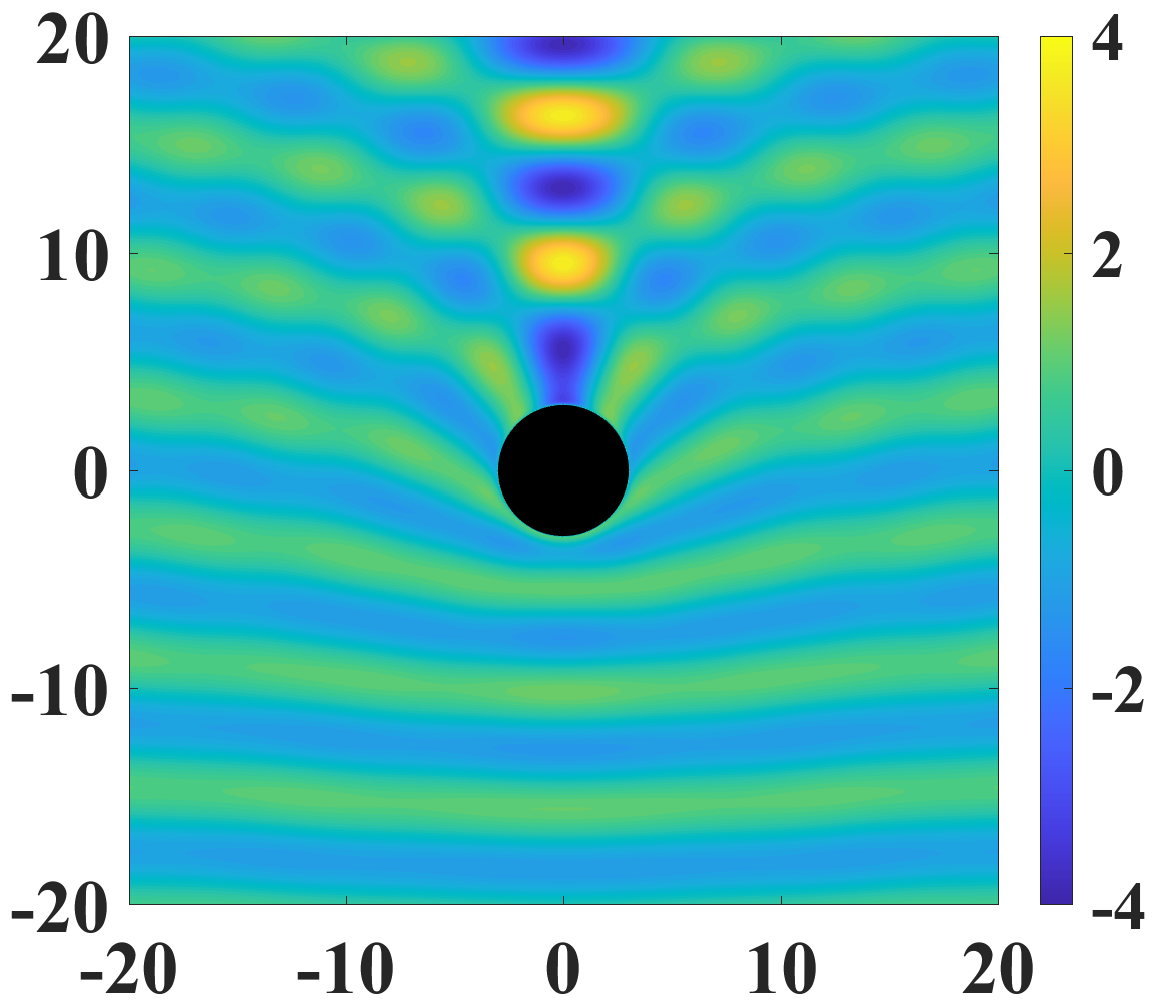}
\caption{$\mathrm{Im}\left( \psi|_{k=1} \right)$}
\label{subfig:Waveform f}
\end{subfigure}
\hspace{0\textwidth}  
\begin{subfigure}[b]{0.24\textwidth}
\centering
\includegraphics[width=\textwidth]{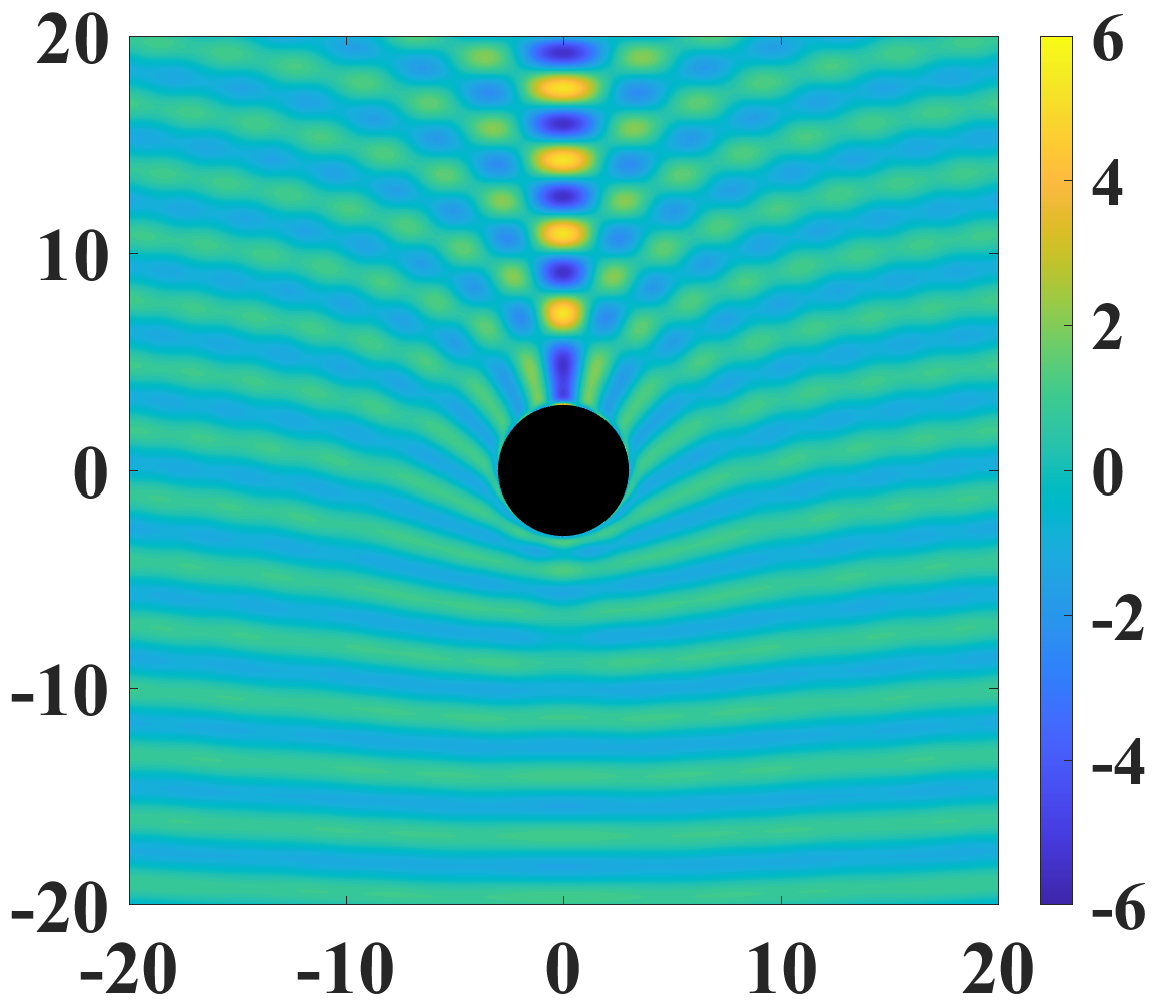}
\caption{$\mathrm{Re}\left( \psi|_{k=2} \right)$}
\label{subfig:Waveform g}
\end{subfigure}
\hspace{0\textwidth}  
\begin{subfigure}[b]{0.24\textwidth}
\centering
\includegraphics[width=\textwidth]{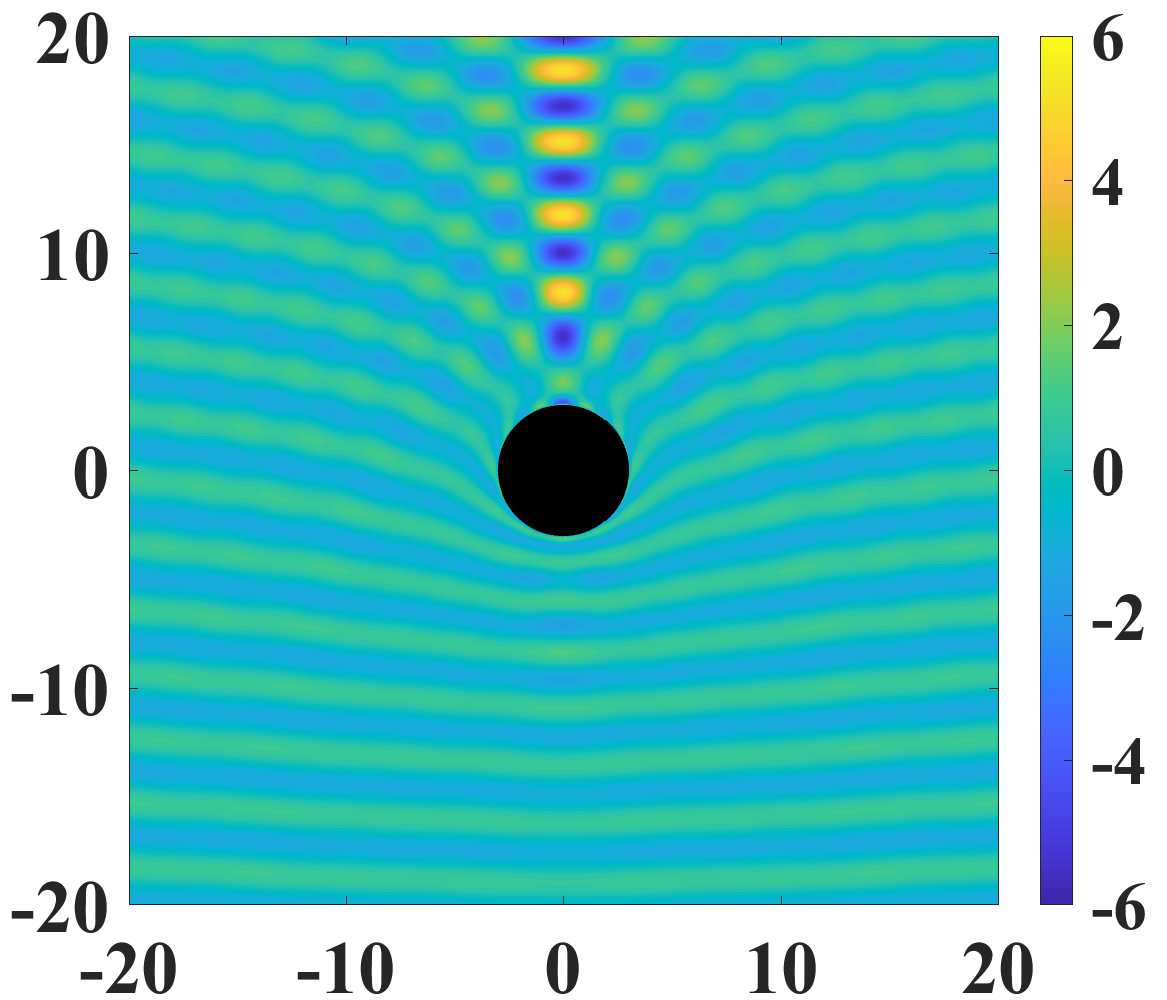}
\caption{$\mathrm{Im}\left( \psi|_{k=2} \right)$}
\label{subfig:Waveform h}
\end{subfigure}

\caption{Waveforms of the RN and conformal anomaly black holes. The first and second rows represent the waveform diagrams for the RN black hole with $Q^2 = 0.5$ and the conformal anomaly black hole with $Q^2=0.5$ and $\tilde{\alpha}=(\tilde{\alpha}_{\mathrm{max}} + \tilde{\alpha}_{\mathrm{min}})/2$, respectively.}
\label{FIG:Waveform3}
\end{figure}

\begin{figure}[htbp]
\centering
\includegraphics[width=\textwidth]{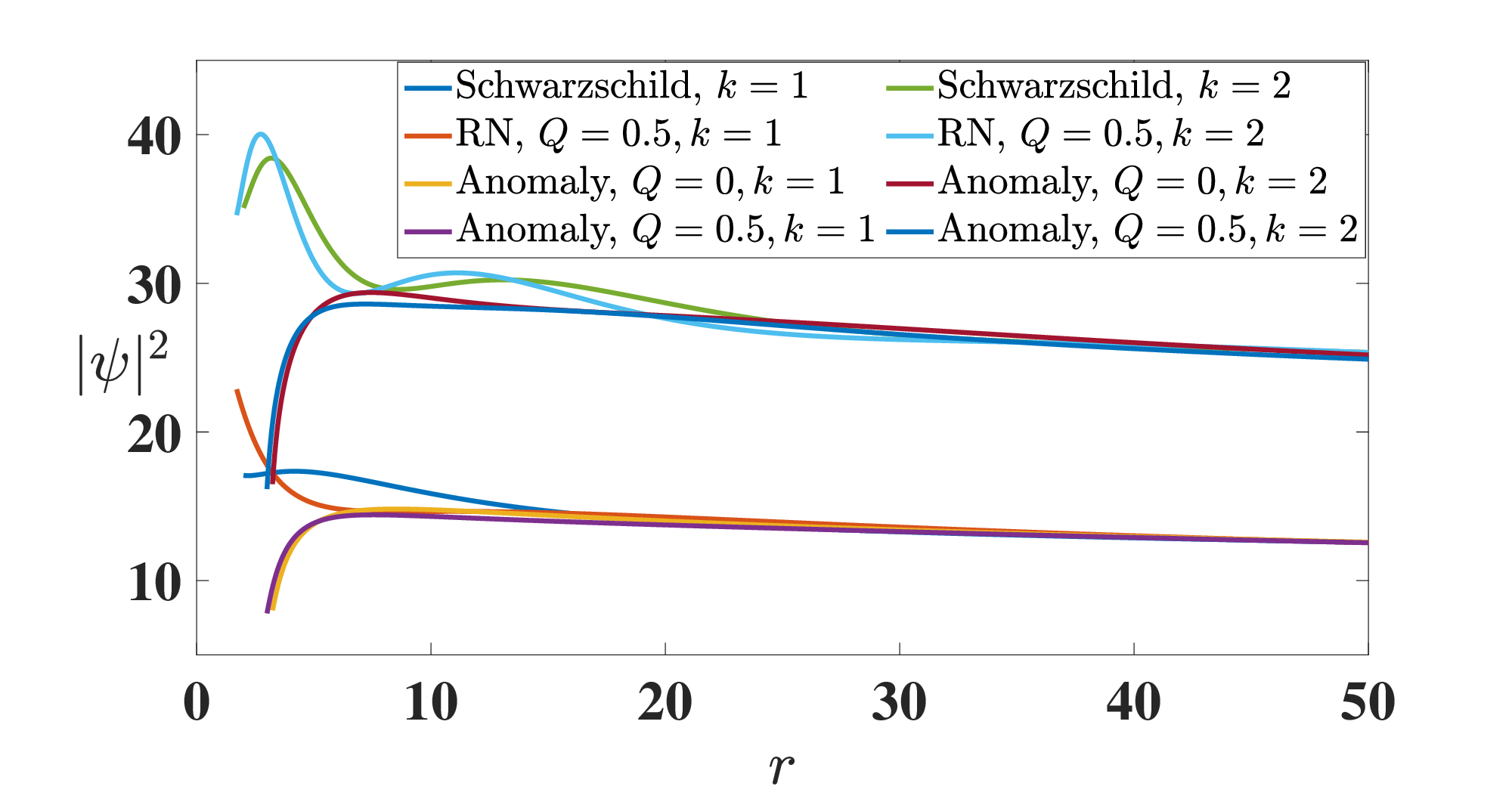}
\caption{Intensity curve of the Poisson spot. The vertical axis represents the intensity of the Poisson spot $|\psi|^2$, and the horizontal axis represents the radial coordinate $r$. The parameters of the curves in the figure correspond to those in in the waveform diagrams above.}
\label{FIG:I_Possion}
\end{figure}

We compute the total wave function within a finite region using the numerical method described above and present the results as waveform diagrams in FIG.\ref{FIG:Waveform1} (comparison between the Schwarzschild and RN black holes), FIG.\ref{FIG:Waveform2} (comparison between the Schwarzschild and conformal anomaly black holes), and FIG.\ref{FIG:Waveform3} (comparison between the RN and conformal anomaly black holes), respectively. 
In generating the figures, we considered eight distinct cases: for wave numbers $k = 1$ and $2$, each applied to the Schwarzschild black hole, the RN black hole with $Q^2 = 0.5$, the conformal anomaly black hole with $Q^2=0$ and $\tilde{\alpha}=(\tilde{\alpha}_{\mathrm{max}} + \tilde{\alpha}_{\mathrm{min}})/2$, and the conformal anomaly black hole with $Q^2=0.5$ and $\tilde{\alpha}=(\tilde{\alpha}_{\mathrm{max}} + \tilde{\alpha}_{\mathrm{min}})/2$. 
From the FIG.\ref{FIG:Waveform1}, FIG.\ref{FIG:Waveform2}, and FIG.\ref{FIG:Waveform3}, it can be seen that all waveform diagrams share similar characteristics: in the the negative $z$-axis region, the waveform closely approximates a plane wave (the ``plane wave'' refers to $e^{ikr\cos{\theta}}$ rather than $e^{ikr_{\ast}\cos{\theta}}$) even at very small values of $-z$; in the positive $z$-axis region, it exhibits a distorted plane wave profile propagating forward, accompanied by a pronounced scattered spherical wave. 
It is particularly noteworthy that a Poisson spot due to interference appears on the $z$-axis at $\theta=0$, surrounded by concentric rings of alternating brightness and darkness in the paraxial region. 
These features bear a strong resemblance to optical diffraction from a small aperture. 
The waveforms we studied are very similar to those for Newtonian scattering and Schwarzschild black hole scattering investigated in Ref.\cite{Li:2025lvl}. 
However, in the waveforms corresponding to different metrics and parameters, the Poisson spot exhibits an apparent ``shift'' along the $z$-axis, and significant differences emerge near the black hole. From FIG.\ref{FIG:Waveform1}, we find that the waveforms of the Schwarzschild and RN black holes differ mainly in the near-horizon region, while the shift of the Poisson spot is not pronounced. From FIG.\ref{FIG:Waveform2} and \ref{FIG:Waveform3}, we find that both the differences in the near-horizon waveforms and the shifts of the Poisson spot are pronounced when the conformal anomaly black hole is compared with the Schwarzschild and RN black holes.
In FIG.\ref{FIG:I_Possion}, we show the intensity of the Poisson spot corresponding to the parameters in the waveform diagrams above as a function of the radial coordinate. 
We argue that the pronounced difference between the Poisson-spot intensities of the RN and conformal anomaly black holes near the horizon in FIG.\ref{FIG:I_Possion} originates mainly from the difference between their metric functions in the near-horizon region. 
This difference in the metric functions affects the effective potential, which in turn modifies the reflection and transmission coefficients and the radial functions of each partial wave. Moreover, near the horizon the tortoise coordinate behaves as $r_\ast \approx \ln(r-r_+)/f'(r_+)$, so different values of $f'(r_+)$ make the same radial interval correspond to different phase increments, causing the on-axis interference fringes to be compressed and shifted to different extents. As $r$ increases, the correction of the conformal anomaly metric relative to the RN metric falls off as $O(\tilde{\alpha}M^2/r^4)$, much faster than the mass and charge terms. Accordingly, the metric functions, potential barriers, and phase shifts of the two black holes approach each other in the far-field region, and the two intensity curves tend to coincide.

While actual observers are in reality located at very large distances from black holes, and the limitations of the PWS method restrict the practical application of the approach developed in this section, the primary influence of a black hole on a scalar wave occurs near its event horizon. Therefore, the properties of scalar wave scattering by a black hole at finite distances studied here still offer valuable insights for practical observations. In the next section, we will employ the SRM to accelerate the convergence of the PWS for the scattered spherical wave and investigate the DCS. The results will be applicable to actual observers.

\section{Absorption and Differential Scattering Cross Sections}
\label{sec:5}
This section investigates the ACS and the DCS of scalar waves by RN black holes and conformal anomaly black holes. The SRM required for studying the DCS has been introduced in Sec.\ref{sec:2}.

\subsection{Absorption Cross Sections}
Both the low-frequency ($k \to 0$)\cite{Das:1996we} and high-frequency ($k \to \infty$)\cite{Crispino:2009ki} limits of the ACS have been theoretically studied, and corresponding results have been established. In the low-frequency limit, it is sufficient to consider the absorption of the s-wave ($\ell=0$). The low-frequency limit of the ACS for massless minimally coupled scalar waves by a spherically symmetric non-extremal black hole ($f'(r_+) \neq 0$) is given by
\begin{equation}
\label{eq:36}
\sigma_{\text{abs}} \left|_{k \to 0}  \right. = A_H \,,
\end{equation}
where $A_H = 4 \pi r_+^2$ is the area of the horizon. For the high-frequency limit, the geodesic (or orbital) equation for massless particles is solved. By applying the circular orbit (photon sphere) condition, the high-frequency limit of the ACS for scalar waves by a black hole can be derived as
\begin{equation}
\label{eq:37}
\sigma_{\text{abs}} \left|_{k \to \infty}  \right. = \frac{\pi r_c^2}{f(r_c)} \,,
\end{equation}
where $r_c$ is the radius of the photon sphere obtained by solving the equation $r_c f'(r_c) - 2 f(r_c) = 0$.

Based on the numerical results, the ACS can be expressed as
\begin{equation}
\label{eq:38}
\sigma_{\text{abs}} = \sum_{\ell=0}^{\infty} \sigma_{\text{abs}\ell} \,,
\end{equation}
where $\sigma_{\text{abs}\ell}$ denotes the ACS of the black hole for scalar waves in the $\ell$-th partial wave mode, and its expression is given by
\begin{equation}
\label{eq:39}
\sigma_{\text{abs}\ell} = \frac{\pi}{k^2} \left( 2\ell + 1 \right) \left( 1 - \left| \frac{B_{\ell {\text{Num}}}}{A_{\ell {\text{Num}}}} \right|^2 \right) \,.
\end{equation}

\begin{figure}[htbp]
\centering
\begin{subfigure}[b]{0.32\textwidth}
\centering
\includegraphics[width=\textwidth]{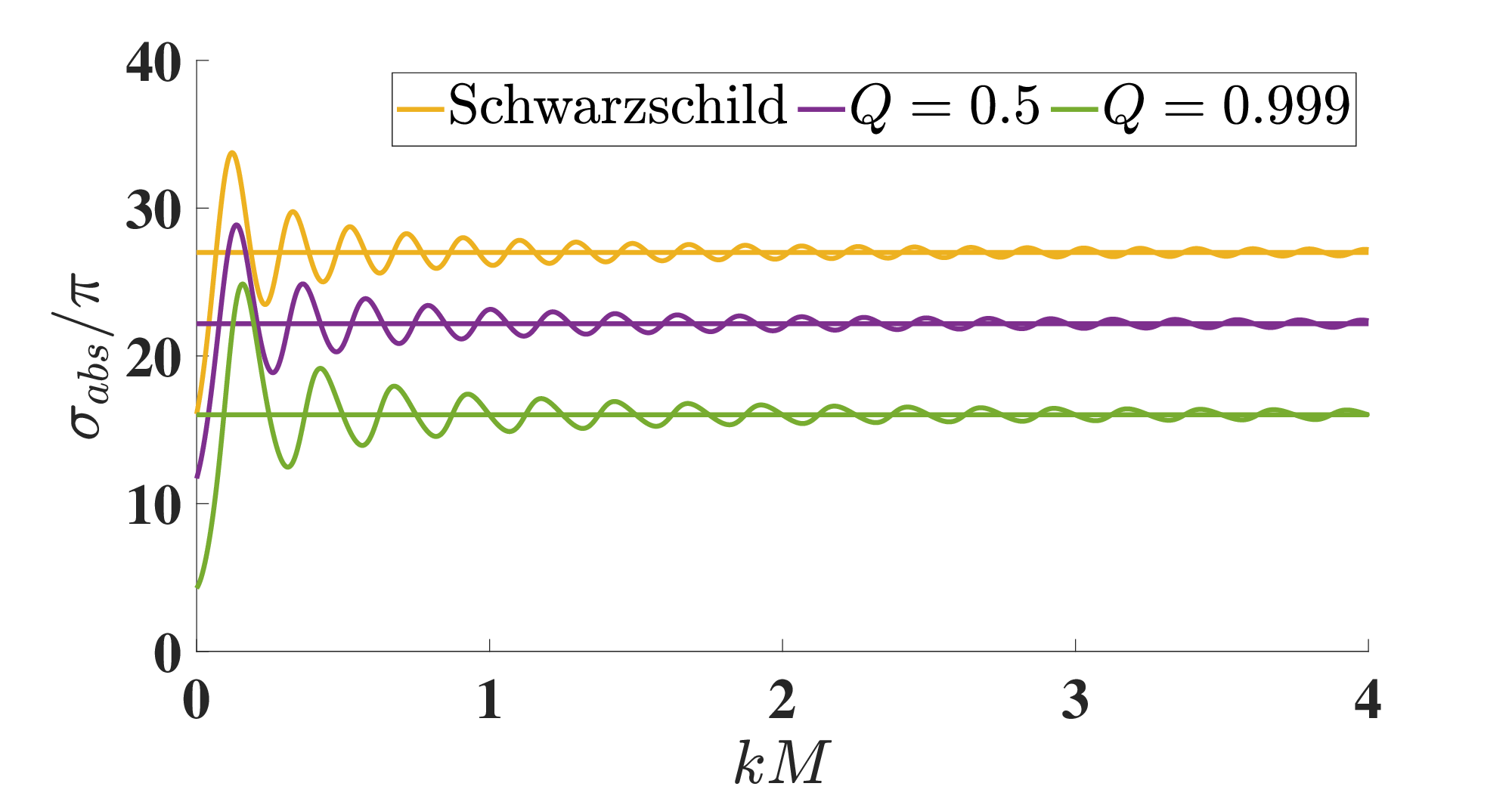}
\caption{}
\label{subfig:ACS_RN}
\end{subfigure}
\hspace{0\textwidth}  
\begin{subfigure}[b]{0.32\textwidth}
\centering
\includegraphics[width=\textwidth]{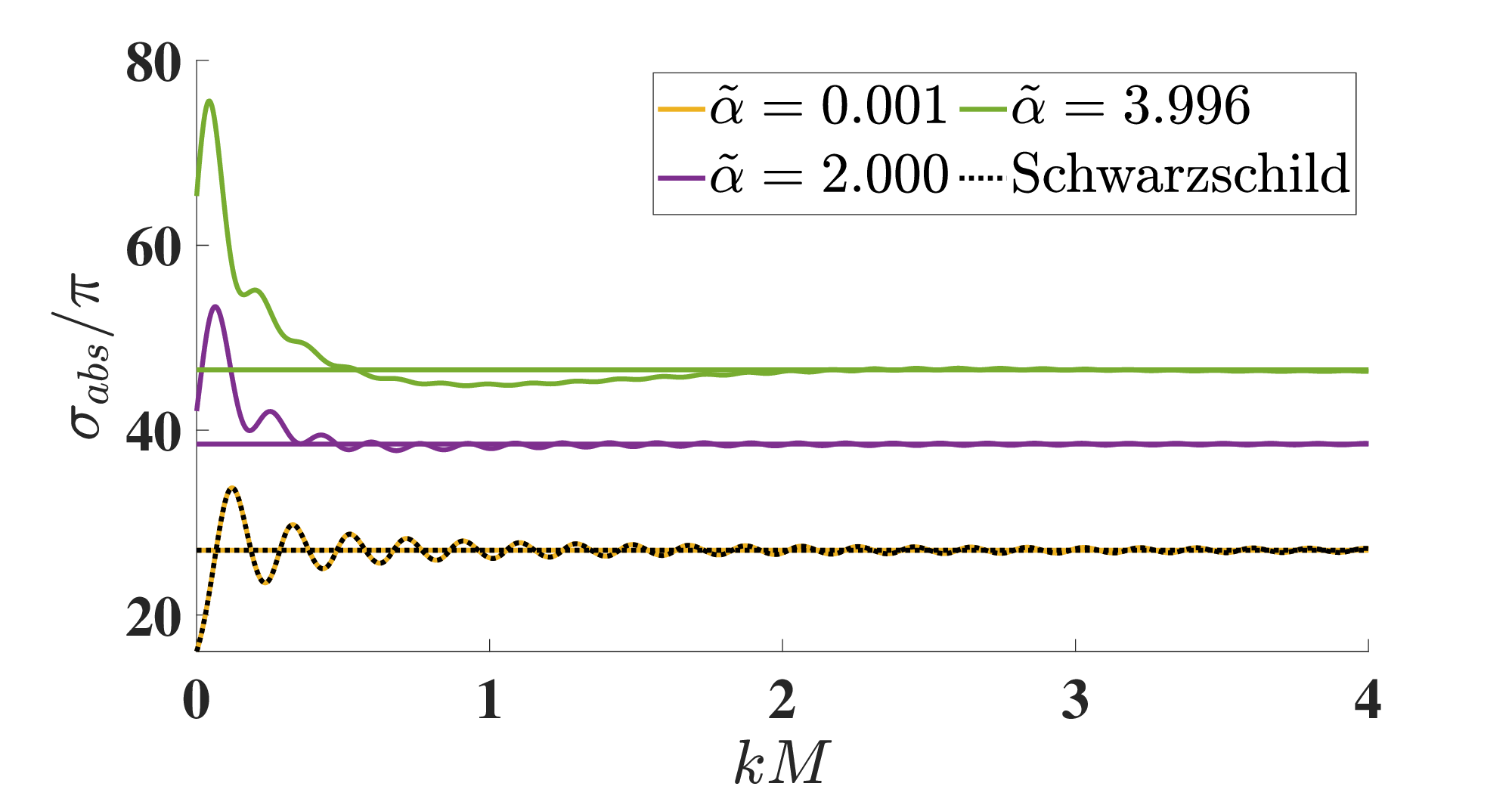}
\caption{}
\label{subfig:ACS_Anomaly_Q_0}
\end{subfigure}
\hspace{0\textwidth}  
\begin{subfigure}[b]{0.32\textwidth}
\centering
\includegraphics[width=\textwidth]{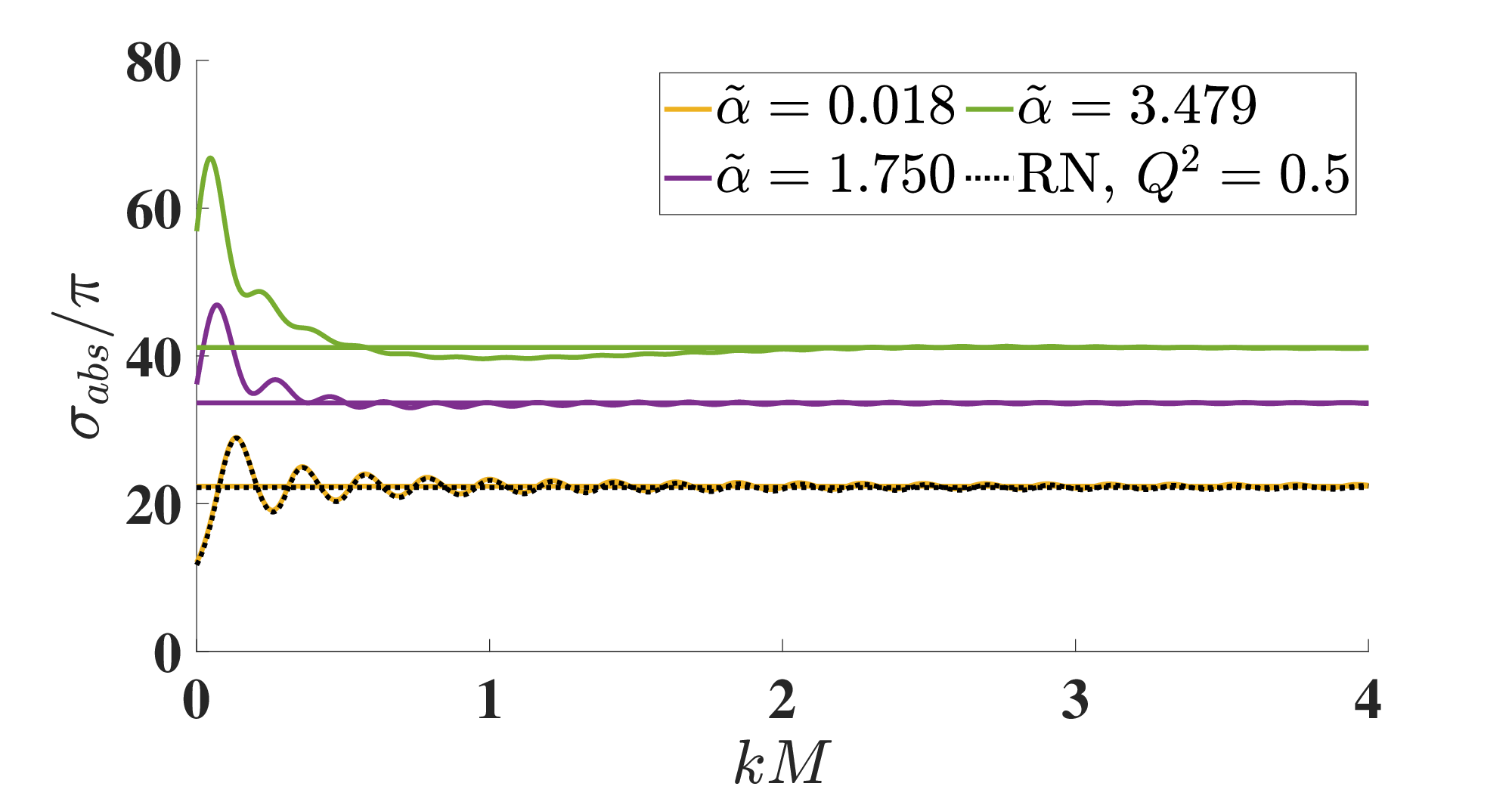}
\caption{}
\label{subfig:ACS_Anomaly_Q_0.5}
\end{subfigure}

\vspace{0\textwidth}  

\begin{subfigure}[b]{0.32\textwidth}
\centering
\includegraphics[width=\textwidth]{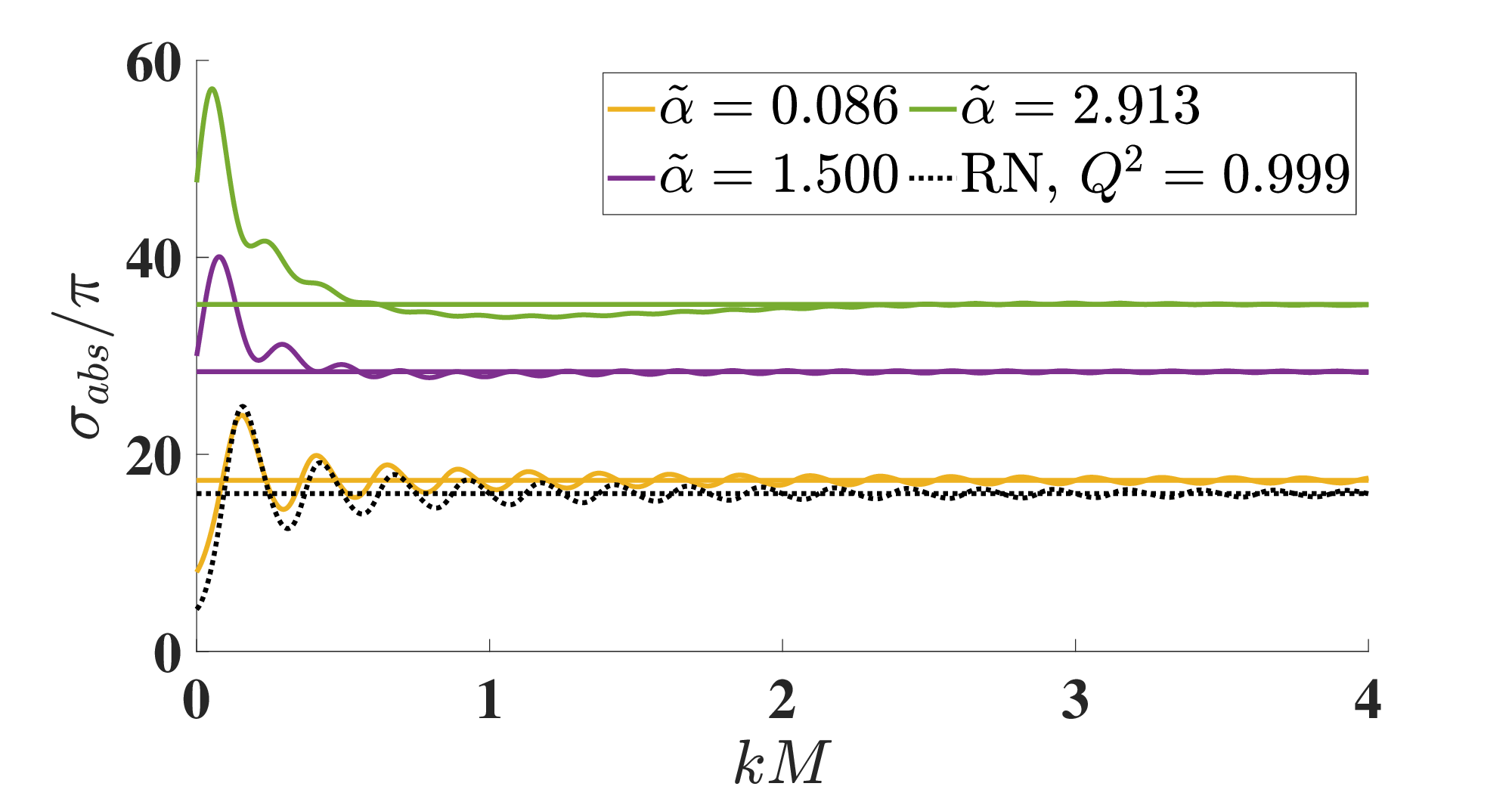}
\caption{}
\label{subfig:ACS_Anomaly_Q_0.999}
\end{subfigure}
\hspace{0\textwidth}  
\begin{subfigure}[b]{0.32\textwidth}
\centering
\includegraphics[width=\textwidth]{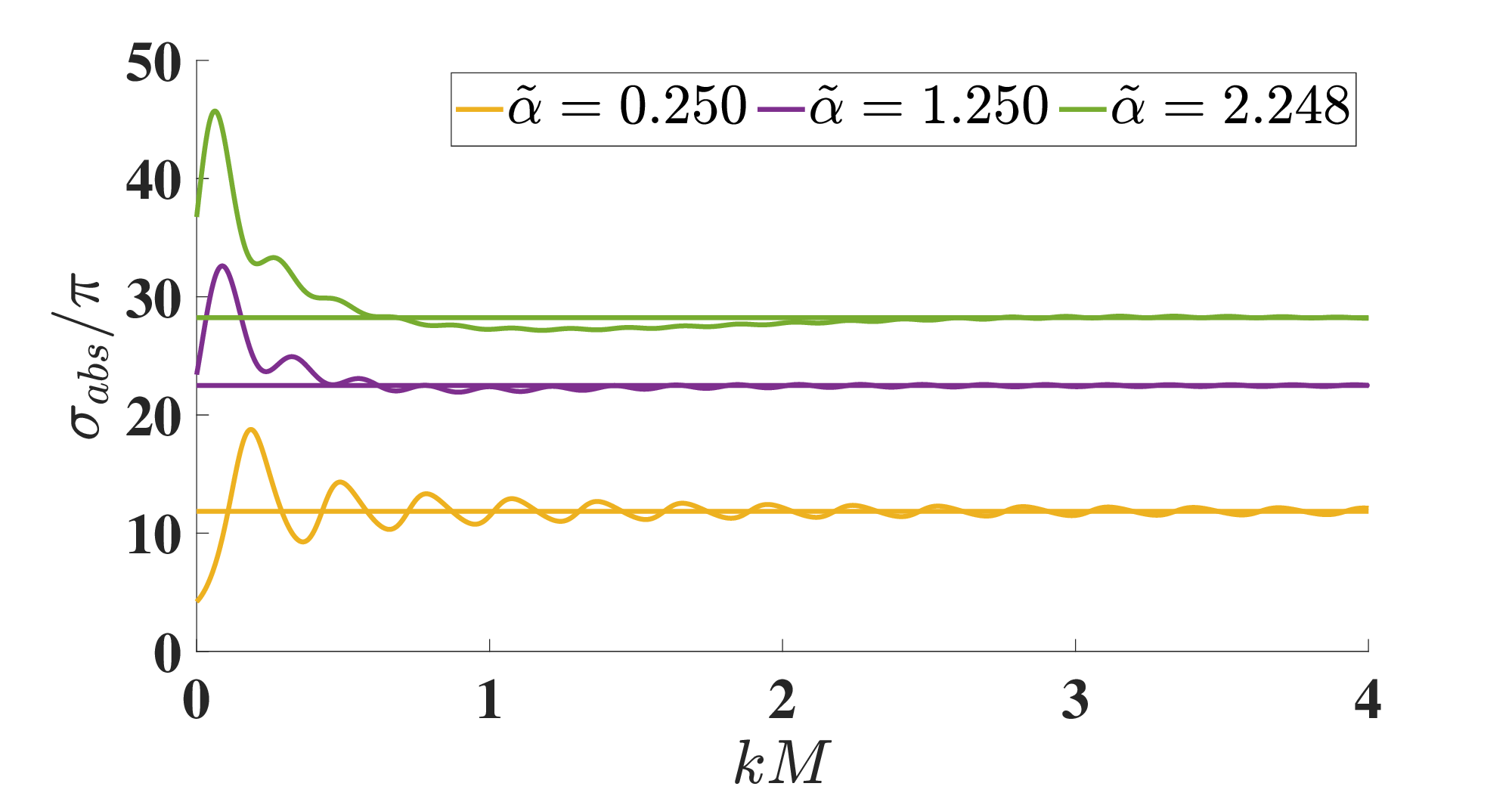}
\caption{}
\label{subfig:ACS_Anomaly_Q_1.5}
\end{subfigure}
\hspace{0\textwidth}  
\begin{subfigure}[b]{0.32\textwidth}
\centering
\includegraphics[width=\textwidth]{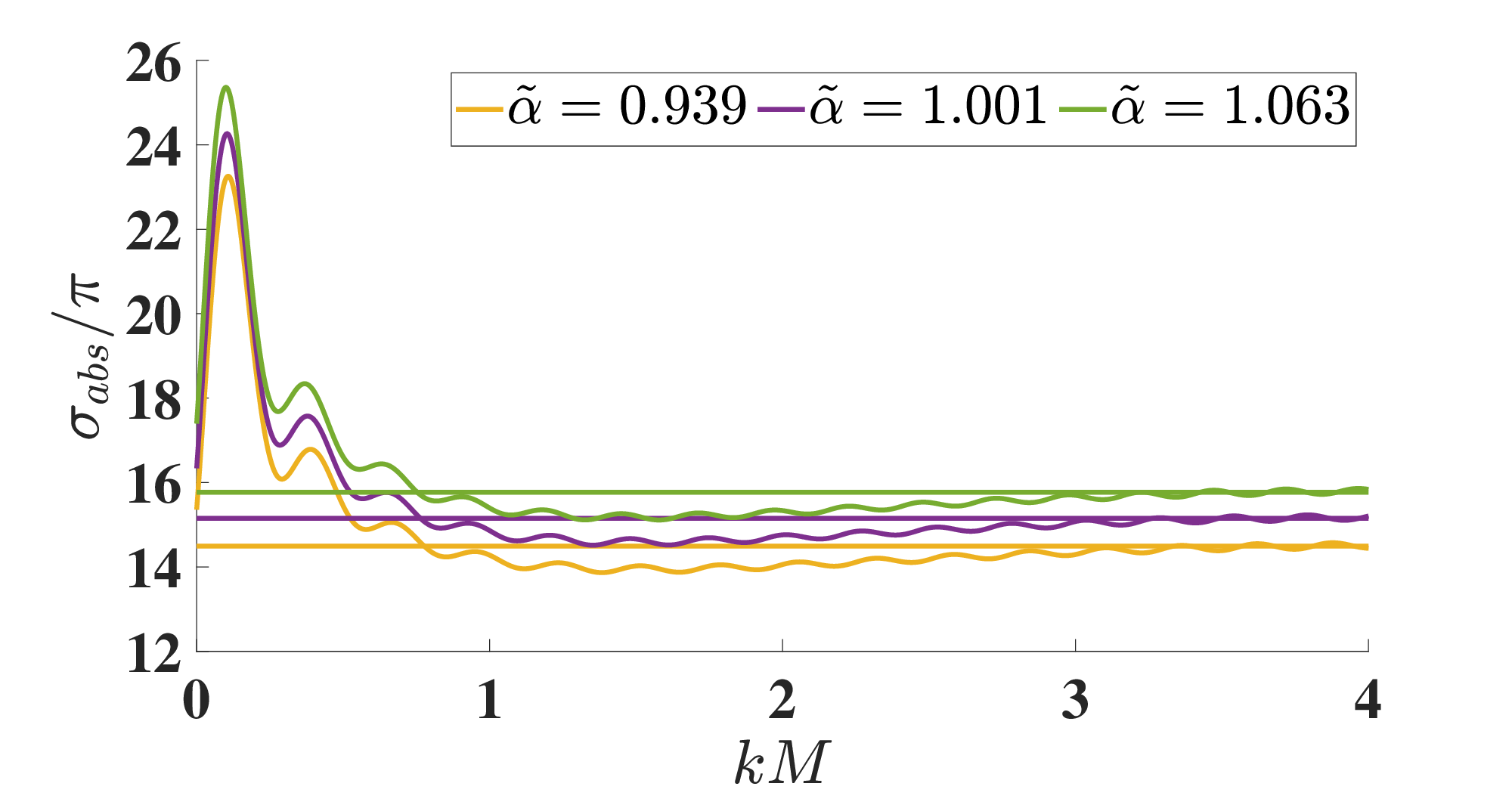}
\caption{}
\label{subfig:ACS_Anomaly_Q_1.998}
\end{subfigure}
\caption{Absorption cross sections of RN and conformal anomaly black holes. This figure shows the ACS of the black holes for scalar waves as a function of the wave frequency $k$. The dashed line in the figure indicates the high-frequency limit of the ACS. Subfigure (a) shows the ACS of the Schwarzschild black hole and the RN black holes for charges $Q^2 = 0.5$ and $0.999$. Subfigures (b) display the ACS curves for conformal anomaly black holes and Schwarzschild black hole, subfigures (c) and (d) display the ACS curves for conformal anomaly black holes and RN black holes with $Q^2 = 0.5$ and $0.999$, and subfigures (e) and (f) display the ACS curves for conformal anomaly black holes with $Q^2 = 1.5$ and $0.999 \times 2$, where the conformal anomaly parameter $\tilde{\alpha}$ is taken as $1.001 \tilde{\alpha}_{\mathrm{min}}$, $(\tilde{\alpha}_{\mathrm{max}} + \tilde{\alpha}_{\mathrm{min}})/2$, and $0.999 \tilde{\alpha}_{\mathrm{max}}$.}
\label{FIG:ACS}
\end{figure}

\begin{figure}[htbp]
\centering
\begin{subfigure}[b]{0.45\textwidth}
\centering
\includegraphics[width=\textwidth]{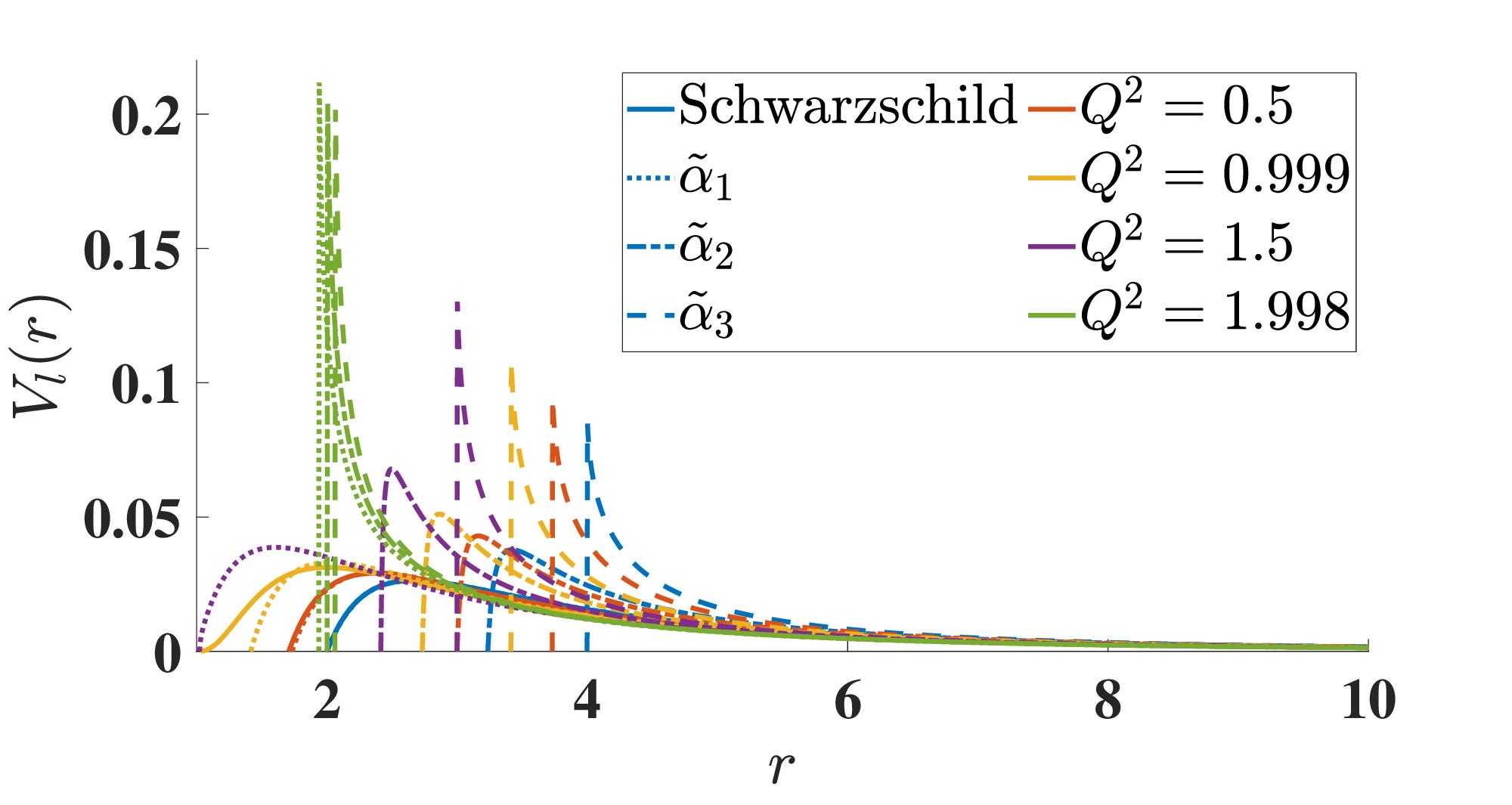}
\caption{}
\label{subfig:Vl_r_l_0_fixQ}
\end{subfigure}
\hspace{0\textwidth}  
\begin{subfigure}[b]{0.45\textwidth}
\centering
\includegraphics[width=\textwidth]{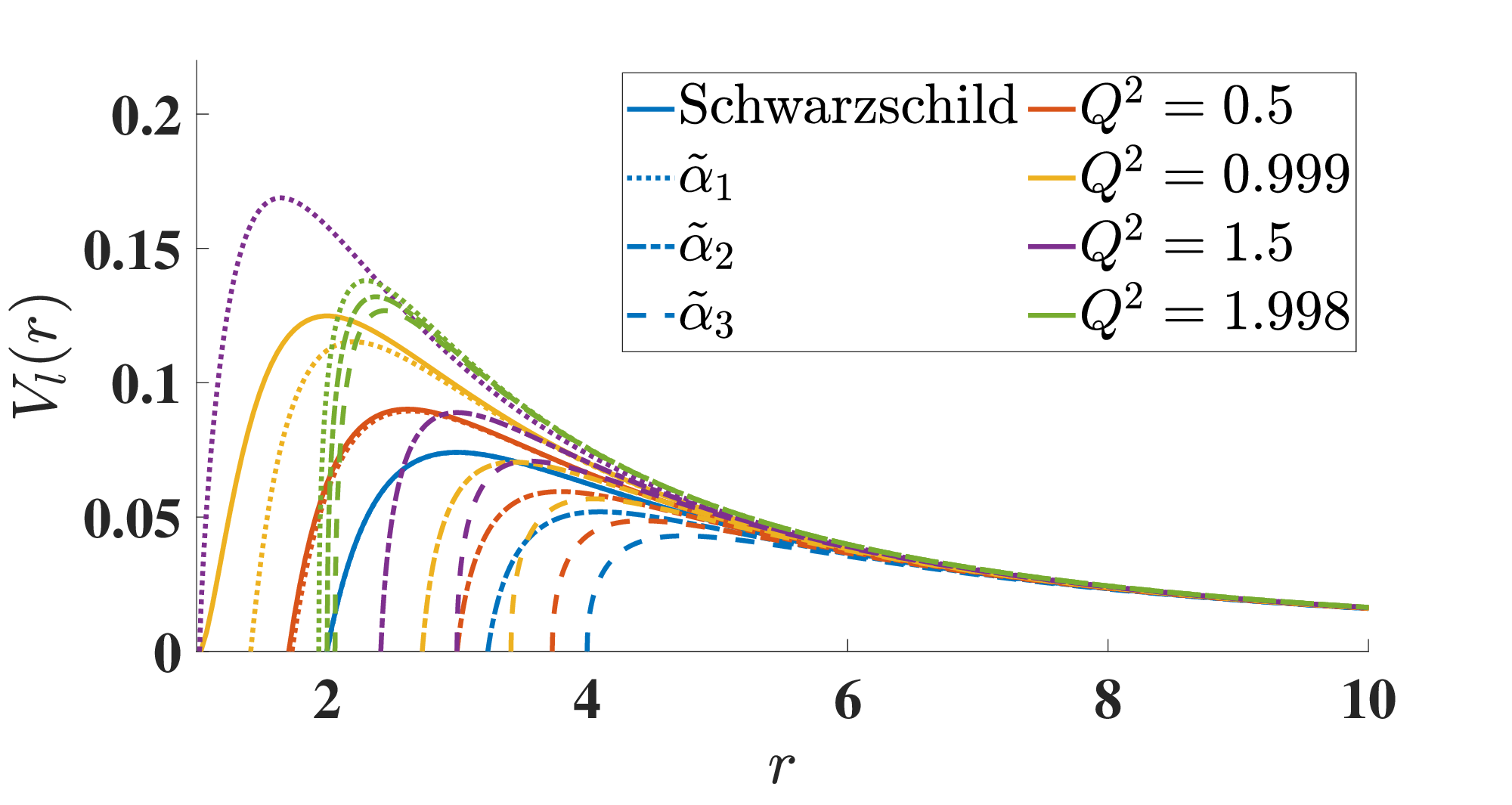}
\caption{}
\label{subfig:Vl_r_l_fixQ}
\end{subfigure}

\vspace{0\textwidth}  

\begin{subfigure}[b]{0.45\textwidth}
\centering
\includegraphics[width=\textwidth]{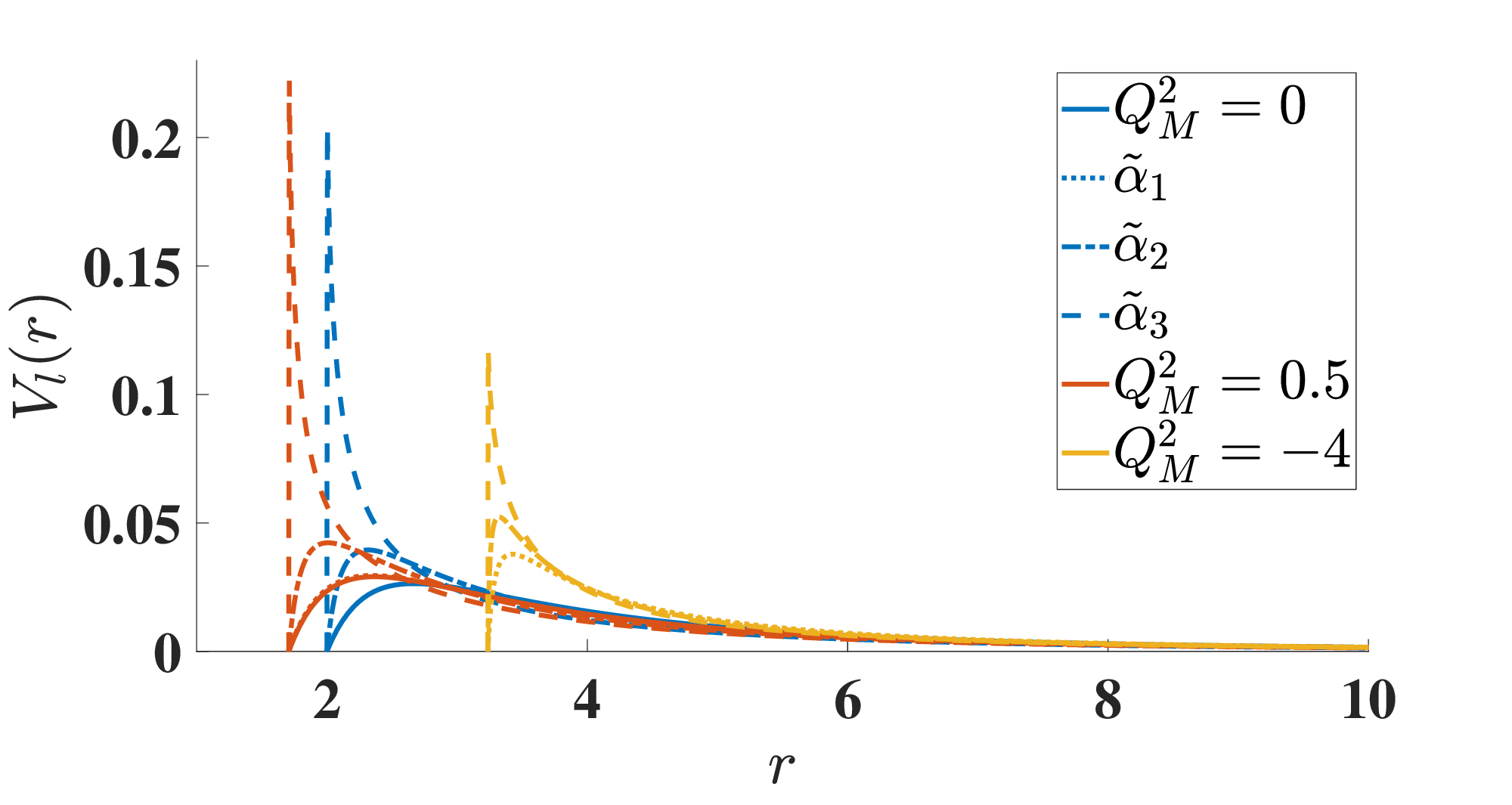}
\caption{}
\label{subfig:Vl_r_l_0_fixQM}
\end{subfigure}
\hspace{0\textwidth}  
\begin{subfigure}[b]{0.45\textwidth}
\centering
\includegraphics[width=\textwidth]{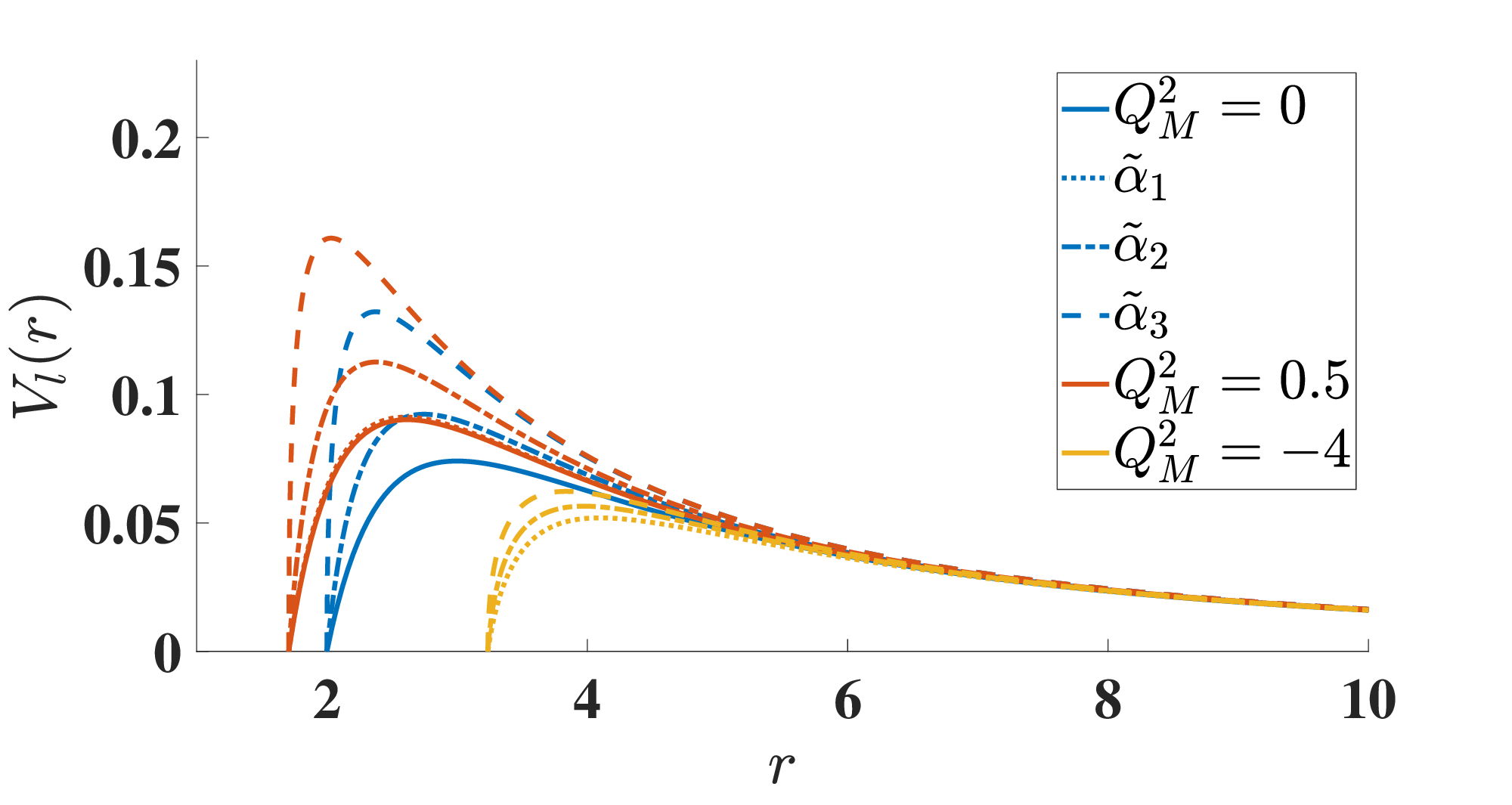}
\caption{}
\label{subfig:Vl_r_l_fixQM}
\end{subfigure}
\caption{Potential function $V_\ell(r)$ as a function of the radial coordinate $r$ for different metrics and parameters. Subfigures.\ref{subfig:Vl_r_l_0_fixQ} and \ref{subfig:Vl_r_l_fixQ}, and subfigures.\ref{subfig:Vl_r_l_0_fixQM} and \ref{subfig:Vl_r_l_fixQM}, correspond to the same metrics and parameters as those in FIG.\ref{FIG:ACS} and FIG.\ref{FIG:ACS_Anomaly_QM}, respectively. In the legends of subfigures.\ref{subfig:Vl_r_l_0_fixQ} and \ref{subfig:Vl_r_l_fixQ}, curves of the same color correspond to the same charge. In the legends of subfigures.\ref{subfig:Vl_r_l_0_fixQM} and \ref{subfig:Vl_r_l_fixQM}, curves of the same color correspond to the same horizon area. In all subfigures, the solid, dotted, dash-dot, and dashed lines correspond to the RN black hole and the conformal anomaly black hole with the minimum, intermediate, and maximum values of $\tilde{\alpha}$ in FIG.\ref{FIG:ACS} and FIG.\ref{FIG:ACS_Anomaly_QM}, respectively. Subfigures.\ref{subfig:Vl_r_l_0_fixQ} and \ref{subfig:Vl_r_l_0_fixQM} show the potential curves for $V_{\ell=0}(r)$. Subfigures.\ref{subfig:Vl_r_l_fixQ} and \ref{subfig:Vl_r_l_fixQM} show the curves for $V_{\ell=1}(r)-V_{\ell=0}(r)$, i.e., considering only the $\ell$-dependent part of the effective potential.}
\label{FIG:Vl(r)_l=0}
\end{figure}

Following the derivation above, we present in FIG.\ref{FIG:ACS} the ACS for scalar waves by RN black holes and conformal anomaly black holes with different parameters. We find that in all cases, the ACS at low-frequency limit is in agreement with the result given by Eq.\eqref{eq:36}. As the frequency increases, the ACS in all cases converges to the high-frequency limit given by Eq.\eqref{eq:37}. From subfigure.\ref{subfig:ACS_RN}, it can be seen that the ACS of the RN black hole decreases as the charge $Q$ increases. 
The curves with the smallest $\tilde{\alpha}$ in Subfigures.\ref{subfig:ACS_Anomaly_Q_0}, \ref{subfig:ACS_Anomaly_Q_0.5} and \ref{subfig:ACS_Anomaly_Q_0.999} indicate that for small $\tilde{\alpha}$, the ACS of the conformal anomaly black hole and the RN black hole are essentially identical. As $\tilde{\alpha}$ increases, the deviation of the ACS of the conformal anomaly black hole from that of the RN black hole becomes increasingly significant. This can be interpreted as the small parameter $\tilde{\alpha}$ being treated as a perturbation. From subfigures.\ref{subfig:ACS_Anomaly_Q_0}--\ref{subfig:ACS_Anomaly_Q_1.998}, it can be seen that for fixed $Q$, as  $\tilde{\alpha}$ increases, the ACS increases. The effect of $\tilde{\alpha}$ on the ACS is opposite to that of the charge for the RN black hole. From all the subfigures of FIG.\ref{FIG:ACS} and in conjunction with the effect of $\tilde{\alpha}$, it can be seen that the effect of charge on the ACS of the conformal anomaly black hole is the same as that on the ACS of the RN black hole. 
These characteristics are consistent with their respective influences on the height of the potential barrier, as can be observed in FIG.\ref{FIG:Vl-r} and subfigures.\ref{subfig:Vl_r_l_0_fixQ} and \ref{subfig:Vl_r_l_fixQ} of FIG.\ref{FIG:Vl(r)_l=0}. Subfigures.\ref{subfig:Vl_r_l_0_fixQ} and \ref{subfig:Vl_r_l_fixQ} of FIG.\ref{FIG:Vl(r)_l=0} respectively show the height and width of the potential barriers $V_{\ell=0}(r)$ and $V_{\ell=1}(r)-V_{\ell=0}(r)$ for the same metrics and parameters as in FIG.\ref{FIG:ACS}.
From subfigure.\ref{subfig:Vl_r_l_0_fixQ}, it can be seen that for the RN black hole, the potential barrier becomes higher and wider as the charge increases. For conformal anomaly black holes with the same charge, as $\tilde{\alpha}$ increases, the potential barrier becomes higher and narrower, starting from the RN barrier with the same charge. From subfigure.\ref{subfig:Vl_r_l_fixQ}, it can be seen that for the RN black hole, the $\ell$-dependent part of the potential barrier becomes higher and wider as the charge increases. For conformal anomaly black holes, as $\tilde{\alpha}$ increases, the $\ell$-dependent part of the potential barrier becomes lower and narrower, starting from the $\ell$-dependent part of the RN barrier with the same charge.
Comparing the figures for the ACS and the potential barrier, it can be seen that in subfigure.\ref{subfig:Vl_r_l_0_fixQ}, the dependence of the height and width of the potential barrier on the metrics and parameters differs significantly from that of the ACS in FIG.\ref{FIG:ACS}. In contrast, in subfigure.\ref{subfig:Vl_r_l_fixQ}, the dependence of the height and width of the potential barrier on the metrics and parameters is the same as that of the ACS in FIG.\ref{FIG:ACS}. This is because, near the low-frequency limit, the ACS is primarily governed by the horizon area of the black hole. As the frequency increases, the contribution from higher-$\ell$ modes gradually becomes non-negligible. Consequently, the dependence of the high-frequency ACS on the metrics and parameters should be the same as that of the $\ell$-dependent part of the potential barrier on the metrics and parameters.
The height and width of the potential barrier directly influences the transmission and reflection coefficients of the scalar wave. Specifically, a higher and wider barrier leads to a larger reflection coefficient and a smaller transmission coefficient, whereas a lower and narrower barrier results in a smaller reflection coefficient and a larger transmission coefficient.
Moreover, we also observe that the ACS approaches the theoretical result given by Eq.\eqref{eq:36} in the low-frequency limit, and that given by Eq.\eqref{eq:37} in the high-frequency limit. This indicates that the numerical results for the ACS of conformal anomaly black holes are consistent with the universal result of the theoretical analysis. From subfigures.\ref{subfig:ACS_Anomaly_Q_0}--\ref{subfig:ACS_Anomaly_Q_1.998}, it can be observed that the ACS curves of conformal anomaly black holes with large $\tilde{\alpha}$ exhibit a pronounced ``dip'' that is suppressed downward and deviates from the oscillations around the high-frequency limit, compared to those of the RN black hole and conformal anomaly black holes with small charge and small $\tilde{\alpha}$. This phenomenon will be analyzed in the following text.

\begin{figure}[htbp]
\centering
\begin{subfigure}[b]{0.32\textwidth}
\centering
\includegraphics[width=\textwidth]{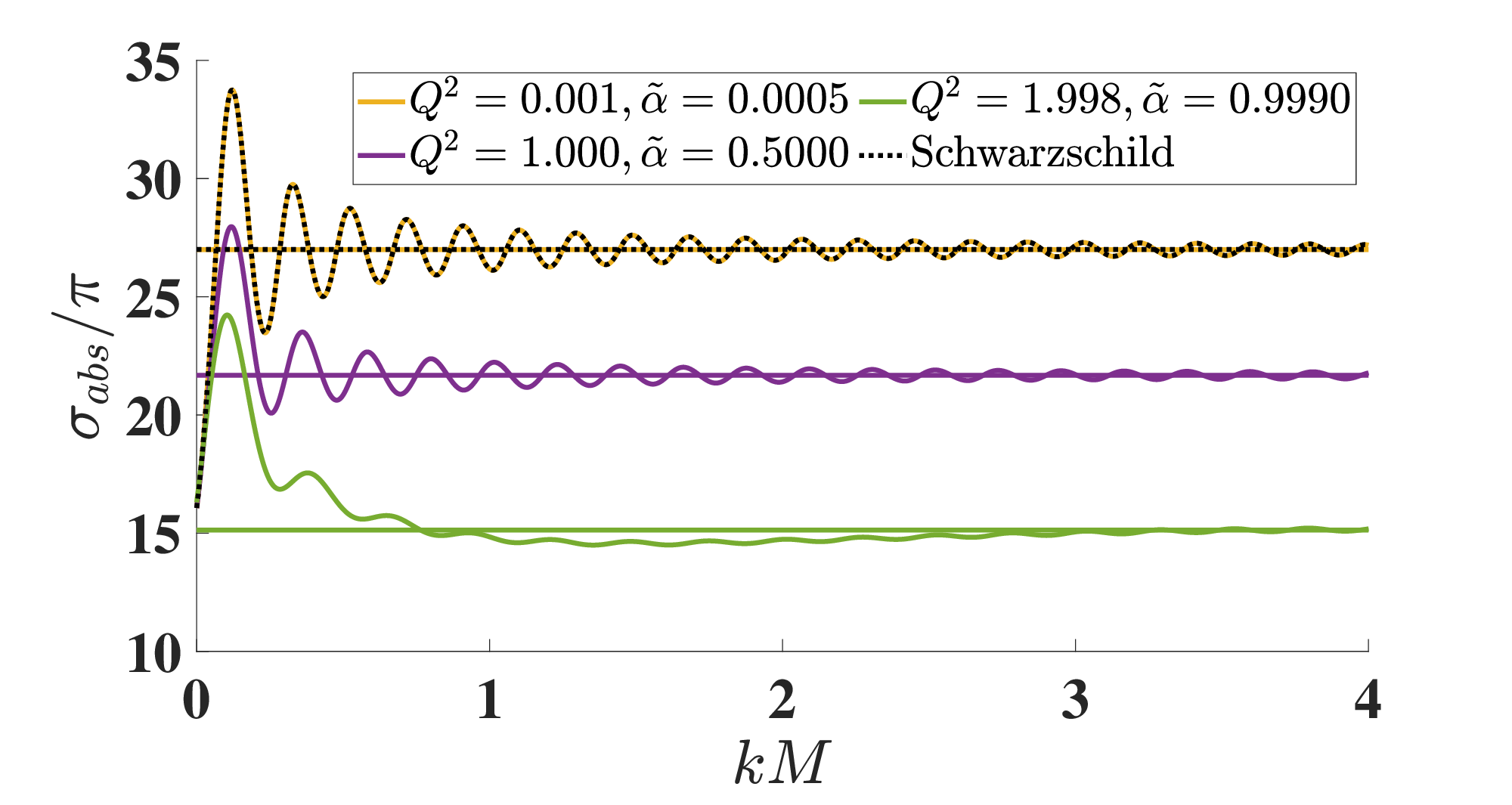}
\caption{}
\label{subfig:ACS_Anomaly_QM_0}
\end{subfigure}
\hspace{0\textwidth}  
\begin{subfigure}[b]{0.32\textwidth}
\centering
\includegraphics[width=\textwidth]{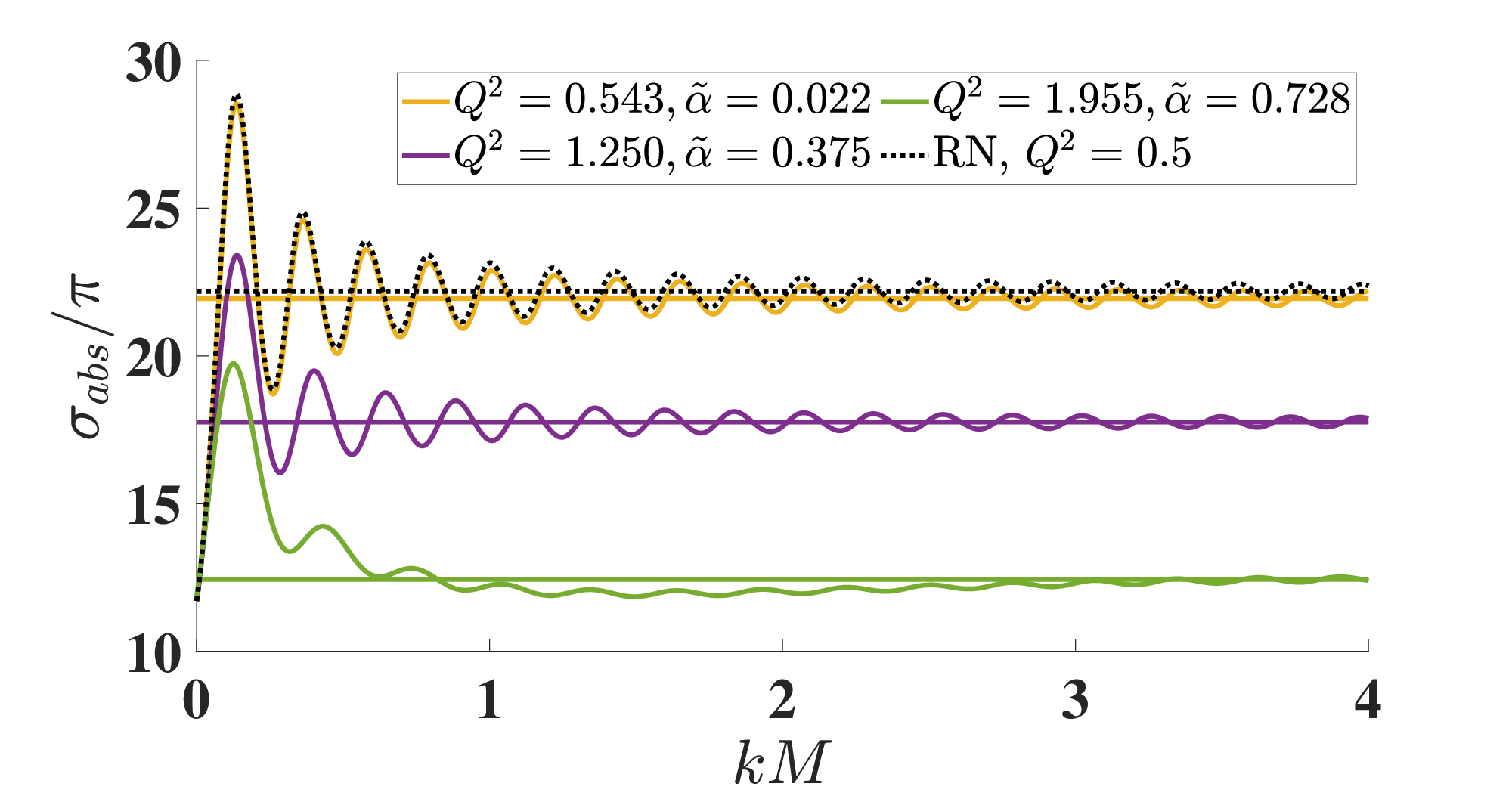}
\caption{}
\label{subfig:ACS_Anomaly_QM_0.5}
\end{subfigure}
\hspace{0\textwidth}  
\begin{subfigure}[b]{0.32\textwidth}
\centering
\includegraphics[width=\textwidth]{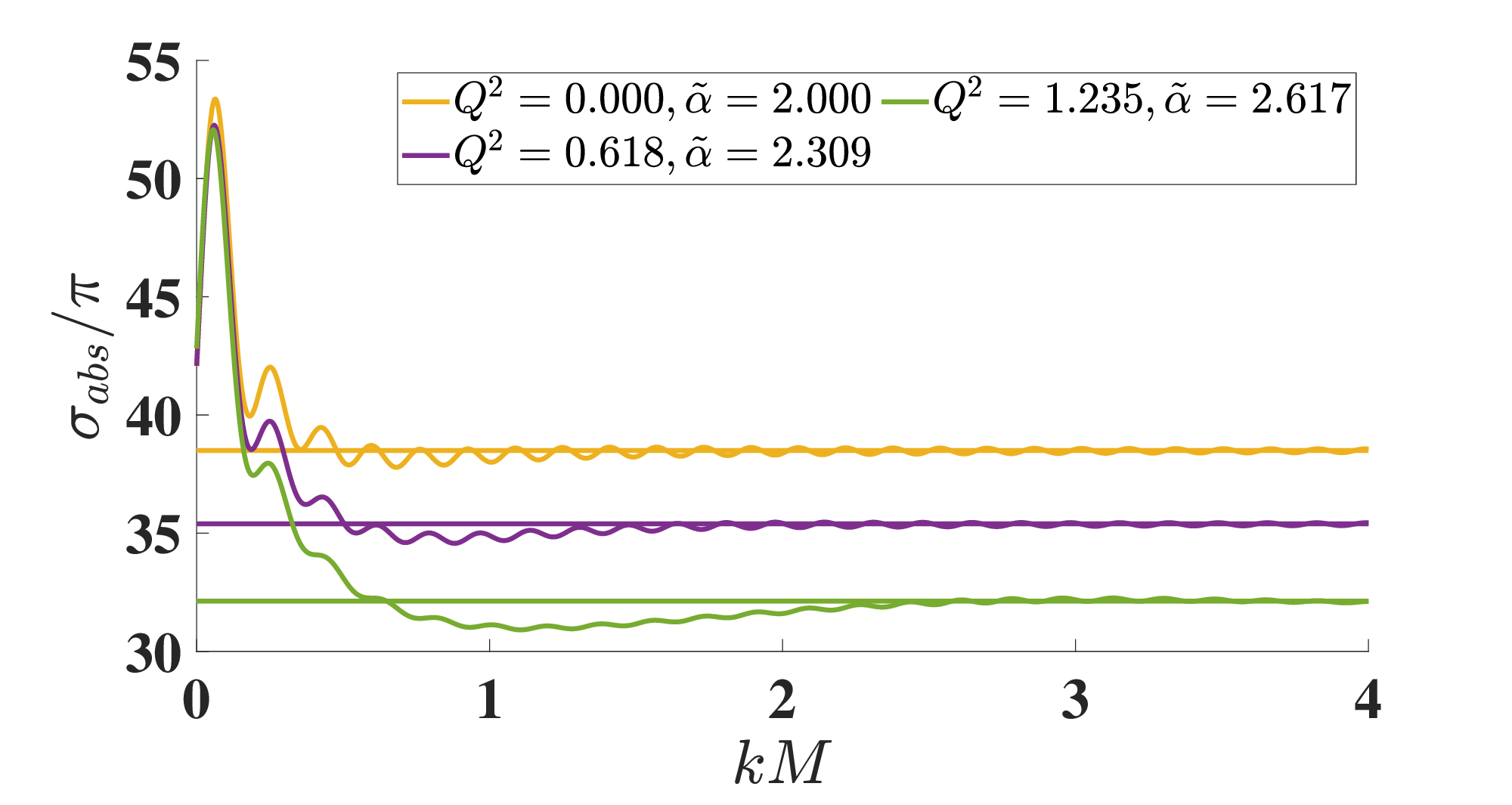}
\caption{}
\label{subfig:ACS_Anomaly_QM_0.999}
\end{subfigure}

\caption{Absorption cross sections of conformal anomaly black holes with fixed horizon area and different parameters $\tilde{\alpha}$. In subfigures (a), (b), and (c), the parameter $Q_M^2$ is taken as $Q_M^2 = 0$, $0.5$, and $-4$, respectively. For the charge $Q$ of the conformal anomaly black hole in subfigures (a) and (b), we take $Q^2 = 1.001 Q_{\mathrm{min}}^2$, $(Q_{\mathrm{min}}^2+Q_{\mathrm{max}}^2)/2$, and $0.999 Q_{\mathrm{max}}^2$, where $Q_{\substack{\mathrm{max} \\ \mathrm{min}}} = M^2 + Q_M^2/2 \pm M \sqrt{M^2-Q_M^2}$. For the charge $Q$ of the conformal anomaly black hole in subfigure (c), we taken $Q^2 = 0$, $Q_{\mathrm{max}}^2/2$, and $0.999 Q_{\mathrm{max}}^2$, where $Q_{\mathrm{max}}^2$ is the same as above.}
\label{FIG:ACS_Anomaly_QM}
\end{figure}

Analysis of Eq.\eqref{eq:28} reveals that the event horizon radius of a conformal anomaly black hole is determined by $M$ and $Q_M$. Consequently, unlike the RN black hole, we can study the ACS for conformal anomaly black holes with different parameters $\tilde{\alpha}$ while holding the horizon area constant. The corresponding ACS as functions of frequency $k$ for these different parameters are presented in FIG.\ref{FIG:ACS_Anomaly_QM}. From this figure, it can be seen that when $\tilde{\alpha} = 0.0005$, the ACS of the conformal anomaly black hole almost coincides with that of the corresponding RN black hole. This indicates that, when $Q_M$ is held fixed, a small variation in $\alpha$ from $0$ to $0.0005$ can still be regarded as a perturbation. Furthermore, when the horizon area of the black hole is kept fixed, a change in the parameter $\tilde{\alpha}$ does not affect the low-frequency limit of the ACS for the conformal anomaly black hole, but it does affect the high-frequency limit. For the fixed $Q_M$, the ACS decreases as the charge and the parameter $\tilde{\alpha}$ increase jointly. This further demonstrates that the effects of the charge and the parameter $\tilde{\alpha}$ on the ACS in conformal anomaly black holes are opposite, and that when they increase simultaneously in the form $Q^2 = Q_M^2 + 2\tilde{\alpha}$, the effect of the charge on the ACS dominates over that of the parameter $\tilde{\alpha}$. Furthermore, in FIG.\ref{FIG:ACS_Anomaly_QM}, we also observe that for large values of $\tilde{\alpha}$, the ACS curve exhibits a ``dip'' similar to that in FIG.\ref{FIG:ACS}.

Similar to how subfigures.\ref{subfig:Vl_r_l_0_fixQ} and \ref{subfig:Vl_r_l_fixQ} of FIG.\ref{FIG:Vl(r)_l=0} display the potential diagrams corresponding to FIG.\ref{FIG:ACS}, subfigures.\ref{subfig:Vl_r_l_0_fixQM} and \ref{subfig:Vl_r_l_fixQM} of FIG.\ref{FIG:Vl(r)_l=0} display the potential diagrams corresponding to FIG.\ref{FIG:ACS_Anomaly_QM}.
From subfigure.\ref{subfig:Vl_r_l_0_fixQM}, it is found that the potential for the $\ell=0$ mode exhibits a sharp peak for near-extremal values of the parameter $\tilde{\alpha}$. 
Based on the preceding analysis of the conformal anomaly metric and the effective potential, we attribute this sharp peak to the term involving $f'(r)$ in the potential. Because $f'(r_+)$ diverges for the extremal parameter $\tilde{\alpha}$, when $\tilde{\alpha}$ takes a near-extremal value, $f'(r)$ should exhibit a rapidly rising spike as $r$ approaches $r_+$. For non-extremal values of $\tilde{\alpha}$, these spikes are all finite.
Furthermore, we also find that the potential curves for the $\ell=0$ mode show that, when the black hole horizon radius is held fixed, the potential barrier becomes higher and wider as the parameter $\tilde{\alpha}$ increases. From subfigure.\ref{subfig:Vl_r_l_fixQM}, it can be seen that the $\ell$-dependent part of the potential also becomes higher and wider as $\tilde{\alpha}$ increases. Therefore, when the black hole horizon radius is held fixed, the potential barrier for any $\ell$-mode exhibits the same trend of becoming higher and wider with increasing $\tilde{\alpha}$. Consequently, for black holes with the same horizon radius (i.e., the same $M$ and $Q_M$), the ACS at general nonzero frequency should satisfy that the ACS for a larger $\tilde{\alpha}$ is smaller than that for a smaller $\tilde{\alpha}$. This feature is consistent with what is shown in FIG.\ref{FIG:ACS_Anomaly_QM}.

The ``dip'' in FIG.\ref{FIG:ACS} and \ref{FIG:ACS_Anomaly_QM} can be explained as follows. In a large $\tilde{\alpha}$, the potential barrier for the $\ell=0$ mode develops a peak. This peak causes the ACS to be consistently suppressed over a relatively small frequency interval. As the frequency increases, the effect of this peak gradually weakens until it becomes negligible. By examining whether a ``dip'' exists in the ACS curves in FIG.\ref{FIG:ACS} and \ref{FIG:ACS_Anomaly_QM} and comparing them with the corresponding potential barrier peaks in FIG.\ref{FIG:Vl(r)_l=0}, we find that the cases where the ACS curve exhibits a ``dip'' all correspond to situations where the height of the $\ell=0$ mode potential barrier exceeds that of the $\ell$-dependent part.

In this subsection, we have investigated the influence of different parameters on the ACS of conformal anomaly black holes. We find that the influence of the parameters on the ACS exhibits a high degree of consistency with their influence on the effective potential. The potential for the $\ell=0$ mode primarily governs the rate of change of the ACS with frequency near the low-frequency limit: the higher and wider the potential barrier, the slower the low-frequency ACS varies. The $\ell$-dependent part of the potential mainly governs the ACS at high frequencies: the higher and wider this part of the barrier, the smaller the high-frequency ACS. Furthermore, we have demonstrated that the ACS of these black holes, in both the low-frequency and high-frequency limits, are consistent with the universal theoretical results. In the next subsection, we will examine the influence of these parameters on the DCS of conformal anomaly black holes.

\subsection{Differential scattering cross section}
\begin{figure}[htbp]
\centering
\begin{subfigure}[b]{0.32\textwidth}
\centering
\includegraphics[width=\textwidth]{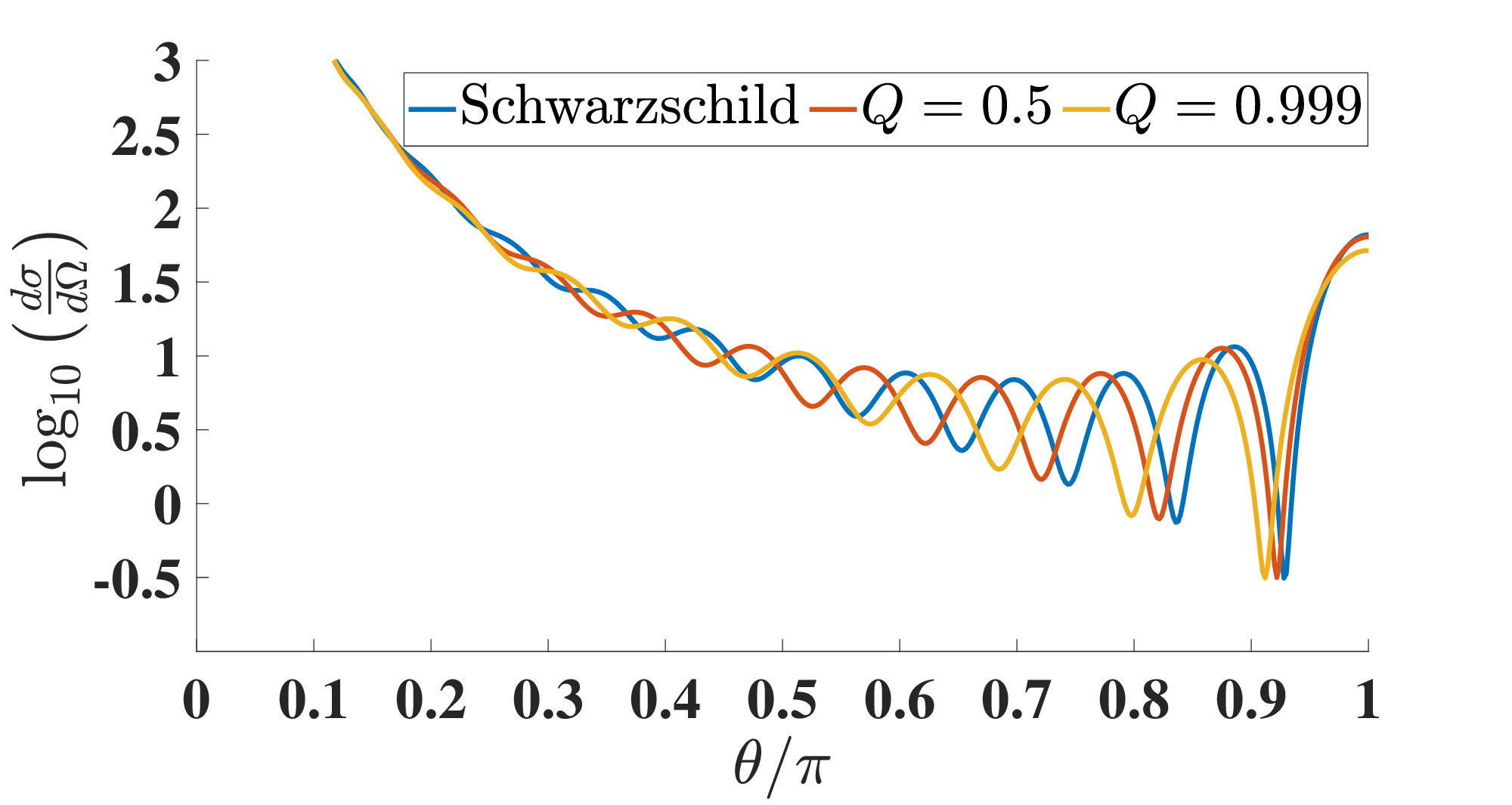}
\caption{}
\label{subfig:DCS_RN_k_2}
\end{subfigure}
\hspace{0\textwidth}  
\begin{subfigure}[b]{0.32\textwidth}
\centering
\includegraphics[width=\textwidth]{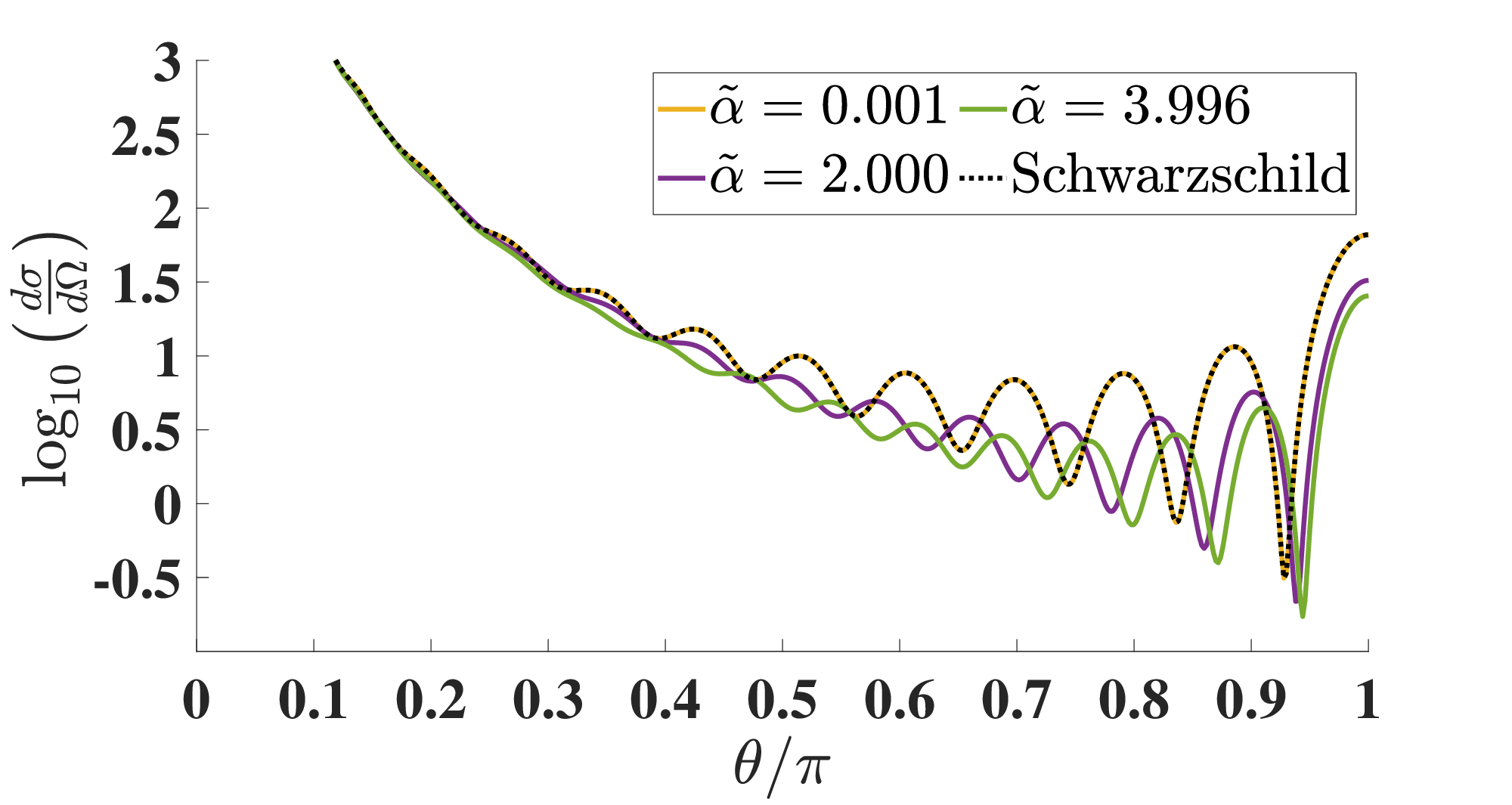}
\caption{}
\label{subfig:DCS_Anomaly_k_2_Q_0}
\end{subfigure}
\hspace{0\textwidth}  
\begin{subfigure}[b]{0.32\textwidth}
\centering
\includegraphics[width=\textwidth]{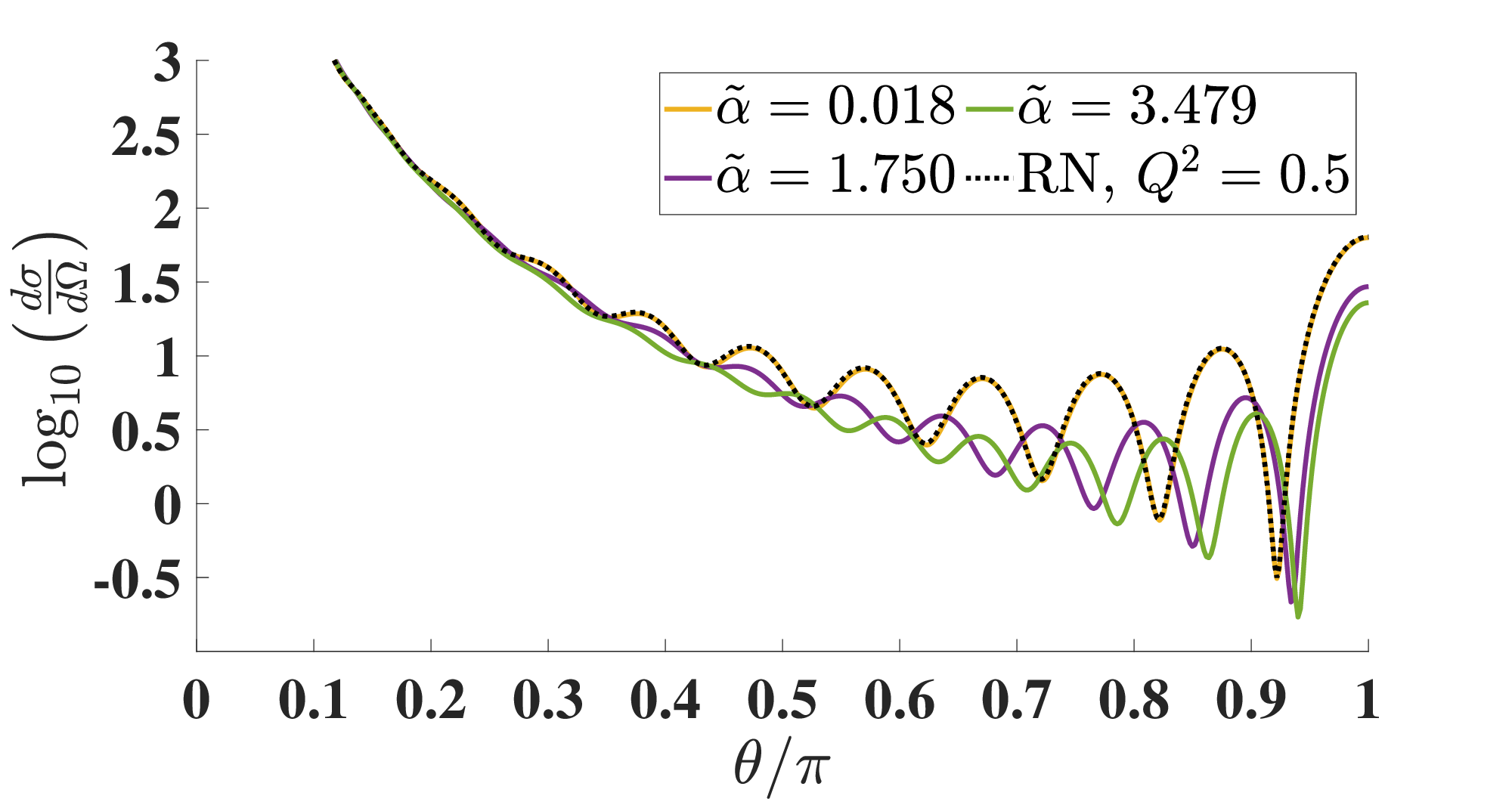}
\caption{}
\label{subfig:DCS_Anomaly_k_2_Q_0.5}
\end{subfigure}

\vspace{0\textwidth}  

\begin{subfigure}[b]{0.32\textwidth}
\centering
\includegraphics[width=\textwidth]{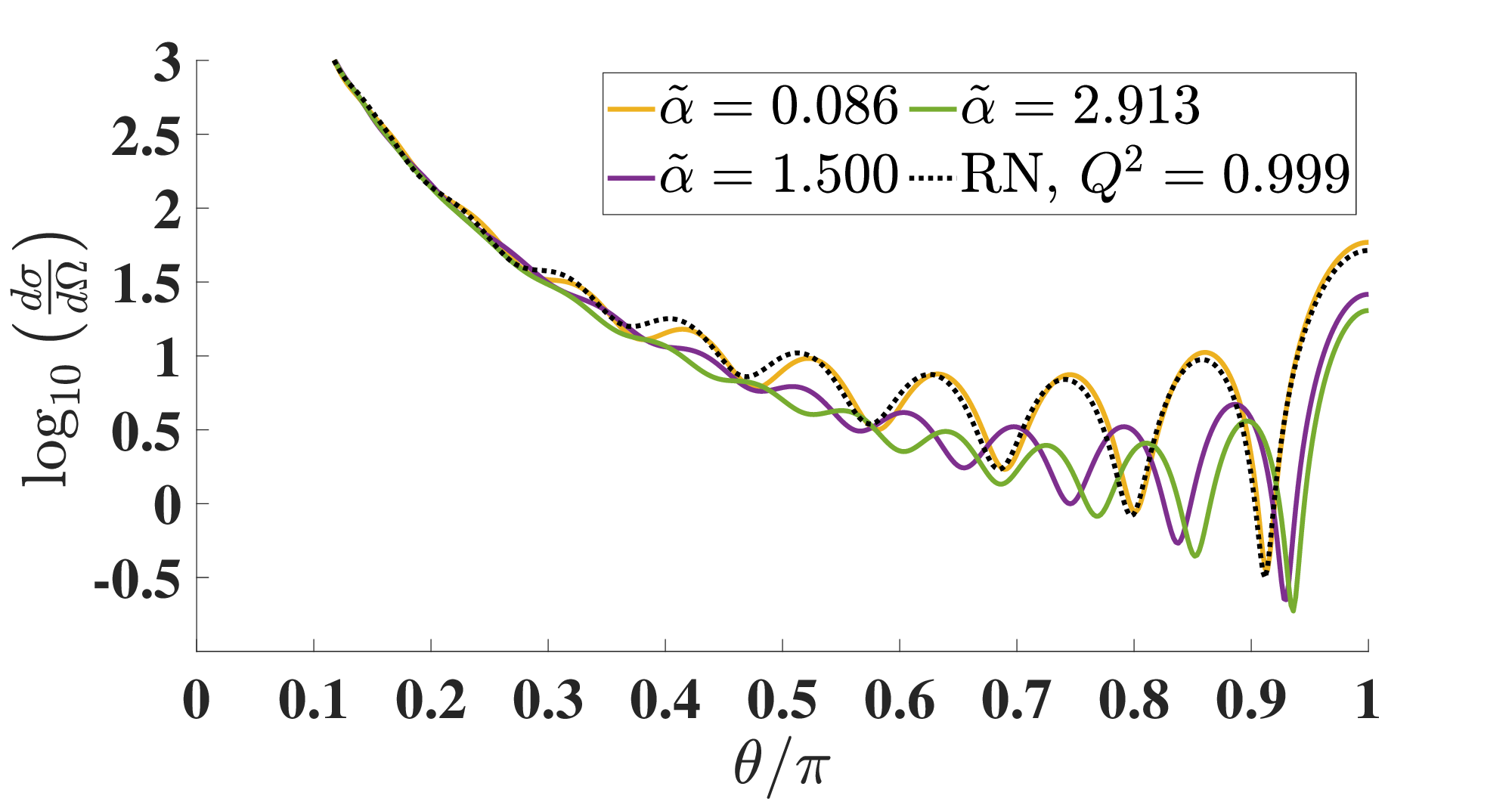}
\caption{}
\label{subfig:DCS_Anomaly_k_2_Q_0.999}
\end{subfigure}
\hspace{0\textwidth}  
\begin{subfigure}[b]{0.32\textwidth}
\centering
\includegraphics[width=\textwidth]{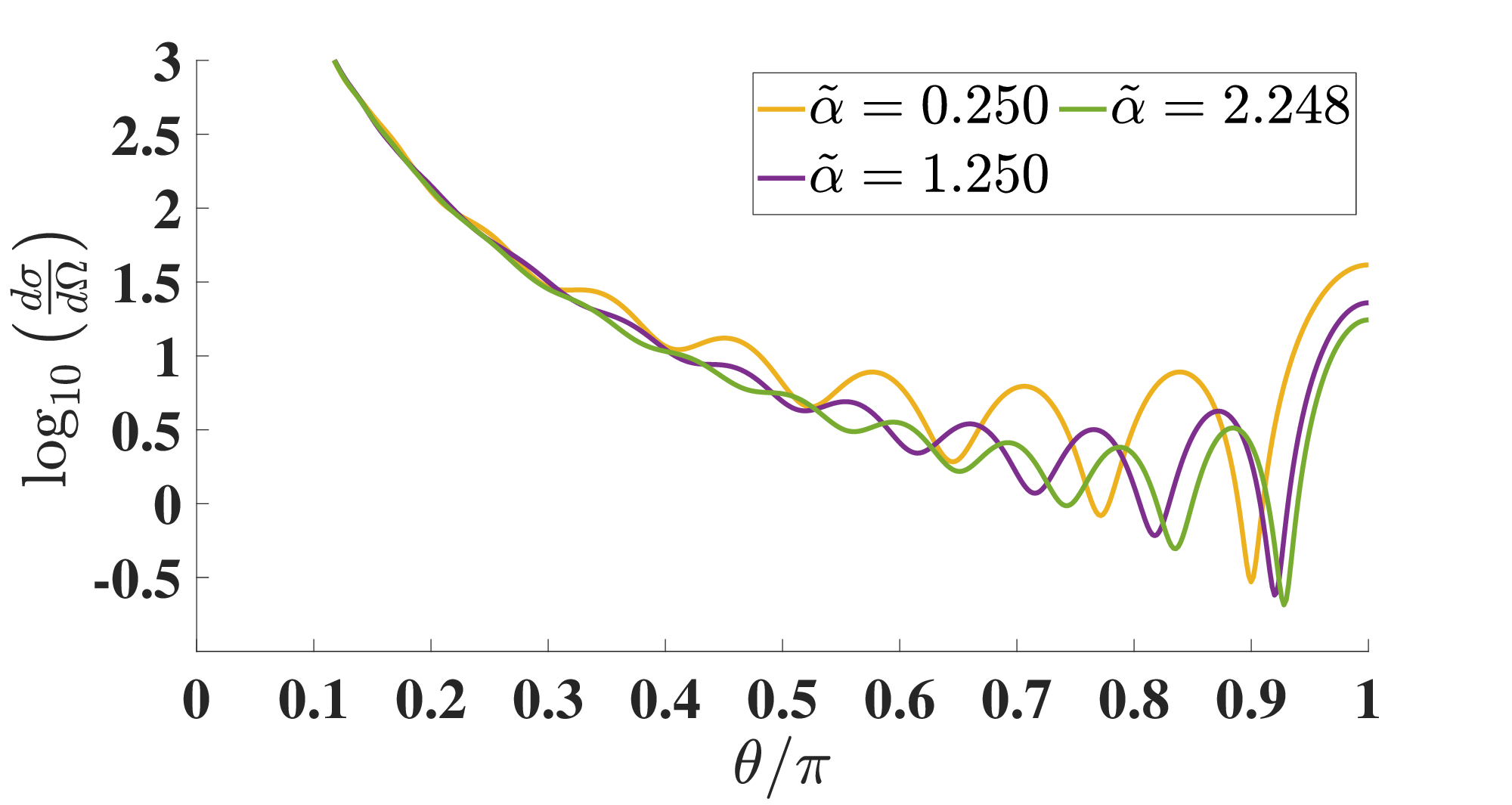}
\caption{}
\label{subfig:DCS_Anomaly_k_2_Q_1.5}
\end{subfigure}
\hspace{0\textwidth}  
\begin{subfigure}[b]{0.32\textwidth}
\centering
\includegraphics[width=\textwidth]{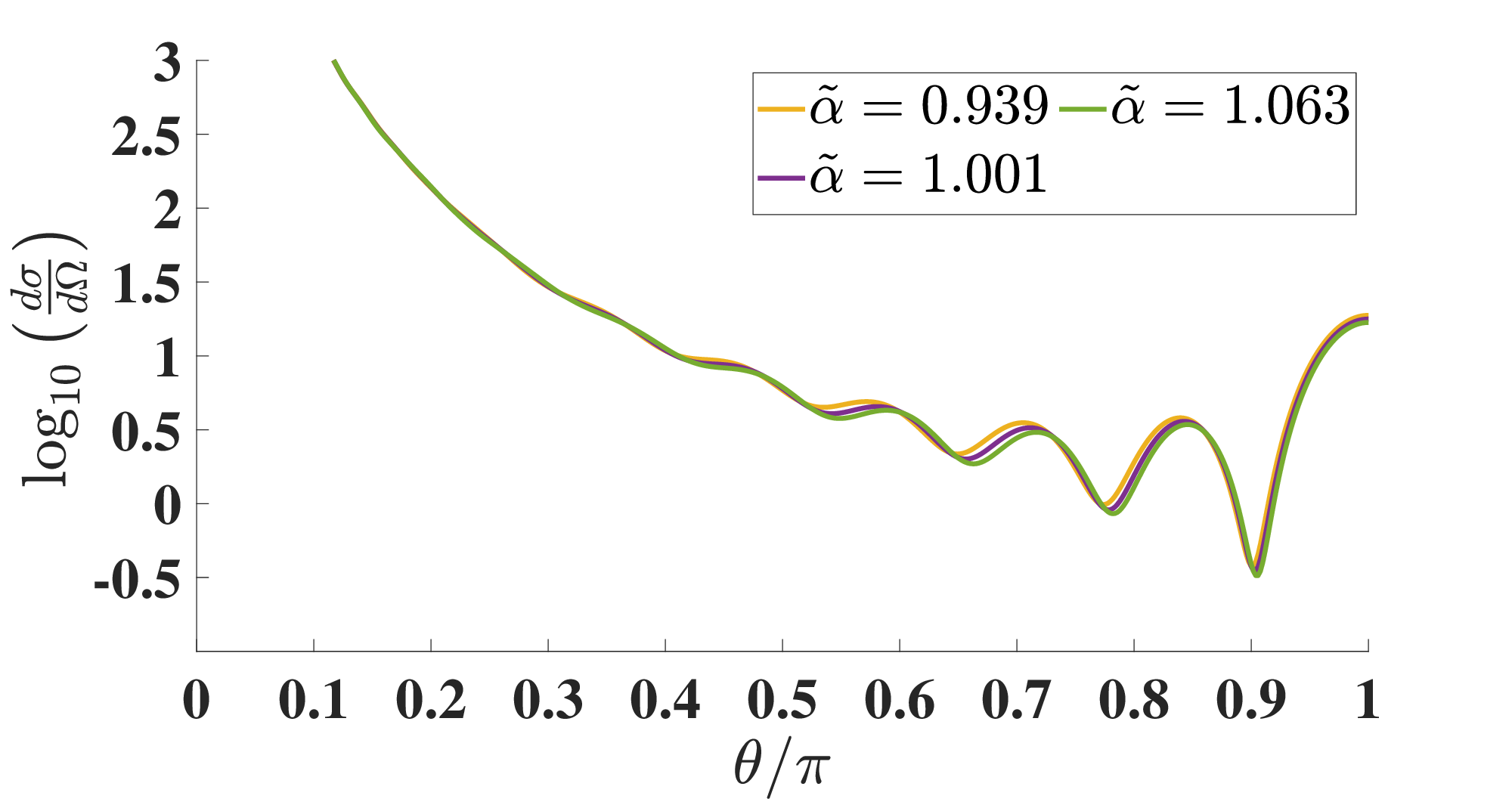}
\caption{}
\label{subfig:DCS_Anomaly_k_2_Q_1.998}
\end{subfigure}
\caption{Differential scattering cross sections of RN and conformal anomaly black holes. This figure shows the DCS of the black holes for scalar waves as a function of the $\theta$. In this figure, we take $k = 2$. The choice of metric, parameter $\tilde{\alpha}$, and their ordering in each subfigure are identical to those in the corresponding subfigure of FIG.\ref{FIG:ACS}.}
\label{FIG:DCS}
\end{figure}

\begin{figure}[htbp]
\centering
\begin{subfigure}[b]{0.32\textwidth}
\centering
\includegraphics[width=\textwidth]{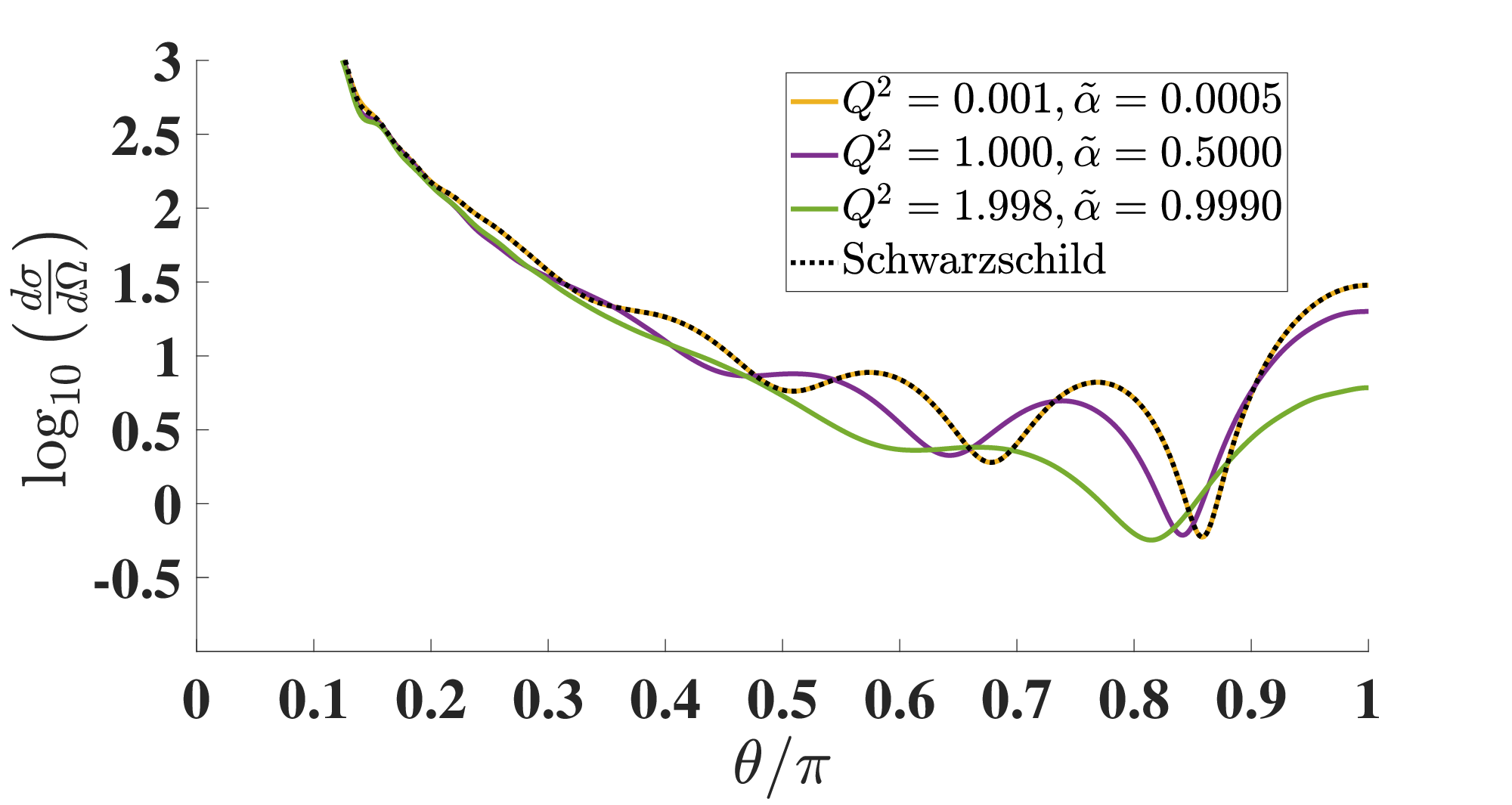}
\caption{}
\label{subfig:DCS_Anomaly_k_1_QM_0}
\end{subfigure}
\hspace{0\textwidth}  
\begin{subfigure}[b]{0.32\textwidth}
\centering
\includegraphics[width=\textwidth]{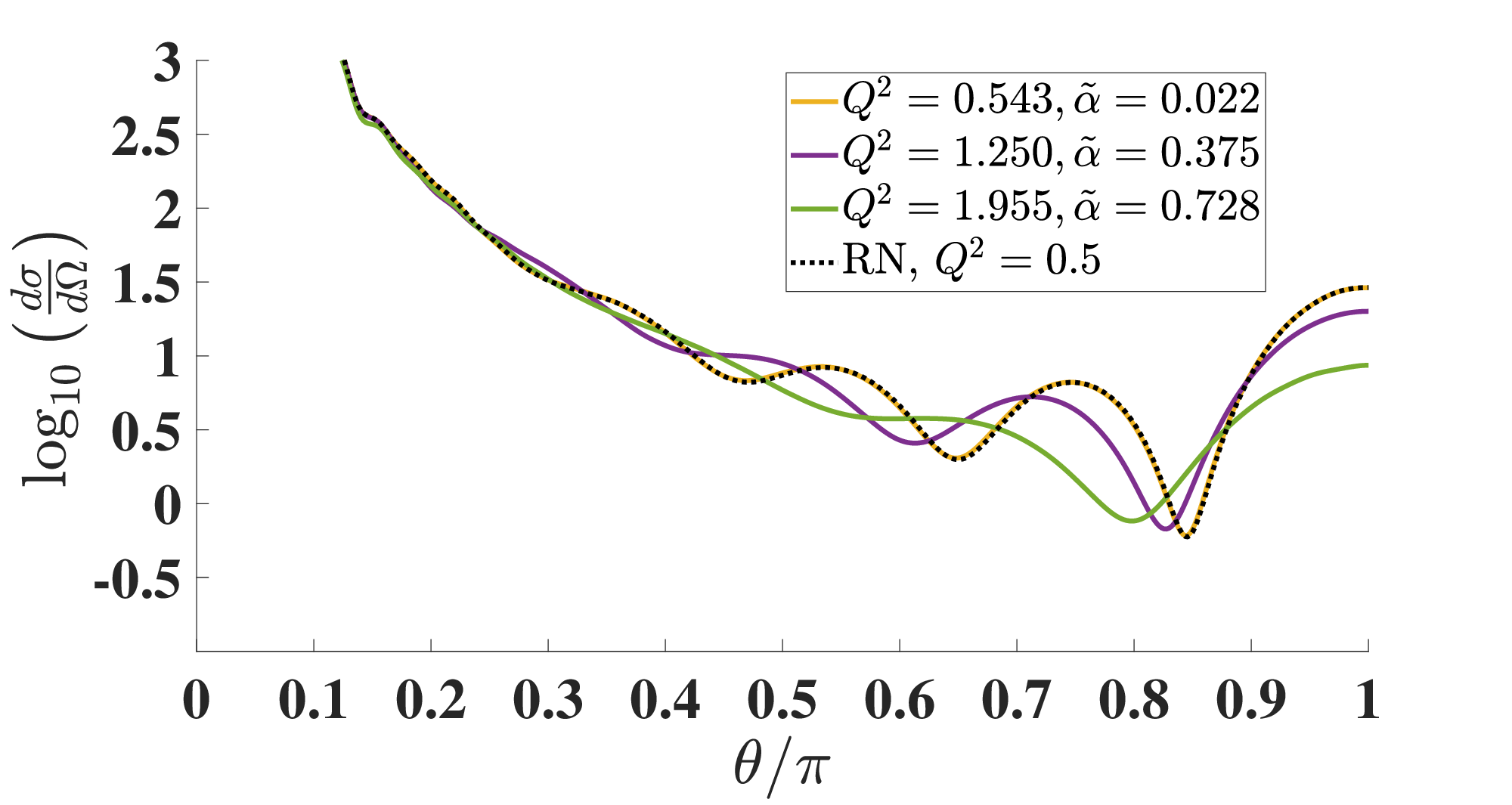}
\caption{}
\label{subfig:DCS_Anomaly_k_1_QM_0.5}
\end{subfigure}
\hspace{0\textwidth}  
\begin{subfigure}[b]{0.32\textwidth}
\centering
\includegraphics[width=\textwidth]{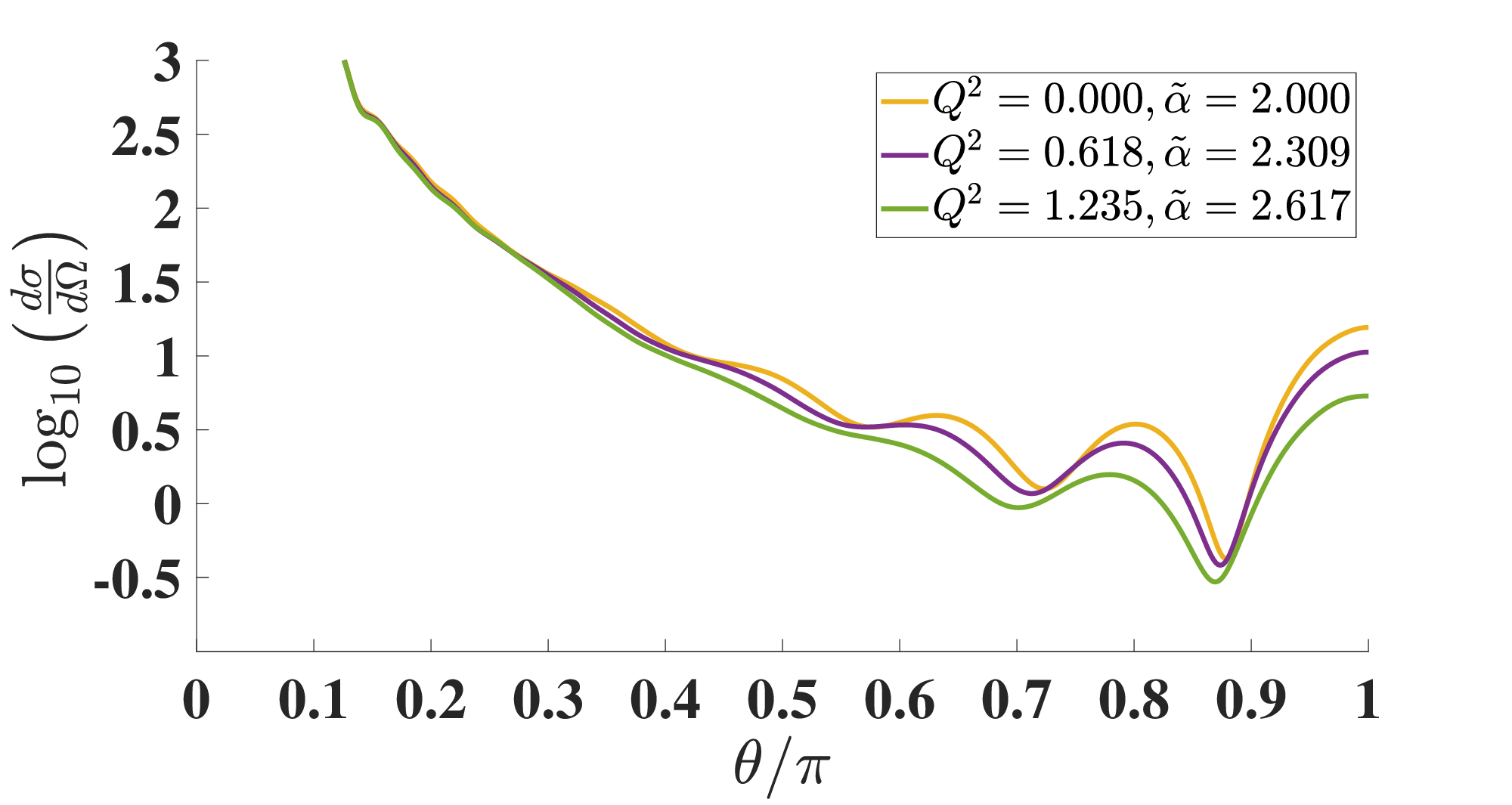}
\caption{}
\label{subfig:DCS_Anomaly_k_1_QM_-4}
\end{subfigure}

\vspace{0\textwidth}  

\begin{subfigure}[b]{0.32\textwidth}
\centering
\includegraphics[width=\textwidth]{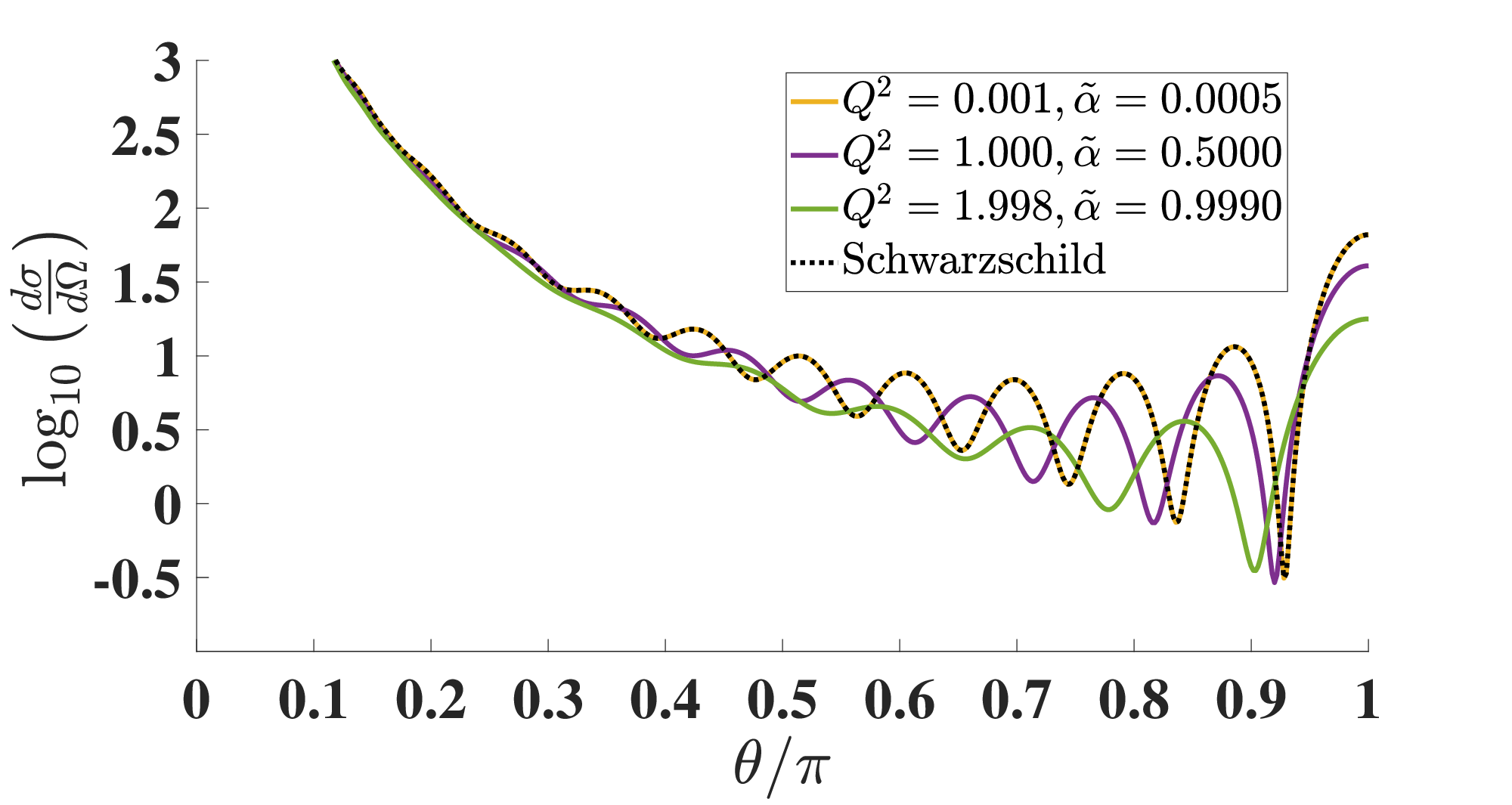}
\caption{}
\label{subfig:DCS_Anomaly_k_2_QM_0}
\end{subfigure}
\hspace{0\textwidth}  
\begin{subfigure}[b]{0.32\textwidth}
\centering
\includegraphics[width=\textwidth]{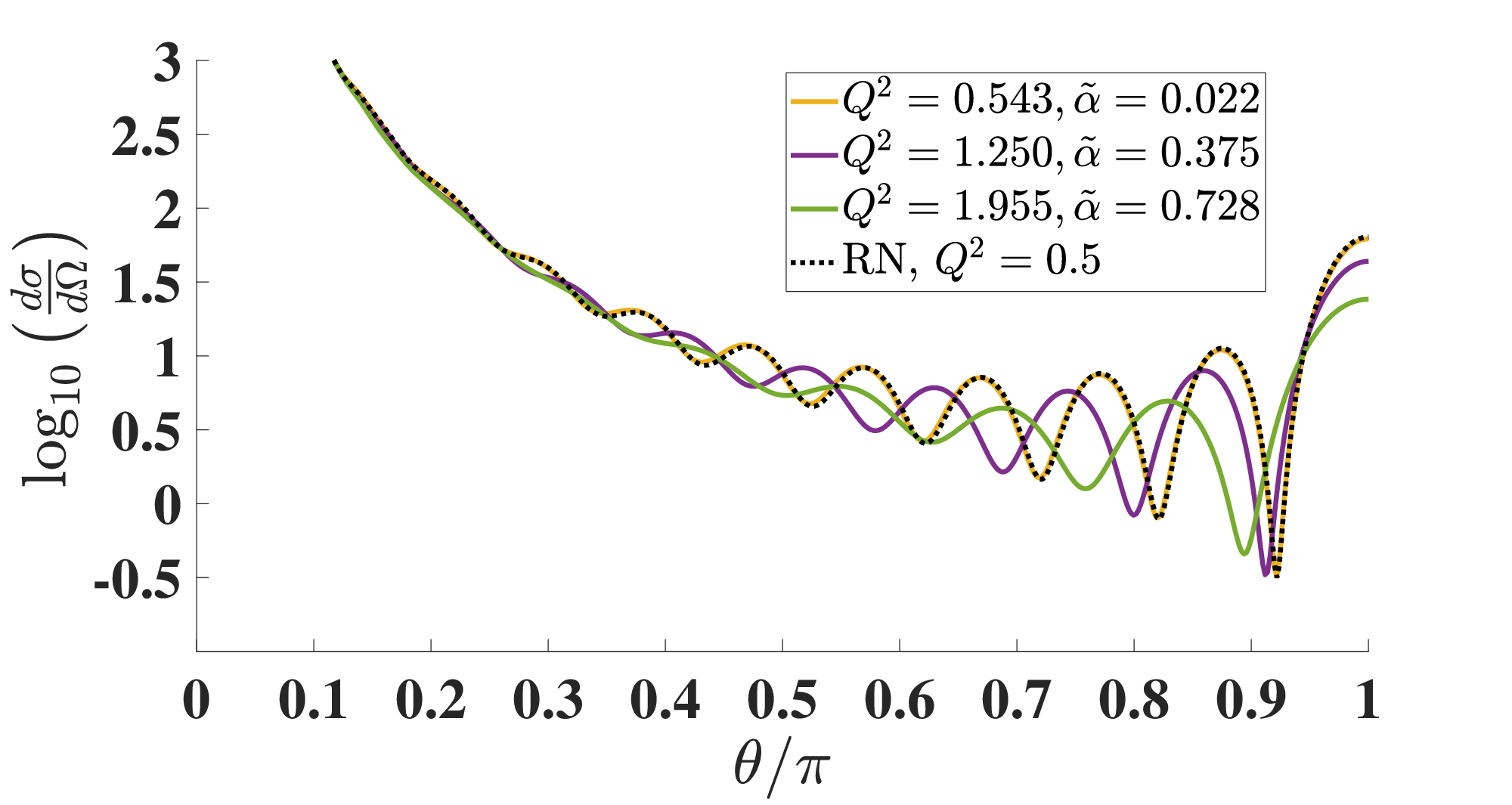}
\caption{}
\label{subfig:DCS_Anomaly_k_2_QM_0.5}
\end{subfigure}
\hspace{0\textwidth}  
\begin{subfigure}[b]{0.32\textwidth}
\centering
\includegraphics[width=\textwidth]{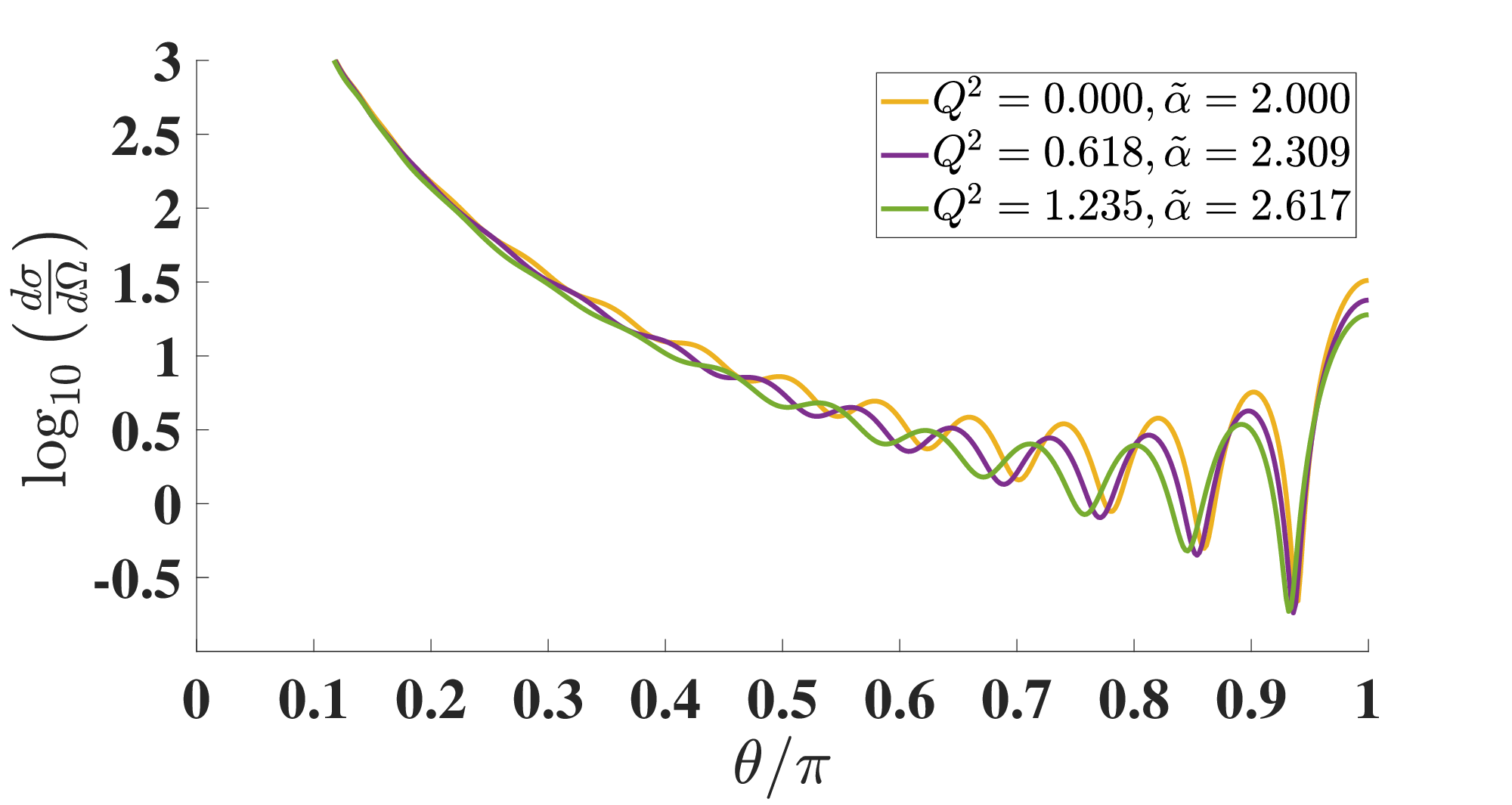}
\caption{}
\label{subfig:DCS_Anomaly_k_2_QM_-4}
\end{subfigure}
\caption{Differential scattering cross sections of RN and conformal anomaly black holes with fixed horizon area. The metric and parameter $\tilde{\alpha}$ in each subfigure of a given row correspond to those in FIG. \ref{FIG:ACS_Anomaly_QM}. For the first row, we take $k = 1$; for the second row, we take $k = 2$.}
\label{FIG:DCS_Anomaly_QM}
\end{figure}

In this subsection, we have calculated and analyzed the DCS for RN black holes and conformal anomaly black holes with the same parameters ($\tilde{\alpha}$ and $Q$) as those for the ACS in the previous subsection, using the SRM introduced in Sec.\ref{sec:2} to accelerate the convergence of the angular factor $f(\theta)$. FIG.\ref{FIG:DCS} illustrates the effect of charge on the DCS of the RN black hole, as well as the DCS of conformal anomaly black holes with the same charge but different values of the parameter $\tilde{\alpha}$. FIG.\ref{FIG:DCS_Anomaly_QM} presents the effect on the DCS when the parameter $\tilde{\alpha}$ and the charge $Q$ vary jointly, while the parameter $Q_M$(i.e., the horizon radius) is held fixed. Regarding FIG.\ref{FIG:DCS} and FIG.\ref{FIG:DCS_Anomaly_QM}, it should be noted that the DCS obtained using the SRM is valid only in the large-angle region, which is precisely suitable for investigating glory scattering.

From FIG.\ref{FIG:DCS}, it can be seen that pronounced glory scattering is present in all six cases. By examining subfigure.\ref{subfig:DCS_RN_k_2}, we find that the width of the glory peak broadens as the charge $Q$ increases, while its height decreases with increasing $Q$. 
By examining subfigures.\ref{subfig:DCS_Anomaly_k_2_Q_0}, \ref{subfig:DCS_Anomaly_k_2_Q_0.5}, and \ref{subfig:DCS_Anomaly_k_2_Q_0.999}, we find that the DCS of the conformal anomaly black hole with a small parameter $\tilde{\alpha}$ can also be regarded as the DCS of the RN black hole with an additional perturbation. This feature is similar to that observed for the ACS. 
From subfigures.\ref{subfig:DCS_Anomaly_k_2_Q_0}--\ref{subfig:DCS_Anomaly_k_2_Q_1.998}, we find that for the DCS of conformal anomaly black holes with the same charge $Q$, the width of the glory peak narrows as $\tilde{\alpha}$ increases, while its height decreases with increasing $\tilde{\alpha}$. The effect of the parameter $\tilde{\alpha}$ on the DCS is no longer simply the same as or opposite to that of the charge $Q$. 

FIG.\ref{FIG:DCS_Anomaly_QM} displays the effect on the DCS of conformal anomaly black holes when the parameter $\tilde{\alpha}$ and the charge $Q$ vary jointly, while the horizon area is held fixed. The scalar wave frequency is taken as $k = 1$ for the first three subfigures and $k = 2$ for the last three subfigures. It can be seen from the subfigures.\ref{subfig:DCS_Anomaly_k_1_QM_0}, \ref{subfig:DCS_Anomaly_k_1_QM_0.5}, \ref{subfig:DCS_Anomaly_k_2_QM_0}, and \ref{subfig:DCS_Anomaly_k_2_QM_0.5} that the DCS of the conformal anomaly black hole with a small $\tilde{\alpha}$ can still be regarded as the DCS of the RN black hole plus a perturbation. 
By comparing subfigures \ref{subfig:DCS_Anomaly_k_1_QM_0} and \ref{subfig:DCS_Anomaly_k_2_QM_0}, \ref{subfig:DCS_Anomaly_k_1_QM_0.5} and \ref{subfig:DCS_Anomaly_k_2_QM_0.5}, and \ref{subfig:DCS_Anomaly_k_1_QM_-4} and \ref{subfig:DCS_Anomaly_k_2_QM_-4}—each pair sharing the same parameters except for the frequency—it can be found that the width of the glory peak is proportional to the scalar wave frequency $k$. Moreover, the height of the peak increases with frequency. 
By examining the six subfigures individually, we find that, when both the scalar wave frequency and the black hole horizon area are held fixed, the width of the glory peak broadens as $\tilde{\alpha}$ and $Q$ jointly increase, while the height of the glory peak decreases. This characteristic is the same as the effect of charge on the DCS of the RN black hole. 

Combining the analysis of FIG.\ref{FIG:DCS} and FIG.\ref{FIG:DCS_Anomaly_QM}, we find that for the RN black hole, the width of the glory peak in the DCS broadens as $Q$ increases, while its height decreases with increasing $Q$; for the conformal anomaly black hole, the width of the glory peak narrows as $\tilde{\alpha}$ increases, while its height decreases with increasing $\tilde{\alpha}$. The effect of $Q$ on the glory peak of the conformal anomaly black hole is the same as that on the glory peak of the RN black hole. Further incorporating the analysis of the potential barrier, when $\tilde{\alpha}$ increases to its extremal value, the potential barrier becomes infinitely height at the event horizon. Such an infinitely height potential barrier is analogous to that in Newtonian and Coulomb potentials. However, neither Newtonian nor Coulomb potential scattering exhibits glory scattering. Additionally, for small $\tilde{\alpha}$, the DCS of the RN black hole and that of the conformal anomaly black hole are nearly identical. Therefore, we consider that glory scattering arise from the scattering of scalar waves by a black hole within a finite region that excludes an immediate neighborhood of the outer event horizon.

\section{Conclusions} \label{sec:6}
A black hole is essentially the only object in the real universe that can block a gravitational wave, and wave-optics effects beyond the geometric-optics shadow remain largely unexplored for black hole solutions outside GR. Motivated by this, we have carried out a systematic study of the scattering of massless scalar waves by static conformal anomaly black holes. This background is a one-parameter generalization of the RN solution in which the anomaly correction is concentrated in the strong-field region, decaying as $O(\tilde{\alpha}M^{2}/r^{4})$, and in which the charge may exceed the RN bound, ranging over $Q^{2}<2M^{2}$. It therefore provides an ideal laboratory for asking how wave-optics observables respond to the near-horizon geometry rather than to the asymptotic charges alone.

The starting point of the analysis is a technical one, but it determines everything that follows. Since the PWS has to be truncated in any practical calculation, we first asked how large $\ell_{\max}$ must be for the wave field at a finite distance $r$ to be reconstructed faithfully. The effective potential provides the answer: its height grows as $\ell^{2}$, whereas its width is set almost entirely by $\ell$ and is insensitive both to the metric and to the parameters, so that high-$\ell$ modes are reflected before they can penetrate into the strong-field region and decay rapidly inside the barrier. This leads to the criterion $\ell_{\max}\gtrsim kr$, which we confirmed numerically by showing that the contribution of the omitted modes is negligible. The finite-distance waveform calculation is thereby a controlled approximation rather than an uncontrolled truncation.

With this criterion in hand, we computed the full scalar waveforms at finite distance. All the backgrounds we examined --- Schwarzschild, RN and conformal anomaly --- share the same qualitative wave-optics structure: on the negative $z$-axis the field reduces to the incident plane wave, while on the positive $z$-axis a distorted plane wave coexists with a pronounced scattered spherical wave, and a bright Poisson spot appears on the axis at $\theta=0$, surrounded by concentric alternating bright and dark rings in the paraxial region. The resemblance of this pattern to optical diffraction by a small aperture is striking, and it shows that the coherent wave nature of the field survives the presence of an absorbing horizon. The geometric-optics shadow is thus not the whole story: the wave does bend around the black hole, and the Poisson spot is precisely the wave-optics counterpart of the shadow, as well as the sharpest theoretical evidence that a black hole cannot be treated as a merely absorbing disk.

What makes the Poisson spot more than a curiosity, however, is that its on-axis intensity carries information about where the wave has been. Following $|\psi|^{2}$ along the axis as a function of the radial coordinate, we find that it is strongly sensitive to the spacetime geometry in the immediate vicinity of the horizon and essentially blind to it far away: close to $r_{+}$, the curves for a conformal anomaly black hole and for the RN black hole with the same $M$ and $Q$ differ markedly, whereas in the far field they converge. The origin of this behavior is transparent. Near the horizon the tortoise coordinate behaves as $r_{\ast}\simeq \ln (r-r_{+})/f'(r_{+})$, so that different values of $f'(r_{+})$ map the same radial interval onto different phase increments and hence compress and shift the on-axis interference fringes by different amounts; conversely, because the anomaly correction decays as $O(\tilde{\alpha}M^{2}/r^{4})$, the two intensity curves must coincide asymptotically. The Poisson-spot intensity therefore acts as a genuine near-horizon probe, complementary to asymptotic observables, which resolve only $M$ and $Q$. To our knowledge, this is the first diffraction observable that isolates the near-horizon metric function in this way.

The cross sections exhibit an analogous structure once the effective potential is decomposed into its $\ell=0$ part and its $\ell$-dependent part, because these two parts control two different frequency regimes. For the ACS, the $\ell=0$ part governs how fast the cross section departs from the universal low-frequency limit $\sigma_{\text{abs}}\to A_{H}$, and it produces a characteristic downward ``dip'' whenever it develops a peak higher than the $\ell$-dependent part; the $\ell$-dependent part, in turn, governs the magnitude of the ACS at high frequency, where the universal geodesic capture result $\sigma_{\text{abs}}\to \pi r_{c}^{2}/f(r_{c})$ is recovered. A further advantage of the conformal anomaly background is that its horizon radius is determined by $Q_{M}^{2}=Q^{2}-2\tilde{\alpha}$ rather than by $\tilde{\alpha}$ alone, so that black holes with different values of $\tilde{\alpha}$ can be compared at fixed horizon area. Such a comparison disentangles the two effects cleanly: the charge and the anomaly parameter act in opposite directions on the ACS, with the charge dominating when the two are increased jointly through $Q^{2}=Q_{M}^{2}+2\tilde{\alpha}$.

For the DCS, we employed the SRM to accelerate the convergence of the angular factor series, which is what makes the glory peak accessible in the first place. Increasing $Q$ broadens and lowers this peak, whereas increasing $\tilde{\alpha}$ narrows and lowers it. Combined with the behavior of the potential barrier, these results allow us to localize the origin of glory scattering. In the extremal limit $\tilde{\alpha}\to\tilde{\alpha}_{\max}$ the barrier becomes infinitely high at the horizon --- a configuration akin to the Newtonian and Coulomb potentials, which nevertheless exhibit no glory --- while for small $\tilde{\alpha}$ the DCS reduces continuously to that of the RN black hole. Glory scattering must therefore be generated by the scattering of the wave in a finite region that excludes a small neighborhood of the outer event horizon, rather than by the near-horizon region itself.

Taken together, these results establish the wave-optics channel --- diffraction fringes and the Poisson spot --- as a genuine complement to the geometric-optics shadow and to the quasinormal-mode spectrum for characterizing black hole spacetimes beyond GR. They also show that, in such spacetimes, the near-horizon geometry leaves an imprint on scattering observables that is qualitatively distinct from the imprint of the asymptotic charges, so that wave scattering can in principle discriminate between two black holes that are indistinguishable at infinity.

Several directions are open for future work. The finite-distance waveforms presented here are limited by the $\ell\gtrsim kr$ truncation to the strong-field region, $r\lesssim O(10^{2})M$, and should be regarded as a proof of principle rather than a directly observable prediction; extending them to observationally relevant distances requires either an asymptotic resummation of the PWS or a complex-angular-momentum (CAM) treatment, which would simultaneously give access to the small-angle DCS and to the on-axis Poisson-spot intensity at large $r$. Beyond that, the analysis should be extended to electromagnetic and gravitational perturbations. Since the radial equation retains the Regge-Wheeler form for any spin and the purely ingoing condition at the horizon is universal, we expect both the diffraction pattern and the near-horizon sensitivity identified here to persist qualitatively; quantifying them is a prerequisite for assessing whether the Poisson-spot signature can be imprinted on ringdown signals or on GWs lensed by a foreground black hole. Finally, it would be interesting to apply the same analysis to rotating and to other quantum-corrected black hole solutions, in order to test whether the near-horizon diffraction signature can discriminate among different departures from GR that share the same asymptotic charges.

\section*{Acknowledgements}
This work is supported by the National Natural Science Foundation of China (NSFC) under Grant nos. 12275106 and 12235019.

%

\end{document}